\documentclass[aps,prc,superscriptaddress,longbibliography, showpacs,floatfix,letterpaper,preprintnumbers,nofootinbib,notitlepage,twocolumn]{revtex4-2}
\usepackage{subcaption, caption}
\usepackage[colorlinks=true, linkcolor=blue, citecolor=blue, urlcolor=blue]{hyperref}
\usepackage{amssymb}
\usepackage[title]{appendix}
\usepackage{color}
\usepackage{graphicx} 
\usepackage{bm}
\usepackage{amsmath,mathrsfs}
\usepackage{braket}
\usepackage{tabularx}
\usepackage{makecell}
\usepackage{xfrac}
\usepackage{soul}
\usepackage{lineno}
\usepackage{slashed}
\usepackage{appendix}
\usepackage{tabularx,enumitem,paralist}
\usepackage{cancel} 
\usepackage{xcolor} 
\usepackage{booktabs}
\usepackage{float} 
\usepackage{orcidlink}
\usepackage{subcaption}
\newcommand{\polylog}[1]{\operatorname{Li}_{#1}}
\DeclareMathOperator{\arcsinh}{arcsinh}

\DeclareMathOperator{\sgn}{sgn}

\begin{document}
\title{Jet transport coefficients associated with fermionic two-point correlators in a weakly-coupled plasma}

\author{Shay Duddy\,\orcidlink{0009-0005-0903-9627}}
\affiliation{Department of Physics, University of Regina, Regina, Saskatchewan S4S 0A2, Canada}

\author{Lukas Opitz\,\orcidlink{0009-0008-0131-4307}}
\affiliation{Department of Physics, University of Regina, Regina, Saskatchewan S4S 0A2, Canada}

\author{Amit Kumar\,\orcidlink{0000-0003-1099-6422}}
\email[Corresponding author: ]{amit.kumar@uregina.ca}
\affiliation{Department of Physics, University of Regina, Regina, Saskatchewan S4S 0A2, Canada}

\author{Gojko Vujanovic\,\orcidlink{0000-0001-5397-6662}}
\email[Corresponding author: ]{gojko.vujanovic@uregina.ca}
\affiliation{Department of Physics, University of Regina, Regina, Saskatchewan S4S 0A2, Canada}

\date{\today}

\begin{abstract}
A new set of jet-medium transport coefficients stemming from jet-medium exchanges involving Glauber quarks encoded in $\hat{\mathcal{F}}_i$ [\href{https://journals.aps.org/prc/10.1103/PhysRevC.111.054913}{Phys. Rev. C 111, 054913 (2025)}, \href{https://journals.aps.org/prc/abstract/10.1103/qsb9-qbhb}{Phys. Rev. C 113, 055207 (2026)}] are obtained by computing the tree-level leading-order $2\to2$ scattering rates and their moments using the approach developed in Refs.~[\href{https://arxiv.org/pdf/2608.17160}{arxiv\:2608.17160}, \href{https://arxiv.org/pdf/2608.17161}{arxiv\:2608.17161}]. Sizeable deviations away from the leading logarithmic dependence of jet-medium transport coefficients and scattering rate are observed here, as previously mentioned in Refs.~[\href{https://arxiv.org/pdf/2608.17160}{arxiv\:2608.17160}, \href{https://arxiv.org/pdf/2608.17161}{arxiv\:2608.17161}]. 
A closed-form expression for $\hat{\mathcal{F}}_i$ accurate to $\lesssim 10$\% or better is obtained, enabling our approach to be used within Monte Carlo simulations of jet-medium interactions. Monte Carlo simulations incorporating $\hat{\mathcal{F}}_i$ are sensitive to flavor hydrodynamization dynamics as the medium created in nucleus-nucleus collisions transitions from the early-time Glasma dynamics to the quark-gluon plasma fluid.
\end{abstract}

\maketitle
\section{Introduction}

High-energy photons, jets, and heavy quarks are among the most important probes of the strongly interacting hot QCD matter produced in ultra-relativistic heavy-ion collisions. Their production and subsequent in-medium modification are sensitive to the entire space-time evolution of the nuclear medium, from the primordial pre-equilibrium Glasma to the hydrodynamical quark-gluon plasma (QGP). The interactions of these energetic probes with the medium are conventionally characterized by a set of jet-medium transport coefficients, which govern the longitudinal drag, momentum diffusion, and transverse momentum broadening experienced by propagating partons.

Previous studies have investigated jet transport coefficients governing parton momentum drag and diffusion using a wide range of theoretical approaches. Within perturbative thermal QCD and effective kinetic theory, leading-logarithmic expressions for drag and diffusion coefficients have been derived in e.g. Ref.~\cite{Moore:2004tg}. The collisional energy loss $dE/dx$ for both light and heavy quarks has been extensively studied in the high-energy limit \cite{Bjorken1982,THOMA1991128,THOMA1991491,Braaten:1991we,Peigne:2007sd,Peigne:2008nd,Peng:2024zvf}. In addition, the transverse momentum broadening coefficient has been calculated using hard-thermal-loop (HTL) resummation~\cite{Caron-Huot:2010qjx}, lattice QCD~\cite{Kumar:2020wvb,Laine:2013apa,Panero:2013pla,Majumder:2012sh,Benzke:2012sz}, holographic AdS/CFT methods~\cite{Liu:2006ug,Lin:2006au,Avramis:2006ip}, and phenomenological extractions from experimental data~\cite{JET:2013cls,Ehlers:2024miy,JETSCAPE:2021ehl}.

In deeply inelastic scattering off a large nucleus, parton energy loss associated with in-medium quark-quark correlators was studied in Ref.~\cite{Schafer:2007xh}. That work systematically considered all possible quark-quark and quark-antiquark scattering processes, with the interaction with the medium modeled through quark exchange characterized by the momentum scaling $k^+ \gg k^-, k_{\perp}$. That calculation was carried out at $\mathcal{O}(\alpha^2_s)$ and included coherent multiple scattering effects, part of the Landau-Pomeranchuk-Migdal (LPM) interference. More recently \cite{Kumar:2025egh,Kumar:2025asj}, these calculations have been extended to all possible medium-induced single-scattering kernels in the Glauber limit, characterized by the medium parton having $k_{\perp} \gg k^+, k^- $. Studying jet-medium interactions in the Glauber momentum region $k_\perp > k^+, k^-$ provides a natural framework for describing the soft interactions between an energetic jet and the QCD medium, where transverse momentum transfer dominates over longitudinal momentum exchange. This kinematic regime is particularly relevant for jet modification at mid-rapidity in relativistic heavy-ion collisions, where high-$p_{\rm T}$ partons propagate through a medium characterized by momenta on the order of the medium temperature $T$. Interactions with Glauber quarks lead to new jet-medium transport coefficients $\hat{\mathcal{F}}_i$ \cite{Kumar:2025egh,Kumar:2025asj}, which are computed herein as moments of the tree-level leading-order $2\to2$ scattering rate.    

This paper explores the determination of jet transport coefficients relevant to the quark-to-gluon conversion processes, and is organized as follows. Section~\ref{sec:fhat} presents operator definitions of the jet transport coefficients in terms of fermion correlators. Section~\ref{sec:comp_fhat} introduces the definitions of the jet transport coefficients within the framework of $2\to 2$ thermal QCD scattering. Section~\ref{sec:rate} provides a detailed discussion of the calculation of $\hat{\mathcal{F}}_{(0)}$ and the differential scattering rate for $2\to2$ processes, including the methodology for performing the phase-space integrals. Section~\ref{sec:E_loss} outlines the calculation of the longitudinal energy loss coefficient in the quark-gluon conversion process. Section~\ref{sec:fhat_L1} gives a detailed derivation of the longitudinal drag transport coefficient $\hat{\mathcal{F}}_{(L,1)}$, while Section~\ref{sec:fhat_T2} computes the transverse momentum broadening coefficient $\hat{\mathcal{F}}_{(T,2)}$. Both Compton and quark-antiquark annihilation processes are considered. A discussion regarding the temperature and energy dependence of transport coefficients is presented in Section~\ref{sec:Fhat_w_T}, with a summary provided in Section~\ref{sec:outlook}.
\onecolumngrid 
\noindent
\begin{figure}[htbp]
\centering
\includegraphics[width=0.72\textwidth]{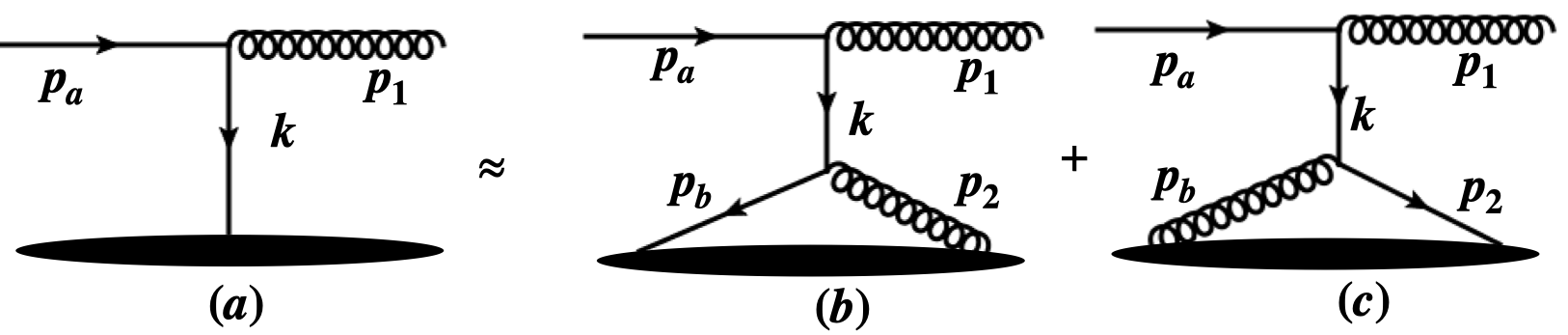}
\caption{Diagrams for quark-to-gluon conversion process in $2\to 2$ thermal QCD scattering.}
\label{fig:fhat_diagrams}
\end{figure}
\twocolumngrid
%
\section{Jet-medium transport coefficients}
\label{sec:fhat}
This section presents jet-medium transport coefficients associated with quark-to-gluon conversion processes. A leading-order process is considered in which an energetic quark traversing a thermal medium exchanges a quark, resulting in the conversion of the quark into a gluon, as illustrated in Fig.~\ref{fig:fhat_diagrams}(a). Therein, the incoming quark carries momentum $p_a$, and the exchanged quark carries momentum $k=(k^+,k^-,\pmb{k}_{\perp})$. The light-cone momenta are defined as $k^+=(k^0+k^z)/\sqrt{2}$ and  $k^-=(k^0-k^z)/\sqrt{2}$. In general, the higher-twist calculations~\cite{Kumar:2025egh, Kumar:2025asj} are formulated in a frame where the jet parton propagates along the negative $z$ direction. In this setup, the $k^+$ component of the exchanged parton momentum is integrated out through the kinematic constraints, leaving $k^-$ and $\pmb{k}_\perp$ as the relevant nonintegrated momentum variables. With these variables identified, jet transport coefficients are defined as follows. The jet transport coefficient characterizing the change in the longitudinal momentum per unit length is defined as $\hat{\mathcal{F}}_{(L,1)}=\langle k^-\rangle/L^-$, where $L^-$ denotes the light-cone length of the medium. Similarly, the transport coefficient governing transverse momentum broadening is defined as $\hat{\mathcal{F}}_{(T,2)}=\langle k_\perp^2\rangle/L^-$. 
For the process illustrated in Fig.~\ref{fig:fhat_diagrams}(a), these transport coefficients can be expressed in terms of fermionic correlators in the light-cone coordinates as
\begin{align}
 \hat{ \mathcal{F}}_{(0)} & =  \frac{C_F \sqrt{2}\, g^{2}_{s}}{q^-} \int d y^{-} d^2  y_{\perp}  \frac{d^2 k_{\perp}}{\left(2\pi\right)^2}  e^{-i \frac{\pmb{k}^2_{\perp}}{2q^-} y^{-}}  e^{i\pmb{k}_{\perp}\cdot \pmb{y}_{\perp} }  \nonumber \\
& \times \left\langle M \left|\bar{\psi}\left(0\right)\frac{\gamma^+}{4}\psi\left(y^-, \pmb{y}_{\perp}\right) \right| M \right\rangle, \label{eq:F_hat_F0}\\
\hat{ \mathcal{F}}_{(L,1)} & =  \frac{C_F \sqrt{2}\, g^{2}_{s}}{q^-}  \int d y^{-} d^2  y_{\perp}  \frac{d^2 k_{\perp}}{\left(2\pi\right)^2}  e^{-i \frac{\pmb{k}^2_{\perp}}{2q^-} y^{-}}  e^{i\pmb{k}_{\perp}\cdot \pmb{y}_{\perp} } \nonumber \\
&  \times  \left\langle M\left|i\left\{\partial^-\bar{\psi}\left(0\right)\right\}\frac{\gamma^+}{4}\psi\left(y^-, \pmb{y}_{\perp}\right) \right|M\right\rangle, \label{eq:F_hat_L1}
\end{align}
as well as
\begin{align}
\hat{ \mathcal{F}}_{(T,2)} & = \frac{C_F \sqrt{2}\, g^{2}_{s}}{q^-}  \int d y^{-} d^2  y_{\perp}  \frac{d^2 k_{\perp}}{\left(2\pi\right)^2}  e^{-i \frac{\pmb{k}^2_{\perp}}{2q^-} y^{-}}  e^{i\pmb{k}_{\perp}\cdot \pmb{y}_{\perp} } \nonumber \\
&  \times  \left\langle M\left|\left\{\partial_{\perp}\bar{\psi}\left(0\right)\right\}\frac{\gamma^+}{4}\left\{\partial_{\perp}\psi\left(y^-, \pmb{y}_{\perp}\right)\right\} \right|M\right\rangle,
\label{eq:F_hat_T2}
\end{align}
where $ \hat{ \mathcal{F}}_{(0)}$ is the rate of the conversion process illustrated in Fig.~\ref{fig:fhat_diagrams}(b), $q^-=(p^0_a-p^z_a)/\sqrt{2}$ is the light-cone energy of the incoming quark, $g_s$ is the strong coupling, and $C_F=4/3$.

The jet energy loss for a quark-initiated jet within the higher-twist (HT) formalism was derived in Ref.~\cite{Kumar:2025asj}. In this framework, the medium-induced single-scattering kernels are derived using a systematic twist expansion, in which the interactions with the nuclear medium are represented by Glauber quark and gluon fields. In particular, the quark-to-gluon conversion channel involves a Glauber-quark exchange and generates non-local two-point fermionic correlators \cite{Kumar:2025egh, Kumar:2025asj}. Through the collinear twist expansion, these correlators can be organized in terms of gradients with respect to the longitudinal momentum ($k^-$) and transverse momentum ($k_\perp$), allowing their corresponding moments to be identified with the jet-medium transport coefficients. We find that the resulting fermionic correlators presented in Ref.~\cite{Kumar:2025egh,Kumar:2025asj} are therefore directly related to the longitudinal drag and diffusion coefficients, with the overall normalization factor: $C_F\sqrt{2}/q^-$.

\section{Computing Transport Coefficients}
\label{sec:comp_fhat}
For a quark in the jet with momentum $p^\mu_a$ traversing the medium, following an exchange with an (anti)quark with momentum $k^\mu$, a gluon with momentum $p^\mu_a+k^\mu$ is generated. The expectation value of an observable $\hat{\mathcal{O}}$ in terms of the differential rate is given as 
\begin{align}
	\langle \hat{\mathcal{O}}\rangle &\equiv \int d^4k\frac{d^7R}{d^3p_a\,d^4k} \hat{\mathcal{O}}.
\end{align}
For $2\to 2$ scattering, the expectation value of $\hat{\mathcal{O}}$ is 
\begin{align}
	\langle \hat{\mathcal{O}} \rangle&=\frac{1}{2E_a\left(2\pi\right)^3}\int\frac{d^3p_b}{2E_b\left(2\pi\right)^3}\frac{d^3p_1}{2E_1\left(2\pi\right)^3}\frac{d^3p_2}{2E_2\left(2\pi\right)^3} \nonumber \\
    & \qquad\qquad\qquad \times   \left(2\pi\right)^4\delta^{(4)}\left(p_a+p_b-p_1-p_2\right)f(p_b) \nonumber \\
    & \qquad\qquad\qquad\times  [1\mp f(p_2)]\; \hat{\mathcal{O}} \; \overline{\lvert\mathcal{M}\rvert^2}, 
    \label{eq:Ohat_3p}
\end{align}
where the four-momentum $p^\mu_a=(E_a,\pmb{p}_a)$, $p^\mu_b=(E_b,\pmb{p}_b)$, $p^\mu_1=(E_1,\pmb{p}_1)$ and $p^\mu_2=(E_2,\pmb{p}_2)$ are labeled in Fig.~\ref{fig:fhat_diagrams}(b) and (c). Using the algebraic manipulations outlined in the Appendix A of Ref.~\cite{Opitz:2026sgz} and Chapter 14 of Ref.~\cite{Kapusta_Gale_2023}, Eq.~(\ref{eq:Ohat_3p}) is recast in terms of integrals over $E_b$, $E_2$, and Mandelstam variables $s$ and $t$ as follows
\begin{align}
	\langle \hat{\mathcal{O}} \rangle &=\frac{1}{16\left(2\pi\right)^7\lvert\pmb{p}_a\rvert E_a}\int dE_b\; f(p_b) \int dE_2 \left[1\mp f(p_2)\right]\times\nonumber \\
    &\qquad\times\int \;ds\;dt \frac{\hat{\mathcal{O}}\overline{\lvert\mathcal{M}\rvert^2}}{\sqrt{\Lambda(s,t)}}, \label{OHat} 
\end{align}
where $\Lambda(s,t)$, in the massless limit, is given by
\begin{align}
\Lambda(s,t)&=-\left(E_b-E_2\right)^2s^2-\left(E_a+E_b\right)^2t^2+st\left(s+t\right)\nonumber \\
&\quad-2\left(E_aE_2+E_bE_1\right)st.\label{eq:Lambda}
\end{align}
Thus, for the trivial case $\hat{\mathcal{O}}={\bf 1}$, the observable is just the scattering rate. The transverse momentum broadening coefficient $\hat{\mathcal{F}}_{T,2}$ of the jet uses $\hat{\mathcal{O}}=\lvert\pmb{k}_{\perp}\rvert^2$.  The energy loss coefficient $\langle \Delta E\rangle$, $\hat{\mathcal{O}}= k^0$, while the longitudinal drag coefficient $\hat{\mathcal{F}}_{L,1}$ employs $\hat{\mathcal{O}}=k^+$. Note that in the $2\to 2$ scattering setup, the jet quark is assumed to be travelling in the positive $z$-direction; therefore, the longitudinal drag coefficient $\hat{\mathcal{F}}_{L,1}$ is the $k^+$ moment of the rate, instead of being the $k^-$ moment.

\section{Scattering rate}
\label{sec:rate}

The differential rate of a relativistic $2\to2$ scattering process $a+b \to 1+2$, where an initial jet particle (indexed with $a$) interacts with an initial medium particle (index $b$), resulting in a jet particle (1) and a medium particle (2), is given by
\begin{align}
\frac{d^3R}{d^3p_a}&=\frac{1}{\left(2\pi\right)^3 2E_a}\int\frac{d^3p_b}{\left(2\pi\right)^32E_b}\frac{d^3p_1}{\left(2\pi\right)^32E_1}\frac{d^3p_2}{\left(2\pi\right)^32E_2} \nonumber \\ 
& \text{ } \times f_b(1\mp f_2)\overline{|\mathcal{M}|^2}\left(2\pi\right)^4\delta^{(4)}(p_a+p_b-p_1-p_2) \nonumber \\
&=\frac{1}{16\left(2\pi\right)^8E_a}\int\frac{d^3p_b\;d^3p_1\;d^3p_2}{E_bE_1E_2}f_b \left[1\mp f_2\right] \nonumber \\
& \text{ } \times \overline{|\mathcal{M}|^2}\delta^{(4)}(p_a+p_b-p_1-p_2),
\label{eq:scatt_rate}
\end{align}
where $\overline{|\mathcal{M}|^2}$ is the sum-averaged matrix element square of the process $a+b\to 1+2$, $\delta^{(4)}$ is the momentum-conserving Dirac delta function, and the distribution functions $f_i$ are those of thermalized medium particles, given by either the Bose-Einstein or the Fermi-Dirac statistics. For a thermally equilibrated medium with temperature $T$, and flow velocity $u^\mu$
\begin{align}
	f(p_a)&=\frac{1}{e^{(u\cdot p_{a}-\mu)/T}\pm 1}=f_a,
\end{align}
with $[1+f]$ Bose-enhancement and $[1-f]$ Pauli-blocking factors for the final-state medium particles. The scattering rate can be recast in terms of integrals over energies $E_{b}$, $E_2$, and Mandelstam variables $s$ and $t$, as follows
\begin{align}
	\frac{d^3R}{d^3p_a}&=\frac{1}{16\left(2\pi\right)^7|\pmb{p}_a|E_a}\int_0^\infty dE_b\, f_b\int_0^{E_a+E_b}dE_2 [1\pm f_2] \nonumber \\ 
    & \text{ }\text{ } \times \int_{t_{\min}}^{t_{\max}}dt\int_{s_-}^{s_+}ds\frac{\overline{|\mathcal{M}|^2}(s,t)}{\sqrt{\Lambda(s,t)}}.
\end{align}
where $\Lambda(s,t)$ is given in Eq.~(\ref{eq:Lambda}). The limits $s_{\pm}$ satisfy $\Lambda(s_{\pm},t)=0$, as $\Lambda(s,t)\geq 0$ is required. In general, the upper and lower limits of the $t$ integral are determined using kinematics and set to be $t_{\rm max}=0$, and $t_{\rm min}=-4\,{\rm min}(E_aE_1, E_bE_2)$. However, if the matrix element diverges at $t_{\rm max}=0$, such a divergence in a thermal system signals that additional in-medium physics in the infrared region should be accounted for, such as Hard Thermal Loops (HTL). The present study neglects HTL effects and simply sets an upper limit to $t$ being the asymptotic quark mass \cite{Ghiglieri:2015ala} $t_{\rm max}=-m^2_\infty$. Thus, $m^2_\infty$ is just a cutoff parameter. 

To illustrate how $t$ and $s$ integrals are to be carried out, it is useful to consider a generic integral of the form
\begin{align}
\mathcal{G}\left[h\right] & = \int_{t_{\min}}^{t_{\max}}dt\int_{s_-}^{s_+}ds\frac{h(s,t)}{\sqrt{\Lambda(s,t)}},
\label{eq:G_def}
\end{align}
where $h(s,t)$ is a generic function, which herein will become $\overline{|\mathcal{M}|^2}(s,t)$. The $\Lambda$ function can be factorized and written in the following form
\begin{align}
\Lambda(s,t) &= \left[ \left(E_b-E_2\right)^2-t\right] \nonumber \\
& \times \left\{g^2_{1}(t) - g_{2}(t) - \left[s-g_{1}\left(t\right)\right]^2\right\},
\label{eq:LambdaFactorized}
\end{align}
where 
\begin{align}
g_{1}(t)&=\frac{t^2 - 2t\left(E_aE_2+E_bE_1\right)}{2\left[\left(E_b-E_2\right)^2-t\right]},\label{eq:g_1}\\
g_{2}(t)&=\frac{t^2\left(E_a+E_b\right)^2}{\left(E_b-E_2\right)^2-t}.\label{eq:g_2}
\end{align}
Using the factorized form of $\Lambda$ in Eq.~(\ref{eq:LambdaFactorized}) together with Eq.~(\ref{eq:G_def}), allows to write
\begin{align}
\mathcal{G}[h] &= \int_{t_{\min}}^{t_{\max}} \frac{dt}{\sqrt{\left(E_b-E_2\right)^2-t}}\int_{s_-}^{s_+}  \frac{ds \text{ } h}{\sqrt{g^2_{1} - g_{2} - \left(s-g_{1}\right)^2}}.
\label{eq:G_factorized}
\end{align}
To illustrate how $\mathcal{G}[h]$ is evaluated, consider the case where $h=s/t$, which can occur in $\overline{|\mathcal{M}|^2}(s,t)$. Thus,
\begin{align}
\mathcal{G}\left[\frac{s}{t}\right] \!\!&=\!\! \int_{t_{\min}}^{t_{\max}}\frac{dt}{t\sqrt{\left(E_b-E_2\right)^2-t}}\int_{s_-}^{s_+}\frac{ds \text{ } s}{\sqrt{g^2_{1} - g_{2} - \left(s-g_{1}\right)^2}} \label{eq:G_s_over_t_def} \\
\!\!& =  \!\!\int_{t_{\min}}^{t_{\max}}\frac{dt}{t\sqrt{\left(E_b-E_2\right)^2-t}} \left[ \left. - \sqrt{g^2_{1} - g_{2} - \left(s-g_{1}\right)^2} \right|^{s_+}_{s_-} \right. \nonumber \\
& + \left. \left. g_1\arcsin\left(\frac{s-g_1}{\sqrt{g^2_1-g_2}}\right)\right|^{s_+}_{s_-} \right]. 
\label{eq:G_s_over_t_slimits}
\end{align}
Since the integration limits $s_+$ and $s_-$ satisfy $g^2_1-g_2-(s_{\pm}-g_1)^2=0$, the first term in Eq.~(\ref{eq:G_s_over_t_slimits}) vanishes identically, whereas the second term simplifies to $\pi g_1$. Consequently, after simplifying, Eq.~(\ref{eq:G_s_over_t_slimits}) reduces to the form
\begin{align}
\mathcal{G}\left[\frac{s}{t}\right] \!\!&=\!\! \int_{t_{\min}}^{t_{\max}} dt \frac{ \pi g_1 }{t\sqrt{\left(E_b-E_2\right)^2-t}}\label{eq:G_s_over_t_tmax}\\
&= \left. \frac{\pi \left[2\left(E_b-E_2\right)^2 -2(E_aE_2+E_bE_1) -t \right] }{\sqrt{\left(E_b-E_2\right)^2-t}} \right|^{t_{\rm max}}_{t_{\rm min}}\nonumber\\
&= \frac{-4\pi \text{ } \lfloor s\rfloor }{\sqrt{\left(E_b-E_2\right)^2+m^2_\infty}}, 
\label{eq:G_s_over_t_mD}
\end{align}    
where $\lfloor s\rfloor\equiv\min\left(E_aE_b,E_1E_2\right) $. In Eq.~(\ref{eq:G_s_over_t_tmax}) substituting $t_{\rm max}=0$ leads to a divergence in the denominator when $E_b=E_2$. To regulate this infrared divergence, we introduce the asymptotic quark mass, $m_\infty$, as a cut-off parameter in the denominator. This will be revised once Hard Thermal Loop corrections are accounted for in the future. After straightforward algebraic manipulation, one finds that both kinematic regions, $t_{\rm min}=-4 \text{ } {\rm min}(E_aE_1, E_bE_2)$, yield a simplified expression solely in terms of $\lfloor s\rfloor$.

The choice of Mandelstam integration variables used to perform and evaluate the integrals in $\mathcal{G}[h]$ is dictated by the functional dependence of $\overline{|\mathcal{M}|^2}$. In general, it is advantageous to choose the Mandelstam variable appearing in the denominator of $h$ as the outer integration variable, since the effects of any divergence are then isolated from the inner Mandelstam-variable integral, thus leading to the simplest integration limits and allowing the remaining phase-space constraint to be expressed through the corresponding $\Lambda$-function. The Mandelstam variable appearing in the numerator then naturally defines the inner integration variable. For example, when $h=u/t$, the appropriate order of integration is $\int dt\int du$, with the $\Lambda$ function written as $\Lambda=\Lambda(t,u)$. On the other hand, if the denominator is unity, the integration variables are determined entirely by the numerator. Thus, for $h=su$, the most convenient choice is $\int ds\int du$, with $\Lambda=\Lambda(s,u)$. This strategy considerably simplifies the analytic evaluation of the phase-space integrals encountered throughout this work. The remaining forms of the $\Lambda$-function can be obtained using crossing symmetry and are given by
\begin{align}
\Lambda(s,u)&=-(E_a-E_2)^2s^2-(E_a+E_b)^2u^2+su(s+u) \nonumber \\
&\quad -2(E_aE_1+E_bE_2)su, \\
\Lambda(t,u)&=-(E_b-E_2)^2u^2-(E_a-E_2)^2t^2+tu(t+u) \nonumber \\
&\quad +2(E_aE_b+E_1E_2)tu.
\end{align}
\textbf{\textit{Running of the asymptotic quark mass.}}--- As was discussed in Ref.~\cite{Opitz:2026sgz}, the thermal mass, herein the asymptotic quark mass, which serves as a cutoff for infrared singularities, changes with the energy-momentum exchanged between the jet parton and the thermal parton, namely Mandelstam $t$. Thus, the asymptotic quark mass given by \cite{Ghiglieri:2015ala}
\begin{align}
m^2_\infty=\frac{4\pi\alpha_sT^2}{3}
\end{align}
is not a constant, as the strong coupling runs with Mandelstam $t$. Thus, after solving the the self-consistent equation $t=-4\pi\alpha_s(t) T^2$ to find $t_{\rm min}$ where 
\begin{align}
\frac{g^2_s(t)}{4\pi}=\alpha_s(t)=\frac{12\pi}{(33-2N_f)\ln\left[\frac{t}{\Lambda^2_{QCD}}\right]},
\end{align}
the asymptotic quark mass becomes
\begin{align}
m^2_\infty=\frac{16\pi^2T^2}{27W_0\left(\frac{16\pi^2T^2}{27\Lambda_3^2}\right)}\label{eq:m_infty}
\end{align}
where $W_0$ is the principal branch of the Lambert W-function \cite{Cohl:2014drm}, while $\Lambda_3$ is explained in Ref.~\cite{Opitz:2026sgz}. The expression for the complete tree-level $2\to2$ scattering rate at leading-order involving quark propagators is presented in Appendix~\ref{app:compt_annih_rates}.
\subsection{$\hat{\mathcal{F}}_{(0)}$ from massless quark-antiquark annihilation}
The first process, considered herein, contributing to medium-induced quark conversion in a weakly-coupled plasma is quark–antiquark annihilation. In this process, an energetic quark from the jet interacts with a thermal antiquark from the medium via the exchange of a quark, resulting in the conversion of the quark into a gluon. 

\subsubsection{$\hat{\mathcal{F}}_{(0)}$ without kinematic approximations at tree-level}
The Feynman diagram relevant to the quark conversion process 
is shown in Fig.~\ref{fig:fhat_diagrams}(b), and the associated matrix element for this process is given by
\begin{align}
\begin{split}	
\overline{\lvert\mathcal{M}_{q\bar{q}\to gg}\rvert^2}&=\frac{1024}{9}\pi^2\alpha_s^2\left[\frac{u}{t} \right],
\end{split}
\label{eq:MSquare_annihilation}
\end{align}
where the spin and color of the incoming jet are averaged, while the initial medium particle and both final particles have their color and spin degrees of freedom summed over. The zeroth-order higher-twist jet-medium transport coefficient $\hat{\mathcal{F}}_{(0)}$ \cite{Kumar:2025egh,Kumar:2025asj} represents the scattering rate of this process, which, within the $2\to2$ scattering setup, becomes
\begin{align}
\hat{\mathcal{F}}_{(0)}  &\equiv  \frac{d^3R_t}{d^3p_a} \\
&=\frac{\alpha^2_s}{18\pi^5E_a^2}\int_0^\infty dE_b\, f_b \int_0^{E_a+E_b}\! \! dE_2  [1+ f_2] \text{ } \mathcal{G} \left[\frac{u}{t}\right], \nonumber
\end{align}
where $\mathcal{G}[u/t]$ represents the Mandelstam integral given by
\begin{align}
\begin{split}
\mathcal{G} \left[\frac{u}{t}\right] & \equiv \int_{t_{\min}}^{t_{\max}}dt\int_{u_-}^{u_+}\frac{du \text{ } u}{t\sqrt{\Lambda(t,u)}},
\end{split}
\end{align}
and the integration limits $u_+$ and $u_-$ satisfy $\Lambda(t,u_{\pm})=0$. Applying the algebraic manipulation outlined in the previous section yields
\begin{align}
\mathcal{G}\left[\frac{u}{t}\right] & = \frac{4\pi\lfloor u\rfloor}{\sqrt{\left(E_b-E_2\right)^2+m_\infty^2}},
\label{eq:ans_G_u_over_t}
\end{align}
where $\lfloor u\rfloor \equiv \min(E_aE_2,E_bE_1)$.
To go further, $\lfloor u\rfloor$ must be split as follows: in the region $0<E_2<E_b$, we have $\lfloor u\rfloor=E_aE_2$ and in the region $E_b<E_2<E_a+E_b$, we get $\lfloor u\rfloor=E_bE_1$. According to Appendix~\ref{app:int_over_floor}, which summarizes these various regions of integration, one obtains
\begin{align}
\hat{\mathcal{F}}_{(0)}&=\frac{2\alpha^2_s}{9\pi^4E_a^2} \int_0^\infty\frac{dE_b}{e^{\beta E_b}+1}\label{eq:F0_annihilation_full}\\
& \times \left[ \int_0^{E_b}dE_2 [1+f_2] \frac{ E_aE_2 }{\sqrt{\left(E_b-E_2\right)^2+m_\infty^2}} \right.\nonumber\\
& \left. + \int_{E_b}^{E_a+E_b}dE_2  [1+f_2] \frac{ E_b (E_a+E_b-E_2)}{\sqrt{\left(E_b-E_2\right)^2+m_\infty^2}}  \right].\nonumber
\end{align}
Computing a closed-form expression of these integrals is not easy, and they will, in general, be evaluated numerically in what follows. However, below we devise a systematically improvable expansion allowing us to obtain closed-form expressions.

\subsubsection{Kinematic approximations for $\hat{\mathcal{F}}_{(0)}$}
The integral over $E_2$ can be carried out analytically in the limit $1+f_2=\left[1-e^{-E_2/T}\right]^{-1}\simeq 1+O\left(e^{-E_2/T}\right)$. The first term of this geometric series expansion yields
\begin{align}
\hat{\mathcal{F}}_{(0)} &\simeq \frac{2\alpha^2_sT}{9\pi^4}  \int_0^\infty\frac{dx_b}{e^{x_b}+1}\label{eq:rate_annh_EbIntegral}\\
& \times \left[ \frac{x_b}{x_a} \left\{ \arcsinh\left(\frac{x_{b}}{z}\right) + \arcsinh\left(\frac{x_{a}}{z}\right) \right\} + \frac{z(x_a+x_b)}{x^2_a} \right. \nonumber \\ 
& \qquad- \left. \frac{x_b\sqrt{x^2_a+z^2}}{x^2_a} - \frac{\sqrt{x^2_b+z^2}}{x_a}  \right] +O\left(e^{-x_2}\right),\nonumber
\end{align}
where $x_i=E_i/T$ for $i\in\{a,b,1,2\}$, and $z=m_\infty/T$. Using the auxiliary functions defined in Appendix \ref{app:th_int}, Eq.~(\ref{eq:rate_annh_EbIntegral}) can be expressed in closed-form as
\begin{align}
\hat{\mathcal{F}}_{(0)}&\simeq\frac{2\alpha^2_sT}{9\pi^4} \Bigg[ \frac{B^+_1(z)}{x_a} + \frac{\pi^2}{12x_a}\arcsinh\left(\frac{x_a}{z}\right)\nonumber\\
& - \frac{A^+_0(z)}{x_a} - \frac{\pi^2}{12x^2_a}\sqrt{x^2_a+z^2} \Bigg] +O\left(e^{-x_2}\right).\label{eq:F0_annihilation_wo_Be_Pb}
\end{align}
\begin{figure}[H]
    \centering
    \includegraphics[width=1\linewidth]{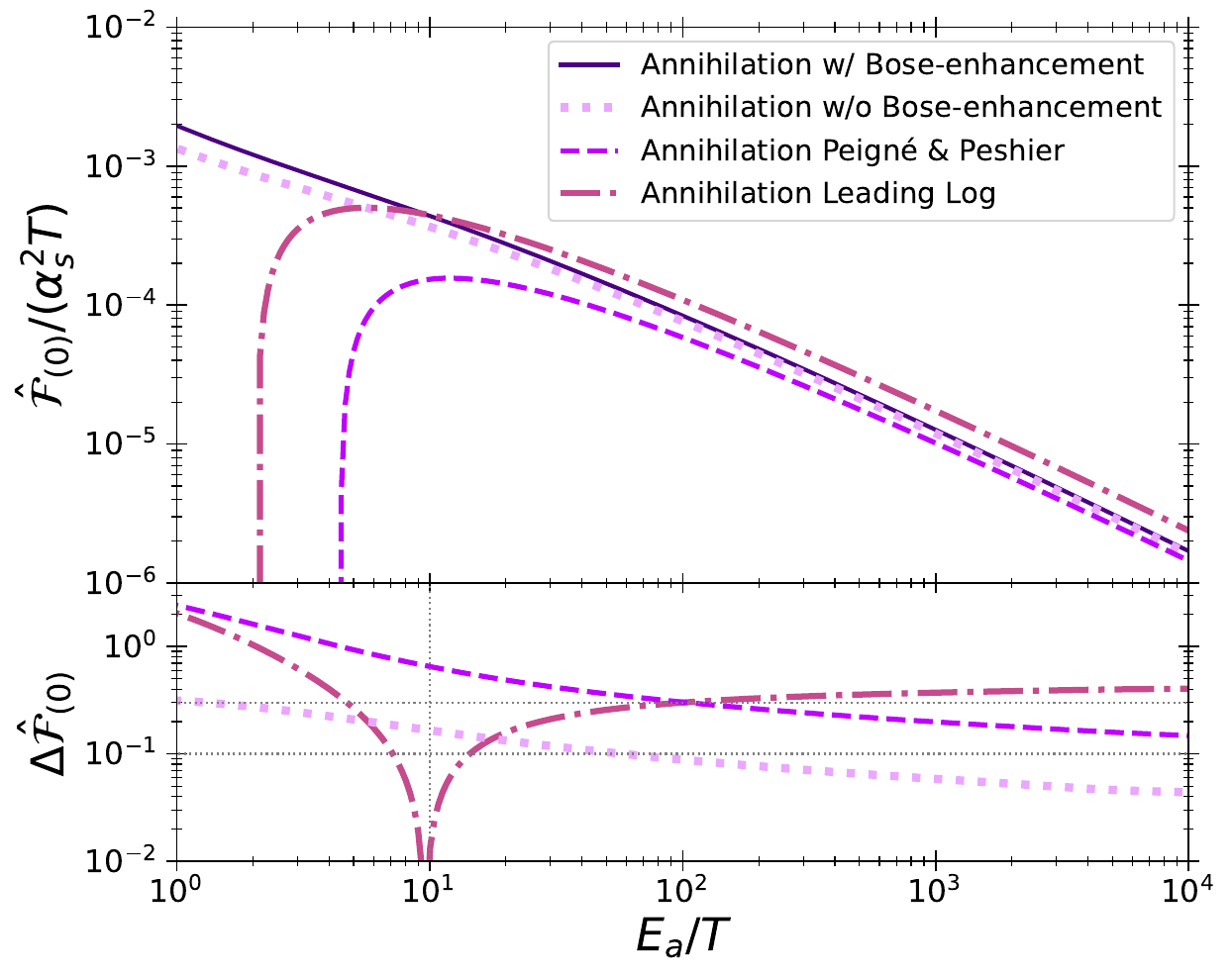}
    \caption{Top panel: $\hat{\mathcal{F}}_{(0)}/(\alpha^2_s T)$ for quark-antiquark annihilation against various approximations. Bottom panel: relative differences between various approaches using Eq.~(\ref{eq:DeltaF}). $m_\infty$ in Eq.~(\ref{eq:m_infty}) at $T=0.2$ GeV is used.}
    \label{fig:F0_annihilation_w_approximations}
\end{figure}
In the leading logarithm (LL) approximation (or leading log for short), $\hat{\mathcal{F}}_{(0)}$ reduces to the following form
\begin{align}
 \hat{\mathcal{F}}^{(LL)}_{(0)} \simeq \frac{\alpha^2_s T}{36\pi^2x_a}\ln\left(\frac{2x_a}{z^2}\right),
\label{eq:fsaLL_Annihilation}  
\end{align}
where any terms missing are not easily recovered, since the approximation used to reach the leading log result (as detailed in Ref.~\cite{Opitz:2026sgz}) is not a systematic expansion. Beyond this, using the Peign\'e \& Peshier \cite{Peigne:2007sd,Peigne:2008nd} approach to compute this process generates
\begin{align}
\hat{\mathcal{F}}^{(PP)}_{(0)}&\simeq\frac{2\alpha_s^2T}{9\pi^4x_a}\left[\frac{\pi^2}{12}\ln\left(\frac{8x_a}{z^2}\right)-\frac{(1+\gamma)\pi^2}{12}+\frac{\zeta'(2)}{2}\right]\nonumber\\
&+O\left(\frac{z}{x_a}\right)
\label{eq:P&P_ann}
\end{align}
where $\gamma\approx0.5772156649$ is the Euler-Mascheroni constant, while the derivative of the Riemann $\zeta$-function, $\zeta'$, gives $\zeta'(2)=\frac{\pi^2}{6}\left[\gamma+\ln\left(2\pi\right) -12\ln\left(A\right)\right]$, with $A\approx 1.282427129$ being the Glaisher–Kinkelin constant \cite{Cohl:2014drm}. The mathematical steps needed to obtain this result are in Ref.~\cite{Opitz:2026sgz}, which herein have been adapted to the matrix element in Eq.~(\ref{eq:MSquare_annihilation}). The relative difference between Eq.~(\ref{eq:F0_annihilation_full}) and various approximations is defined as 
\begin{align}
\Delta \hat{\mathcal{F}}_{(i)}\equiv\left\lvert\frac{\hat{\mathcal{F}}_{(i)} -\hat{\mathcal{F}}^{\rm (approx)}_{(i)}}{\hat{\mathcal{F}}_{(i)}}\right\rvert,
\label{eq:DeltaF}
\end{align}
where $(i)=(0)$ is also depicted in Fig.~\ref{fig:F0_annihilation_w_approximations}. 

Figure \ref{fig:F0_annihilation_w_approximations} presents both $\hat{\mathcal{F}}_{(0)}$ within various approximations in the top panel, as well as $\Delta \hat{\mathcal{F}}_{(0)}$ in the bottom panel, where $\Delta \hat{\mathcal{F}}_{(0)}$ is one way of computing the theoretical systematic uncertainty on the $\hat{\mathcal{F}}_{(0)}$ transport coefficient, to be accounted for in future Bayesian analysis. The leading log result presented in Fig.~\ref{fig:F0_annihilation_w_approximations} deviates significantly from the closed-form result ($10\%\lesssim \Delta\hat{\mathcal{F}}_{(0)}\lesssim 30\%$) for $15\lesssim x_a\lesssim 100$. The leading log affects even more partons with $x_a>100$, as seen by a larger deviation $\Delta\hat{\mathcal{F}}_{(0)}\gtrsim 30\%$. As such, if the leading logarithm is used to describe a jet in a Monte Carlo simulation, the core of the jet will have a large uncertainty, significantly limiting the usefulness of the leading-log approximation in jet-medium Monte Carlo simulations. The result from the Peign\'e \& Peshier approach is somewhat better, as deviations of about $10\lesssim \Delta\hat{\mathcal{F}}_{(0)}\lesssim30\%$ are seen relative to the full closed-form result for $x_a>100$. Modern jet-medium Monte Carlo simulations should aim at systematic uncertainties $\sim 10\%$ and possibly below that, given the precision of the available data. Such a precision is achieved herein by Eq.~(\ref{eq:F0_annihilation_wo_Be_Pb}) whose deviation relative to the full result in Eq.~(\ref{eq:F0_annihilation_full}) is $\Delta\hat{\mathcal{F}}_{(0)}\sim 5-10\%$ ($x_a>10$).

\subsection{$\hat{\mathcal{F}}_{(0)}$ for massless Compton scattering }
Another important channel contributing to medium-induced quark conversion is quark-gluon Compton scattering. Since thermal gluons constitute a significant fraction of the QGP degrees of freedom, this channel also affects the chemistry of the jet and the QGP via fermion-boson conversion. The zeroth-order higher-twist jet-medium transport coefficient $\hat{ \mathcal{F}}_{(0)}$ depicted in Fig.~\ref{fig:fhat_diagrams}(c) is obtained both in the (forward-scattering) leading logarithm approximation and using the complete scattering kinematics.

\subsubsection{$\hat{\mathcal{F}}_{(0)}$ at tree-level}

The matrix element for the $t$-channel quark-gluon Compton scattering process is given by
\begin{align}    \overline{\lvert\mathcal{M}\rvert^2}(s,t) = \frac{1024}{9} \pi^2 \alpha^2_s\left[-\frac{s}{t}\right],
\label{eq:Compton_MSquare}
\end{align}
where the spin and color of the incoming jet quark are averaged. The spin and color degrees of freedom of the outgoing particles and incoming thermal parton are summed. The zeroth-order jet transport coefficient $\hat{\mathcal{F}}_{(0)}$ is the scattering rate for this channel, which has the following form:
\begin{align}
   \hat{\mathcal{F}}_{(0)} = -\frac{\alpha^2_s}{18\pi^5E_a^2}\int_0^\infty dE_b\, f_b\int_0^{E_a+E_b}dE_2 \left[1-f_{2}\right] \mathcal{G} \left[\frac{s}{t} \right],
\end{align}
where $\mathcal{G}[s/t]$ represents the Mandelstam integral given by 
\begin{align}
    \mathcal{G}\left[\frac{s}{t} \right] &=  \frac{-4\pi\lfloor s\rfloor}{\sqrt{\left(E_b-E_2\right)^2+m_\infty^2}}, \label{eq:G_s_over_t}\\
    \lfloor s\rfloor&\equiv \min\left(E_aE_b,E_1E_2\right).\nonumber
\end{align}

The function $\mathcal{G}[s/t]$ is defined in Eq.~(\ref{eq:G_s_over_t_def}), and expressed as a double integral over the Mandelstam variables $t$ and $s$. The intermediate algebraic manipulations, together with the analytical evaluation of the integral, are detailed in Eqs.~(\ref{eq:G_s_over_t_def}--\ref{eq:G_s_over_t_mD}). As $\mathcal{G}\left[\frac{s}{t} \right]\propto\lfloor s\rfloor$, Appendix~\ref{app:int_over_floor} shows how to obtain the three regions contributing to the integral over $E_2$, which are:
\begin{align}
\hat{\mathcal{F}}_{(0)} &= \frac{2\alpha^2_s}{9\pi^4E_a^2}\int_0^\infty \frac{dE_b}{e^{\beta E_b}-1}\nonumber\\
&\left[  \int_0^a dE_2\frac{[1-f_2] E_2\left(E_a+E_b-E_2\right)}{\sqrt{\left(E_b-E_2\right)^2+m_\infty^2}} \right. \nonumber \\
&\quad + \int_a^b dE_2 \frac{[1-f_2] E_aE_b}{\sqrt{\left(E_b-E_2\right)^2+m_\infty^2}} \nonumber \\
&\quad + \left. \int_b^{a+b}dE_2 \frac{[1-f_2]E_2\left(E_a+E_b-E_2\right)}{\sqrt{\left(E_b-E_2\right)^2+m_\infty^2}} \right].
\label{eq:F0_Compton_full}
\end{align}
These integration regions, which account for the fact that $a= \min(E_a,E_b)$ and $b= \max(E_a,E_b)$, affect the boundaries of $E_2$.
\subsubsection{Kinematic approximations to $\hat{\mathcal{F}}_{(0)}$}
In the limit $\left[1+e^{-x_2}\right]^{-1}\simeq 1+O\left(e^{-x_2}\right)$, $\hat{\mathcal{F}}_{(0)}$ is given by
\begin{align}
\hat{\mathcal{F}}_{(0)} \simeq &\frac{2\alpha^2_s T}{9\pi^4}\int_0^\infty\frac{dx_b}{e^{x_b}-1} \left[\left(\frac{1}{2x_a}-\frac{x_b}{x_a^2}\right)\sqrt{x_a^2+z^2} \right. \nonumber \\
& +\left(\frac{x_b}{2x_a^2}-\frac{1}{x_a}\right)\sqrt{x_b^2+z^2} \nonumber \\ 
& + \left(\frac{x_b}{x_a}+\frac{z^2}{2x_a^2}\right) \left\{\arcsinh\left(\frac{x_a}{z}\right)+\arcsinh\left(\frac{x_b}{z}\right)\right\} \nonumber \\
& + \lvert x_a-x_b\rvert \left(\frac{z}{x_a^2}-\frac{1}{2x_a^2}\sqrt{\left(x_a-x_b\right)^2+z^2}\right)\nonumber \\ 
&\left. -\sgn(x_a-x_b)\frac{z^2}{2x_a^2}\arcsinh\left(\frac{x_a-x_b}{z}\right) \right]\nonumber\\
&+O\left(e^{-x_2}\right).
\label{eq:F0_compton_xb}
\end{align}
with $\sgn\left(x_a-x_b\right)=2\Theta\left(x_a-x_b\right)-1$, where $\Theta$ is the Heaviside function. Performing the remaining integral gives
\begin{align}
&\hat{\mathcal{F}}_{(0)} \simeq \frac{2\alpha^2_s T}{9\pi^4} 
\Bigg[\frac{A^-_1(z)}{2x^2_a} -\frac{\pi^2\sqrt{x^2_a+z^2}}{6x^2_a} + \frac{B^-_1(z)}{x_a}+ \frac{z^2B^-_0(z)}{2x^2_a} \nonumber \\
&-\frac{1}{2x_a} \lim_{\epsilon\rightarrow0^+}\left\{\Delta\alpha^-_0(\epsilon)+\sqrt{x^2_a+z^2}\ln\left(\epsilon\right) \right\} \nonumber \\ 
&- \frac{1}{x_a} \lim_{\epsilon\rightarrow 0^+}\Big\{A^-_0\left(\epsilon;z\right)+z\ln\left(\epsilon\right)\Big\} + \frac{\pi^2}{6x_a} \arcsinh\left(\frac{x_a}{z}\right)\nonumber \\
&- \frac{z^2}{2x^2_a} \lim_{\epsilon\rightarrow 0^+}\left\{ \Delta\beta^-_0(\epsilon)+\arcsinh\left(\frac{x_a}{z}\right)\ln(\epsilon) \right\} \nonumber \\
&+\frac{2z}{x^2_a}\left\{\polylog{2}\left( e^{-x_a}\right) -\frac{\pi^2}{12}\right\}+\frac{\Delta \alpha^-_1}{2x^2_a} \Bigg] +O\left(e^{-x_2}\right),
\label{eq:F0_Compton_wo_Be_Pb}
\end{align}
where the functions $\Delta\alpha_i, \Delta\beta_i$ as well as the $A_i$ and $B_i$ are defined in Appendix~\ref{app:th_int}. In the leading logarithm (LL) approximation $\hat{\mathcal{F}}_{(0)}$ yields 
\begin{align}
\hat{\mathcal{F}}^{(LL)}_{(0)}\simeq \frac{\alpha^2_s T}{36\pi^2x_a}\ln\left(\frac{2x_a}{z^2}\right),
\label{eq:fsaLL}  
\end{align}
while Peign\'e \& Peshier \cite{Peigne:2007sd,Peigne:2008nd} approach gives 
\begin{align}
\hat{\mathcal{F}}^{(PP)}_{(0)}&\simeq \frac{2\alpha_s^2T}{9\pi^4x_a}\left[\frac{\pi^2}{6}\ln\left(\frac{4x_a}{z^2}\right)-\frac{\gamma\pi^2}{6}+\zeta'(2)\right]\nonumber\\
&+O\left(\frac{z}{x_a}\right),
\label{eq:PandP_Compton}
\end{align}
where $\gamma$ and $\zeta'(2)$ appear in Eq.~(\ref{eq:P&P_ann}). 
\begin{figure}[H]
    \centering
    \includegraphics[width=1\linewidth]{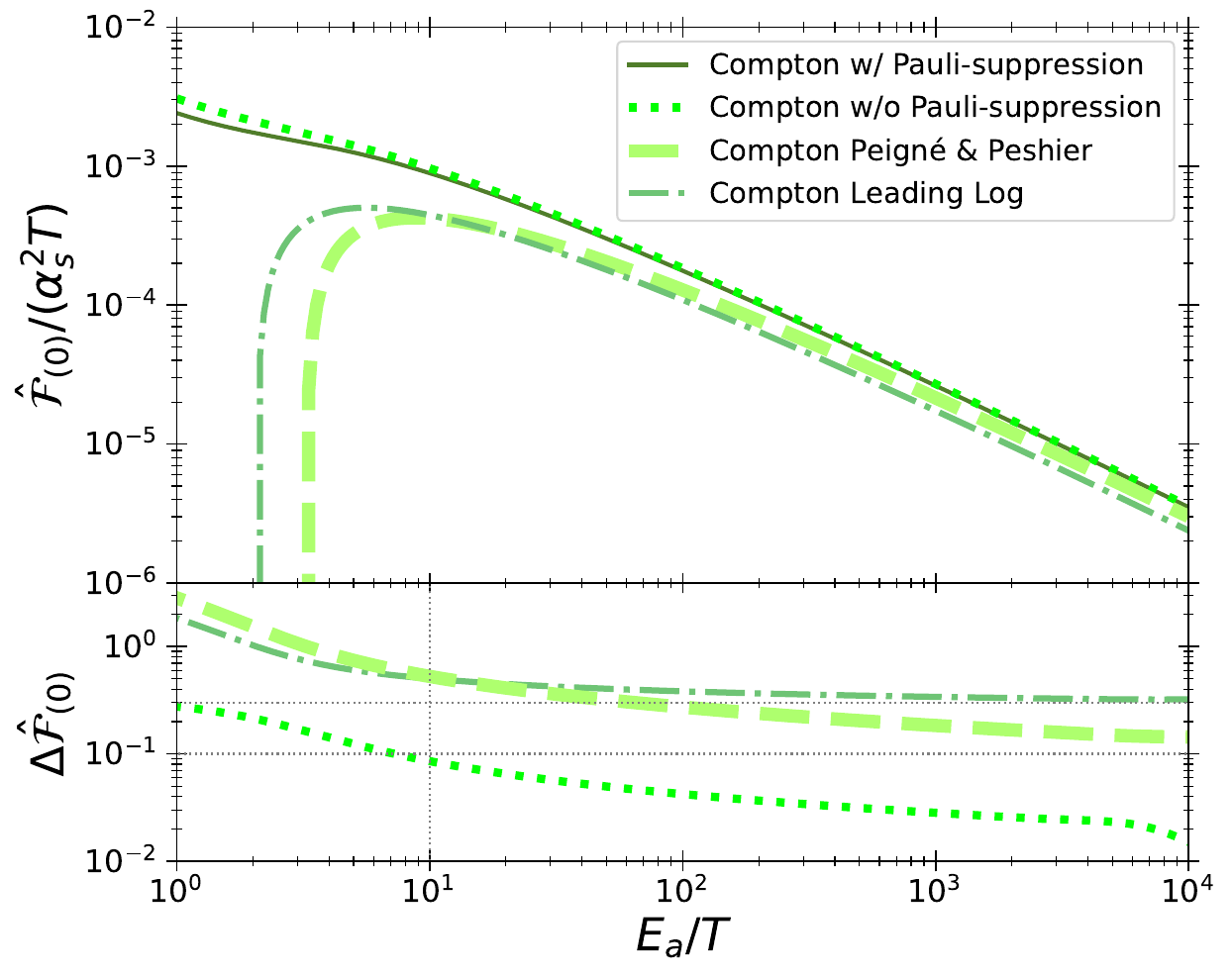}
    \caption{Comparison of $\hat{\mathcal{F}}_{(0)}/(\alpha^2_s T)$ as a function of $E_a/T$ for quark-gluon Compton Scattering and their relative differences with Eq.~(\ref{eq:fsaLL}) and Eq.~(\ref{eq:PandP_Compton}). Here, $m_\infty$ in Eq.~(\ref{eq:m_infty}) at $T=0.2$ GeV is used.}
    \label{fig:F0_Compton_w_approx}
\end{figure}

A comparison between the Compton scattering contribution to $\hat{\mathcal{F}}_{(0)}$ in Eq.~(\ref{eq:F0_Compton_full}) against various approximations is given in Fig.~\ref{fig:F0_Compton_w_approx}. Also, depicted are the relative differences $\Delta \hat{\mathcal{F}}_{(0)}$, as defined in Eq.~(\ref{eq:DeltaF}) for $(i)=(0)$, between Eq.~(\ref{eq:F0_Compton_full}) and various approximations. The behaviour of the leading log and the Peign\'e \& Peshier result mirrors what was found when considering the annihilation process in the previous section: with Eq.~(\ref{eq:F0_Compton_wo_Be_Pb}) having a precision $\Delta \hat{\mathcal{F}}_{(0)}\lesssim 10\%$ for $x_a>10$. 
\begin{figure}[H]
    \centering
\includegraphics[width=1\linewidth]{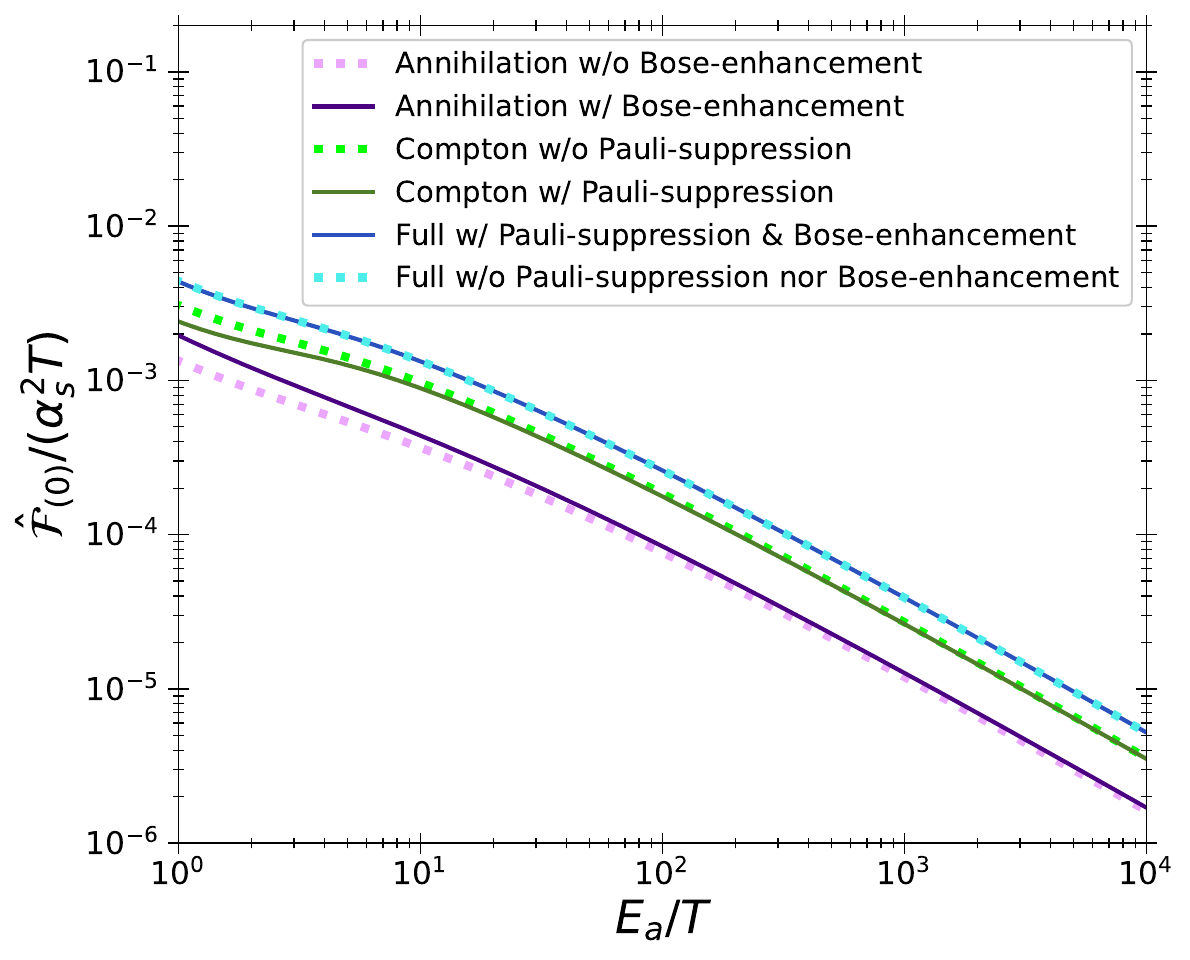}
    \caption{Comparison of $\hat{\mathcal{F}}_{(0)}/(\alpha^2_s T)$ as a function of $E_a/T$ for quark-antiquark annihilation and quark-gluon Compton processes. Here, $m_\infty$ in Eq.~(\ref{eq:m_infty}) at $T=0.2$ GeV is used.}
    \label{fig:Fhat0_annih_compton}
\end{figure}
\begin{figure}[H]
    \centering
    \includegraphics[width=1\linewidth]{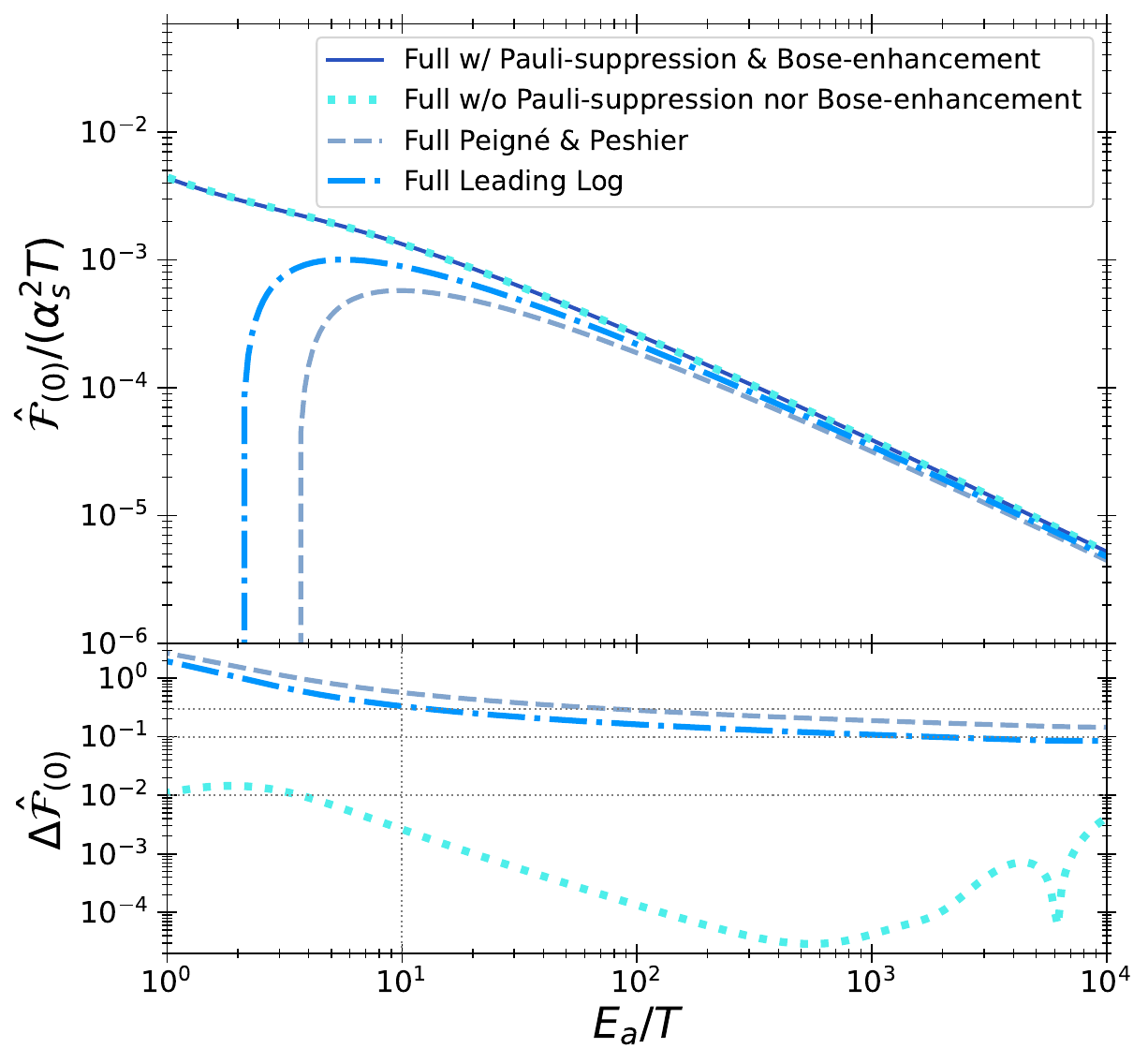}
    \caption{Comparison of $\hat{\mathcal{F}}_{(0)}/(\alpha^2_s T)$ as a function of $E_a/T$ and their combined relative differences. Here, $m_\infty$ in Eq.~(\ref{eq:m_infty}) at $T=0.2$ GeV is used.}
    \label{fig:F0_full_w_approx}
\end{figure}

\subsection{Full $\hat{\mathcal{F}}_{(0)}$ versus approximations}
Fig.~\ref{fig:Fhat0_annih_compton} shows the $x_a=E_a/T$ dependence of $\hat{\mathcal{F}}_{(0)}$ for the quark-antiquark annihilation, quark-gluon Compton scattering, as well as their sum, labelled as ``full" in the figure. The transport coefficient $\hat{\mathcal{F}}_{(0)}$ is scaled by the temperature and the square of the strong coupling constant to highlight the $x_a$ dependence of $\hat{\mathcal{F}}_{(0)}$ at tree-level. The Compton scattering contribution, including the Pauli-blocking factor for the outgoing quark (labelled by the momentum index ``2"), dominates over the annihilation channel and is approximately a factor of two larger.

\begin{figure}[H]
    \centering
\hspace{-0.6cm}    \includegraphics[width=1.05\linewidth]{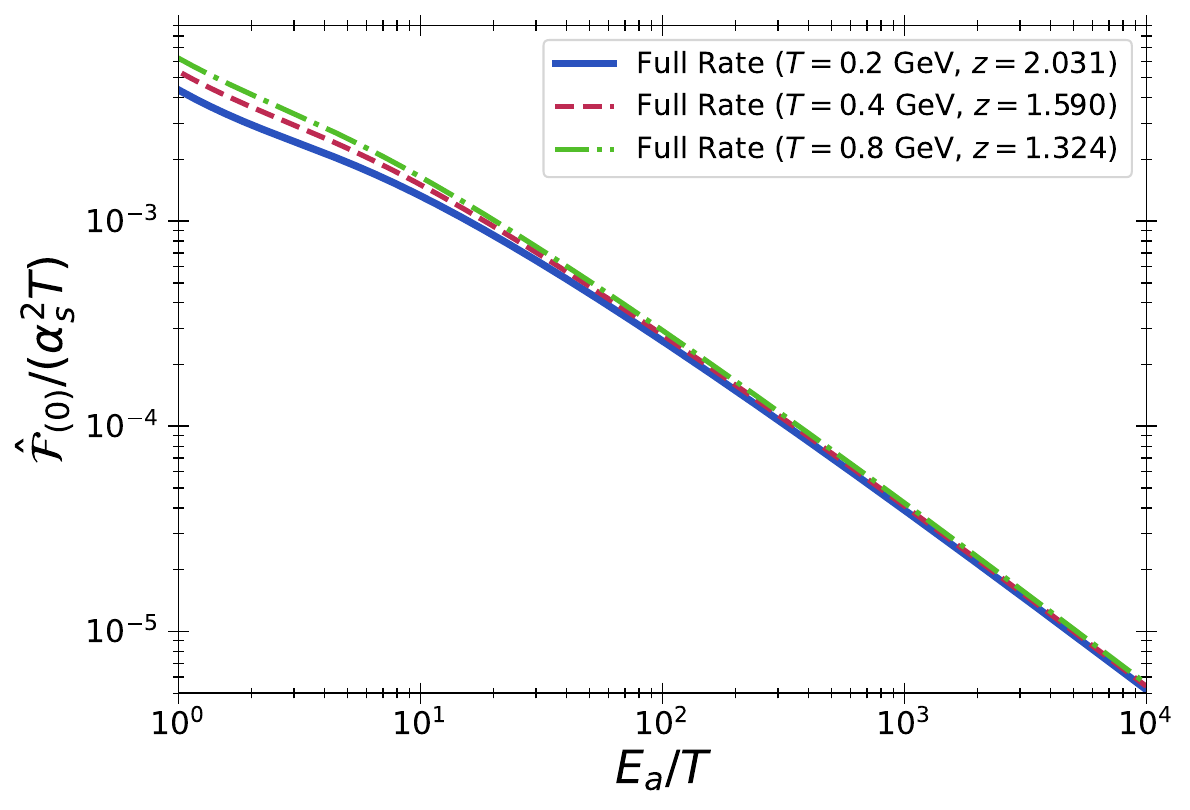}
    \caption{$\hat{\mathcal{F}}_{(0)}/(\alpha_s^2T)$ as a function of $x_a$ for $T=[0.2,0.4,0.8]$ GeV.}
    \label{fig:F0_full_w_T}
\end{figure}

Once the complete annihilation contribution (with Bose-enhancement) is added to the complete Compton scattering (with Pauli-blocking), compensations between the Bose-enhancement and Pauli-blocking are such that the associated total $\hat{\mathcal{F}}_{(0)}$ is well approximated by summing Eq.~(\ref{eq:F0_annihilation_wo_Be_Pb}) and Eq.~(\ref{eq:F0_Compton_wo_Be_Pb}). In fact $\Delta \hat{\mathcal{F}}_{(0)}$, shown in the bottom of Fig.~\ref{fig:F0_full_w_approx}, reveals that a relative $<2\%$ difference exists between the sum of Eq.~(\ref{eq:F0_annihilation_wo_Be_Pb}) and Eq.~(\ref{eq:F0_Compton_wo_Be_Pb}) and the sum of Eq.~(\ref{eq:F0_annihilation_full}) and Eq.~(\ref{eq:F0_Compton_full}) over the entire $x_a$ range. Such a result is ideal for implementation within modern jet-medium Monte-Carlo simulations. Finally, since the quark asymptotic mass depends on temperature according to Eq.~(\ref{eq:m_infty}), $\hat{\mathcal{F}}_{(0)}$ changes with temperature as illustrated in Fig.~\ref{fig:F0_full_w_T}. Further discussion about the temperature dependence of $\hat{\mathcal{F}}_{(0)}$ is found in Section~\ref{sec:Fhat_w_T}.

\section{Energy Loss for conversion process}
\label{sec:E_loss}
The average energy loss for the quark-to-gluon conversion process is defined as the difference in energies between the outgoing gluon and incoming jet quark, namely  
\begin{align}
   \langle \Delta E \rangle = \langle E_b -E_2 \rangle.
\end{align}
Of course, $\langle \Delta E \rangle$ also satisfies the relation $\langle \Delta E \rangle= \langle E_1 - E_a \rangle$, but that form will not be used herein. 
\subsection{Energy loss for annihilation process}
The squared matrix element for the quark-antiquark annihilation channel is given in Eq.~(\ref{eq:MSquare_annihilation}). Thus to obtain $\langle \Delta E \rangle$, one needs
\begin{align}
\hat{\mathcal{O}}=\Delta E \overline{\lvert\mathcal{M}\rvert^2}= \frac{1024\pi^2 \alpha^2_s}{9} \left(E_b-E_2\right)\frac{u}{t},
\end{align}
and thus the appropriate choice of Mandelstam integration variables is $t$ and $u$. Performing the integration over $t$ and $u$, $\langle \Delta E \rangle$ is recast into the following form
\begin{align}
 \langle \Delta E \rangle &= -\frac{\alpha_s^2}{18\pi^5 E_a^2}
\int_0^{\infty} \frac{dE_b}{e^{ \beta E_b}+1}\int_0^{E_a+E_b}  dE_2 \left[1+f_{2}\right] \nonumber \\
&\times \Big\{\left(E_2-E_b\right)\mathcal{G}\left[\frac{u}{t}\right] \Big\},
\label{eq:DeltaE_G_u_t}
\end{align}
where $\mathcal{G}[u/t]$ is defined in Eq.~(\ref{eq:ans_G_u_over_t}). Using  Eq.~(\ref{eq:ans_G_u_over_t}), one expresses
\begin{align}
\langle \Delta E\rangle &=- \frac{\alpha_s^2}{18\pi^5E_a^2}\int_0^{\infty}dE_b f_b\times\label{eq:DeltaE_annih}\\
&\times\int_0^{E_a+E_b}dE_2\left[1+f_2\right]\left\{\textbf{I}+\textbf{II}\right\}\nonumber\\
\textbf{I}&= \left[E_2-E_b\right]\left[4\pi\frac{E_aE_2}{\sqrt{\left(E_b-E_2\right)^2+m_\infty^2}}\right]\nonumber\\
\textbf{II}&=\left[E_2-E_b\right]\left[4\pi\frac{E_aE_b+E_b^2-E_bE_2}{\sqrt{\left(E_b-E_2\right)^2+m_\infty^2}}\right]\nonumber
\end{align}
Note that $\left(E_2-E_b\right)$ changes sign within $E_2\in[0,E_a+E_b]$ (at $E_2=E_b$), which reduces $\langle \Delta E\rangle$ for the case containing $[1+f_2]$, relative to the approximation $1+f_2\simeq 1+O\left(e^{-x_{2}}\right)$. In the limit $[1+f_2]\simeq 1+O\left(e^{-x_{2}}\right)$, the average value of $\Delta E$ is given by
\begin{align}
\langle \Delta E \rangle &\simeq  -\frac{\alpha^2_s T^2}{9\pi^4}\int_0^\infty \frac{dx_b}{e^{x_b}+1}\Bigg[  \frac{x_b}{x_a}\left\{\sqrt{x_a^2+z^2} - \sqrt{x_b^2+z^2}\right\} \nonumber \\
&\qquad\quad + \frac{z^2}{x_a^2}\left\{x_b\arcsinh\left(\frac{x_a}{z}\right) - x_a\arcsinh\left(\frac{x_b}{z}\right)\right\} \Bigg]\nonumber\\
& +O\left(e^{-x_2}\right),
\label{eq:deltaE_annih_xb}
\end{align}
which, using auxiliary functions in Appendix~\ref{app:th_int}, gives
\begin{align}
 \langle \Delta E \rangle &\simeq  -\frac{\alpha^2_s T^2}{9\pi^4} \Bigg[\frac{\pi^2}{12x_a}\sqrt{x^2_a+z^2}  -\frac{A^+_{1}(z)}{x_a}\label{eq:DeltaE_annih_wo_BE}\\
& - \frac{z^2}{x_a}B^+_{0}(z) + \frac{\pi^2z^2}{12x^2_a}\arcsinh\left(\frac{x_a}{z}\right) \Bigg]+O\left(e^{-x_2}\right).\nonumber
\end{align}
In the asymptotic ultraviolet limit, $\langle \Delta E \rangle$ becomes
\begin{align}
\lim_{x_{a}\rightarrow \infty} \langle \Delta E \rangle &= - \frac{\alpha^2_s}{108\pi^2}T^2.
 \label{eq:dEdX_infinite_E_annihilation}   
\end{align}
Thus, $\langle\Delta E\rangle/(\alpha_s^2T^2)$ for the quark reaches a finite asymptotic value even in the limit of infinite quark energy. 

\begin{figure}[H]
    \centering
    \includegraphics[width=1\linewidth]{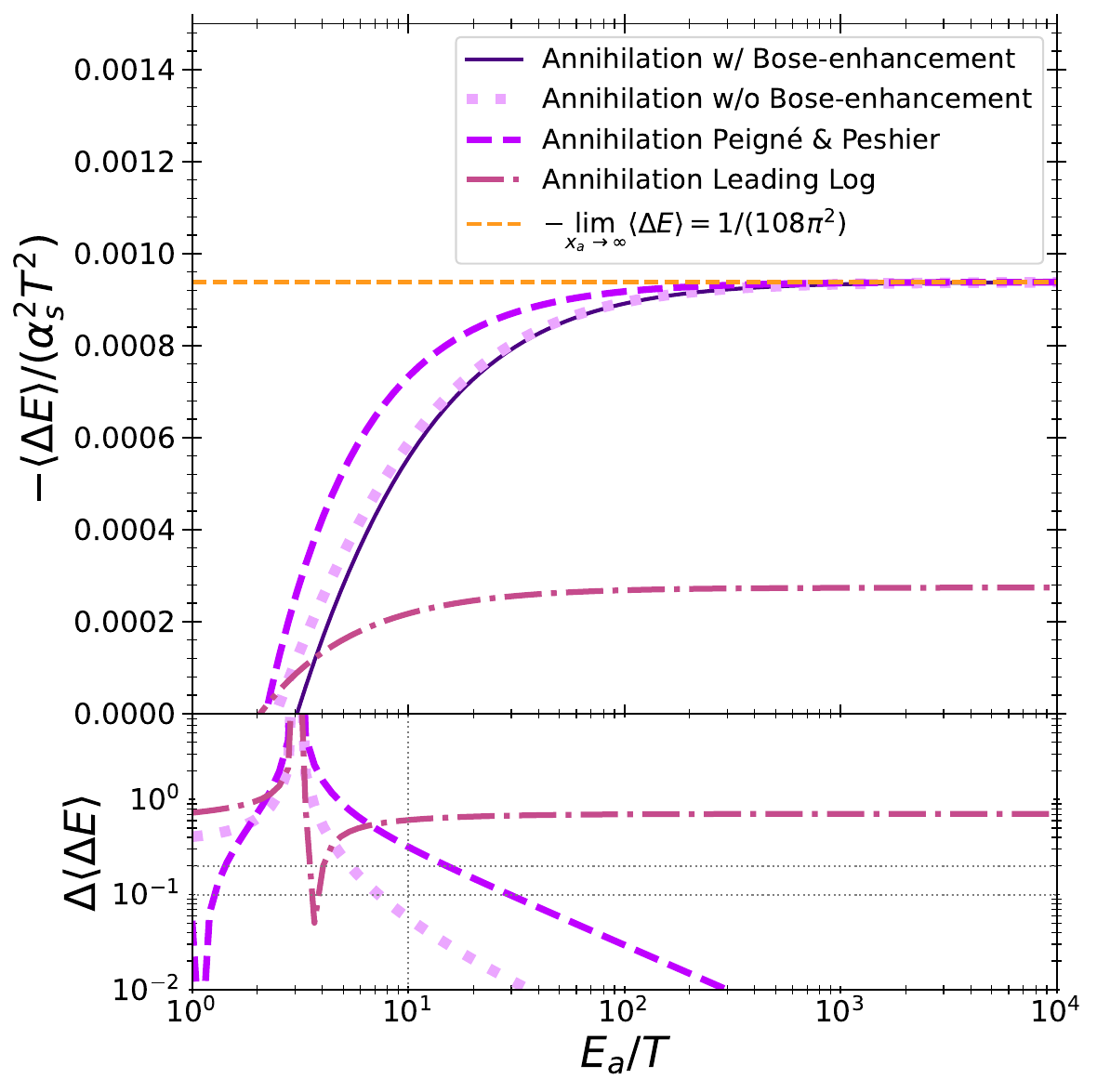}
    \caption{Top panel: Energy loss $\langle\Delta E\rangle/(\alpha_s^2T^2)$ as a function of $E_a/T$ for quark-antiquark annihilation, with $m_\infty$ in Eq.~(\ref{eq:m_infty}) at $T=0.2$ GeV being used. The bottom panel: relative difference between various approximations and the tree-level result given in Eq.~(\ref{eq:DeltaE_annih}).}
    \label{fig:DeltaE_Annihilation}
\end{figure}

As far as the Peign\'e \& Peshier approximation~\cite{Peigne:2007sd,Peigne:2008nd} is concerned, one gets
\begin{align}
    \langle\Delta E\rangle^{(PP)}\simeq-\frac{\alpha_s^2T^2}{9\pi^4}\left[\frac{\pi^2}{12}-\frac{3\zeta(3)}{2x_a}\right]+O\left(\frac{z^2}{x_a}\right).
\label{eq:DeltaE_annih_PP}
\end{align}
When evaluating the leading log for the energy loss, one encounters two separate poles: a collinear pole $\frac{1}{1-\cos(\theta)}$ as $\theta\to 0$, and an infrared pole $\frac{1}{e^{x_b}-1}$ as $x_b\to0$. The matrix element does not provide enough compensation to regulate the infrared pole. On the other hand, the collinear pole stems from the leading log approximation itself, as shown in Ref.~\cite{Opitz:2026sgz}. To obtain a finite result, the plus-distribution~\cite{Peskin:1995ev} regularization is used
\begin{align}
    \int_0^1dx\frac{f(x)}{1-x}\to\int_0^1dx\frac{f(x)}{(1-x)_+}\equiv\int_0^1dx\frac{f(x)-f(1)}{1-x},
\label{eq:plus_dist}
\end{align}
where $f(x)$ does not contain divergences. Doing so yields
\begin{align}
\langle\Delta E\rangle^{(LL)}\simeq-\frac{\alpha_s^2T^2}{18\pi^4}\ln^2(2)\left(1-\frac{z^2}{2x_a}\right).
\label{eq:DeltaE_annih_LL}
\end{align}
Figure~\ref{fig:DeltaE_Annihilation} depicts $\langle \Delta E\rangle$ including its various approximations. 
The leading log provides a poor approximation for $\langle \Delta E\rangle$. Unlike the leading log approximation for $\hat q$ for $gg\to gg$ scattering, which provides a good approximation to the complete tree-level $\hat q$ \cite{Opitz:2026zql}, the leading log approximation of energy loss $\langle \Delta E\rangle$ for gluon-gluon scattering already started showing its limitations in Ref.~\cite{Opitz:2026zql}, and these limitations are also seen here. As we will see in the next section, the leading log approximation is a poor predictor of  $\langle \Delta E\rangle$ for both quark-antiquark annihilation and quark-gluon Compton scattering. The Peign\'e \& Peshier approximation performs better; however, given its deviation from the full result in Eq.~(\ref{eq:DeltaE_G_u_t}), shown in the bottom panel of Fig.~\ref{fig:DeltaE_Annihilation}, it is of limited usefulness within a Monte Carlo simulation of jets in the QGP. Note that
\begin{align}
\Delta \langle\Delta E \rangle \equiv \left| \frac{\langle\Delta E \rangle -\langle\Delta E \rangle^{({\rm approx})} }{\langle \Delta E\rangle}\right|
\end{align}
where $\langle\Delta E \rangle$ is given by Eq.~(\ref{eq:DeltaE_G_u_t}), while $\langle\Delta E \rangle^{({\rm approx})}$ is either Eq.~(\ref{eq:DeltaE_annih_wo_BE}), Eq.~(\ref{eq:DeltaE_annih_PP}), or Eq.~(\ref{eq:DeltaE_annih_LL}).

\subsection{Energy loss for Compton scattering}
For $2\to 2$ Compton scattering in Fig.~\ref{fig:fhat_diagrams} (c), energy loss operator $\hat{\mathcal{O}}$ is
\begin{align}
\hat{\mathcal{O}}=\Delta E \overline{\lvert\mathcal{M}\rvert^2}=- \frac{1024\pi^2 \alpha^2_s}{9} (E_b-E_2)\frac{s}{t}. 
\end{align}
After performing the integration over $t$ and $s$, $ \langle \Delta E \rangle$ becomes
\begin{align}
 \langle \Delta E \rangle &= -\frac{\alpha_s^2}{18\pi^5 E_a^2}
\int_0^{\infty} dE_b\, f_b \int_0^{E_a+E_b} dE_2 \left[1-f_{2}\right] \nonumber \\
&\times \left\{\left(E_b-E_2\right)\mathcal{G}\left[\frac{s}{t}\right] \right\},
\label{eq:DeltaE_G_s_t}
\end{align}
where $\mathcal{G}[s/t]$ is defined in Eq.~(\ref{eq:G_s_over_t}). Under the approximation $[1-f_2]\simeq 1+O(e^{-x_2})$, $\langle \Delta E\rangle$ is given as
\begin{align}
 \langle \Delta E \rangle &\simeq -\frac{\alpha^2_s T^2}{9\pi^4}\int_0^\infty \frac{dx_b}{e^{x_b}-1}   \nonumber  \\
& \times \Bigg[  \frac{\left(4z^2+3x_ax_b+x^2_a\right)}{3x^2_a}\sqrt{x_a^2+z^2} \nonumber \\
&\qquad -\frac{\left(4z^2+3x_ax_b+x^2_b\right)}{3x^2_a}\sqrt{x_b^2+z^2}   \nonumber \\
&\qquad - \frac{ \sgn\left(x_a-x_b\right)  \left\{4z^2 + (x_a-x_b)^2 \right\} }{3x^2_a} \sqrt{\left(x_a-x_b\right)^2+z^2}\nonumber \\
&\qquad - \frac{z^2(x_a-x_b)}{x_a^2}\left\{\arcsinh\left(\frac{x_a}{z}\right) +\arcsinh\left(\frac{x_b}{z}\right)\right\} \nonumber \\
&\qquad + \frac{|x_a-x_b| z^2}{x^2_a} \arcsinh\left(\frac{x_a-x_b}{z}\right) \nonumber \\ 
&\qquad + \sgn\left(x_a-x_b\right) \frac{4z^3}{3x^2_a}\Bigg] +O\left(e^{-x_2}\right),
\label{eq:deltaE_compton_xb}
\end{align}
whose closed-form expression is
\begin{align}
\langle \Delta E \rangle &\simeq -\frac{\alpha^2_sT^2}{9\pi^4}\Bigg[
\frac{\pi^2}{6x_a}\sqrt{x^2_a+z^2} - \frac{A_1^-(z)}{x_a}-\frac{A_2^-(z)}{3x_a^2} \nonumber \\
&-\left(\frac{1}{3} +\frac{4z^2}{3x^2_a} \right) \lim_{\epsilon\rightarrow 0^+}\left\{\Delta \alpha^{-}_{0}(\epsilon)+\sqrt{x^2_a +z^2} \ln(\epsilon)\right\} \nonumber \\
&-\frac{4z^2}{3x_a^2}\lim_{\epsilon\rightarrow 0^+}\left\{A_0^-(\epsilon;z)+z\ln(\epsilon)\right\}+\frac{z^2B_1^-(z)}{x_a^2} \nonumber \\
&-\frac{z^2B_0^-(z)}{x_a} + \frac{\pi^2 z^2}{6x^2_a}\arcsinh\left(\frac{x_a}{z}\right) \nonumber \\
& + \frac{z^2}{x_a} \lim_{\epsilon\rightarrow 0^+}\left\{\Delta\beta^-_{0}(\epsilon) + \arcsinh\left(\frac{x_a}{z}\right)\ln(\epsilon)  \right\} \nonumber  \\
& + \frac{8z^3}{3x^2_a}\ln\left(1-e^{-x_a}\right)- \frac{z^2}{x^2_a}\Delta\beta^{-}_{1} - \frac{2}{3x_a}\Delta\alpha^{-}_{1} \nonumber \\
&  -\frac{1}{3x^2_a}\Delta\alpha^-_2 \Bigg]+O\left(e^{-x_2}\right),
\label{eq:DeltaE_Compton_wo_BE}
\end{align}
where auxiliary functions defined in Appendix~\ref{app:th_int} have been used. In the asymptotic limit $x_{a}\rightarrow \infty$, the transport coefficient $\langle \Delta E \rangle$ takes form
\begin{align}
    \lim_{x_{a}\rightarrow \infty}  \langle \Delta E \rangle &=  -\frac{2\alpha^2_sT^2}{9\pi^4} \lim_{x_{a}\rightarrow \infty}\left[ \int^{x_{a}}_{0} dx_{b} \frac{x_b}{e^{x_{b}}-1}\right]\nonumber\\
    &= - \frac{\alpha^2_s}{27\pi^2}T^2.
    \label{eq:dEdX_infinite_E_compton}
\end{align}
As the result in Eq.~(\ref{eq:dEdX_infinite_E_compton}) is similar to that in Eq.~(\ref{eq:dEdX_infinite_E_annihilation}), so is its interpretation. 
\begin{figure}[H]
    \centering
    \includegraphics[width=1\linewidth]{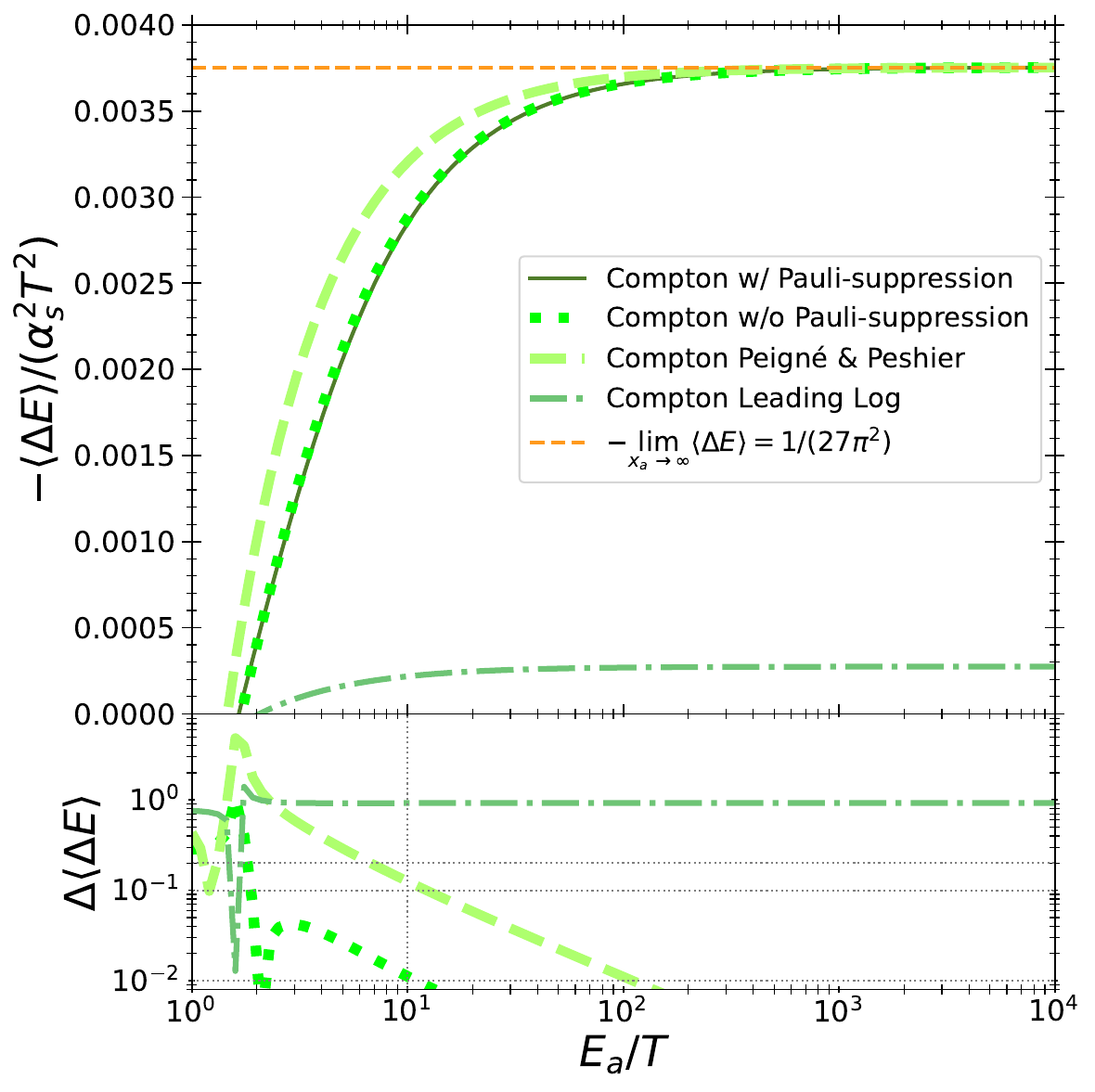}
    \caption{Top panel: Energy loss $\langle \Delta E\rangle/(\alpha_s^2T^2)$ as a function of $E_a/T$ for quark-gluon Compton scattering, where $m_\infty$ in Eq.~(\ref{eq:m_infty}) at $T=0.2$ GeV is used. Bottom panel: relative difference between various approximations and the tree-level result in Eq.~(\ref{eq:DeltaE_G_s_t}).}
    \label{fig:DeltaE_compton}
\end{figure}
The Peign\'e \& Peshier approach~\cite{Peigne:2007sd,Peigne:2008nd} yields
\begin{align}
    \langle\Delta E\rangle^{(PP)}\simeq-\frac{\alpha_s^2T^2}{9\pi^4}\left[\frac{\pi^2}{3}-\frac{4\zeta(3)}{x_a}\right]+O\left(\frac{z^2\ln(x_a)}{x_a}\right), 
\end{align}
while the (plus-distribution regularized) leading logarithm approximation gives
\begin{align}
    \langle\Delta E\rangle^{(LL)}\simeq-\frac{\alpha_s^2T^2}{18\pi^4}\ln^2(2)\left(1-\frac{z^2}{2x_a}\right).
\end{align}
As we have seen for quark-antiquark annihilation, the leading logarithm approach to Compton scattering is well below the $\langle \Delta E\rangle$ obtained from Eq.~(\ref{eq:DeltaE_G_s_t}). As depicted in Fig.~\ref{fig:DeltaE_compton}, using Eq.~(\ref{eq:DeltaE_Compton_wo_BE}) in Monte Carlo simulations of jet-medium interactions ensures that theoretical systematic uncertainties for $x_a>10$ deviate by $\lesssim 1$\% from the strict leading-order approximation in perturbation theory used in Eq.~(\ref{eq:DeltaE_G_s_t}). For $x_a\leq 10$, the strict leading-order calculation starts becoming ill-suited to describe jet-medium interactions in Monte Carlo simulations.  

\begin{figure}[H]
    \centering
    \includegraphics[width=1\linewidth]{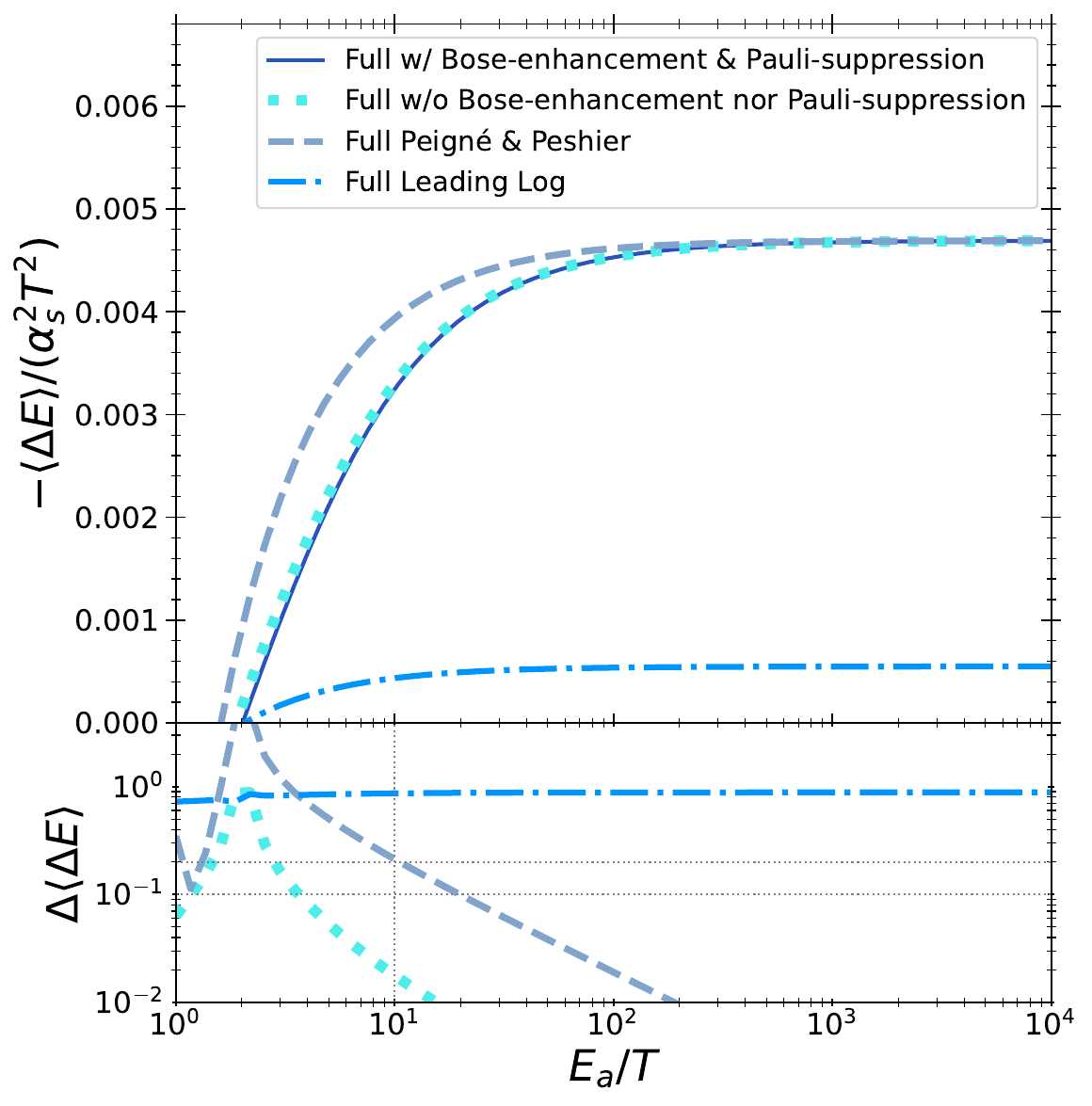}
    \caption{Comparison of $\langle \Delta E\rangle$ as a function of $E_a/T$ (top panel) and their combined relative differences with various approximation methods (bottom panel). $m_\infty$ in Eq.~(\ref{eq:m_infty}) at $T=0.2$ GeV is used throughout.}
    \label{fig:DeltaE_all}
\end{figure}

Figure~\ref{fig:DeltaE_all} depicts the tree-level leading-order sum of Compton scattering and annihilation. Inspecting the bottom panel of Fig.~\ref{fig:DeltaE_all} shows the advantage of Eq.~(\ref{eq:DeltaE_annih_wo_BE}) and Eq.~(\ref{eq:DeltaE_Compton_wo_BE}) in reproducing full tree-level massless quark energy loss via quark exchange within $\lesssim 2$\% of the full numerical result encapsulated in Eq.~(\ref{eq:DeltaE_annih}) and Eq.~(\ref{eq:DeltaE_G_s_t}).

%

\section{Longitudinal Drag Coefficient $\hat{\mathcal{F}}_{(L,1)}$}
\label{sec:fhat_L1}
The longitudinal drag coefficient is another important jet transport coefficient that encodes the jet energy loss in the medium. In our $2\to 2$ scattering setup, the jet quark is assumed to be travelling in the positive $z$-direction. Using light-cone coordinates, the longitudinal drag coefficient --- $\hat{\mathcal{F}}_{(L,1)}$ --- characterizes the average value of the positive light-cone component of the exchanged parton momentum in the in-medium scattering process, i.e. $\hat{\mathcal{F}}_{(L,1)}=\langle k^+\rangle$. To calculate this coefficient from tree-level $2\to 2$ scattering, the exchanged parton momentum $k^+$ is expressed in terms of the final phase-space integration variables as 
\begin{align}
    k^+=\frac{k^0+ k^z}{\sqrt{2}}=-\left[ \frac{2(E_2-E_b)}{\sqrt{2}}-\frac{t}{2\sqrt{2}E_a}\right]
    \label{eq:def_kplus},
\end{align}
where the outgoing jet parton carries momentum $p^\mu_1=p^\mu_a+k^\mu$, while the outgoing medium parton carries $p^\mu_2=p^\mu_b-k^\mu$.
\subsection{Longitudinal drag coefficient for quark-antiquark annihilation}
The $t$-channel processes, shown in Fig.~\ref{fig:fhat_diagrams}(b), and the corresponding squared matrix element in Eq.~(\ref{eq:MSquare_annihilation}) are used to determine the longitudinal drag coefficient via the $ k^+$-weighted matrix element squared. Constructing this object gives
\begin{align}
  k^+ \overline{\lvert\mathcal{M}\rvert^2}=- \frac{512\pi^2 \alpha^2_s}{9\sqrt{2}E_a} \left[ 4E_a(E_2-E_b) \frac{u}{t} - u \right], 
\end{align}
where Eq.~(\ref{eq:def_kplus}) has been used. The above equation suggests that the appropriate choice of Mandelstam integration variables herein is $t$ and $u$. Performing the integration over $t$ and $u$, $\hat{\mathcal{F}}_{(L,1)}$ can be recast as follows
\begin{align}
\hat{\mathcal{F}}_{(L,1)} &=- \frac{\alpha_s^2}{36\sqrt{2}\pi^5E_a^3}
\int_0^{\infty} dE_b\, f_b \int_0^{E_a+E_b}  dE_2 \left[1+f_{2}\right] \nonumber \\
&\times \Bigg\{4E_a\left(E_2-E_b\right)\mathcal{G}\left[\frac{u}{t}\right]-\mathcal{G}\left[u\right]\Bigg\},\\
\mathcal{G} \left[u\right] &= \frac{8\pi}{3} \lfloor E\rfloor \left\{ \lfloor E\rfloor^2 - 3 \lfloor u\rfloor   \right\}\label{eq:G_u}
\end{align}
where $\mathcal{G}[u/t]$ is defined in Eq.~(\ref{eq:ans_G_u_over_t}). Thus,
\begin{align}
\hat{\mathcal{F}}_{(L,1)}&= -\frac{\alpha_s^2}{36\sqrt{2}\pi^5E_a^3}\times \nonumber\\
&\times \Bigg[\int_0^{E_a}dE_bf_b \int_0^{E_a+E_b}dE_2\left[1+f_2\right]\left\{\textbf{I}+\textbf{II}+\textbf{III}\right\}\nonumber\\
&+\int_{E_a}^\infty dE_bf_b \int_{0}^{E_a+E_b}dE_2\left[1+f_2\right]\left\{\textbf{I}+\textbf{III}+\textbf{IV}\right\}
\Bigg],\label{eq:FL1_annih}
\end{align}
where,
\begin{align}
&\textbf{I}= 4E_a\left(E_2-E_b\right)\times \left[4\pi\frac{E_aE_2}{\sqrt{\left(E_b-E_2\right)^2+m_\infty^2}}\right]\nonumber\\
&\qquad -\frac{8\pi}{3}\left[E_2^3-3E_aE_2^2\right],\nonumber\\
&\textbf{II} = 4E_a\left(E_2-E_b\right)\times \left[4\pi \frac{E_b\left(E_a+E_b-E_2\right)}{\sqrt{\left(E_b-E_2\right)^2+m_\infty^2}}\right]\nonumber\\
&\qquad -\frac{8\pi}{3}\left[E_b^3-3\left(E_b\left(E_a+E_b-E_2\right)\right)\right], \nonumber\\
&\textbf{III}= 4E_a\left(E_2-E_b\right)\times \left[4\pi \frac{E_b\left(E_a+E_b-E_2\right)}{\sqrt{\left(E_b-E_2\right)^2+m_\infty^2}}\right] \nonumber\\
&\qquad -\frac{8\pi}{3}\left[\left(E_a+E_b-E_2\right)^3-3E_b\left(E_a+E_b-E_2\right)^2\right] \nonumber,\\
&\textbf{IV}= 4E_a\left(E_2-E_b\right)\times\left[4\pi \frac{E_aE_2}{\sqrt{\left(E_b-E_2\right)^2+m_\infty^2}}\right] \nonumber\\
&\qquad -\frac{8\pi}{3}\left[E_a^3-3E_a^2E_2\right] \nonumber
\end{align}

\begin{figure}[H]
    \centering
    \includegraphics[width=1\linewidth]{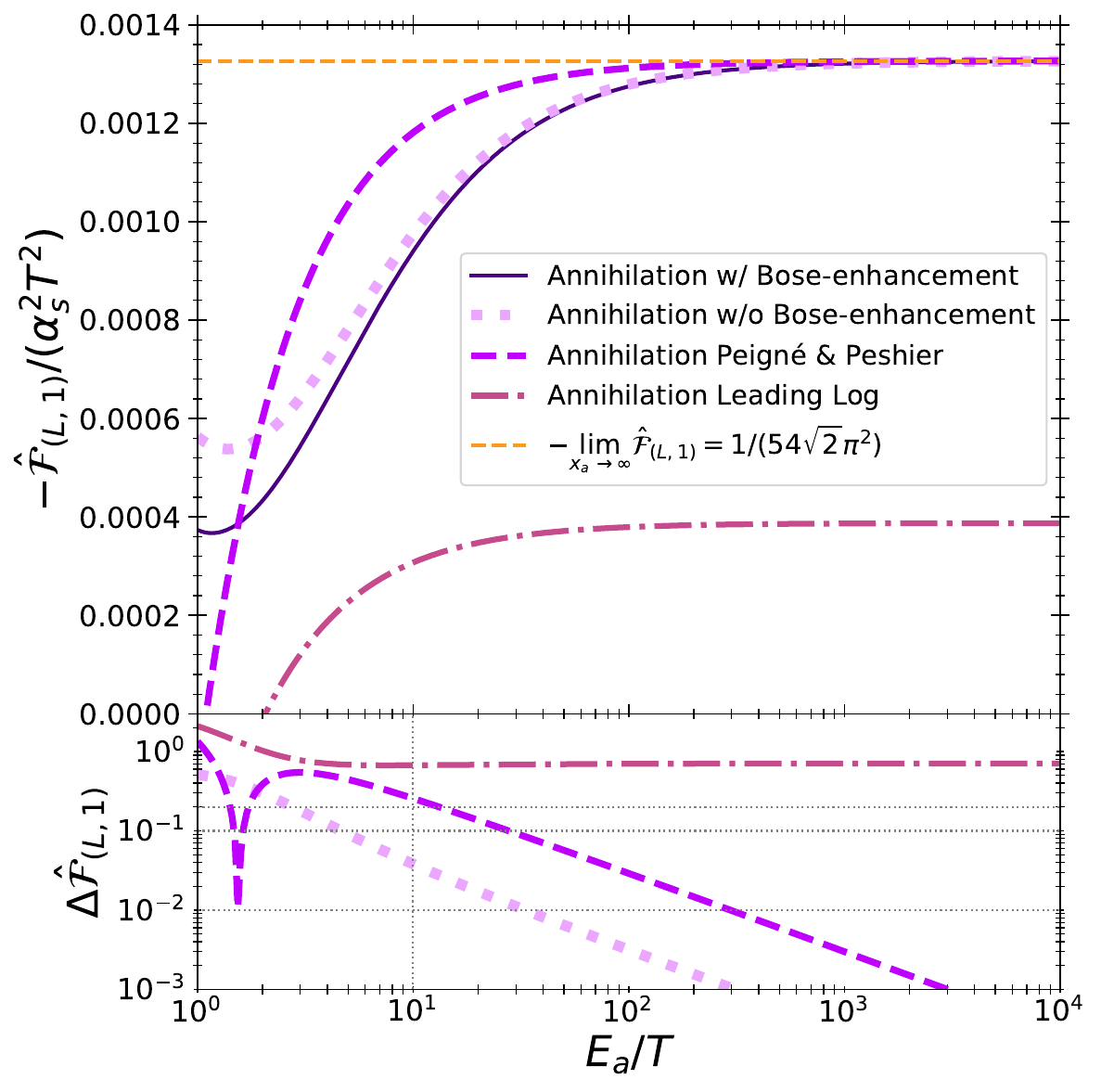}
    \caption{Top panel: $\hat{\mathcal{F}}_{(L,1)}$ for quark-antiquark annihilation, with $m_\infty$ in Eq.~(\ref{eq:m_infty}) evaluated at $T=0.2$ GeV. Bottom panel: relative difference as obtained by Eq.~(\ref{eq:DeltaF}).}
    \label{fig:FL1_Ann}
\end{figure}

Inspecting terms \textbf{I} through \textbf{IV}, one notices that $E_2-E_b$ changes sign over the integration range $E_2\in[0,E_a+E_b]$, and the Bose-enhancement $[1+f_2]$ in Eq.~(\ref{eq:FL1_annih}) does not cause $|\hat{\mathcal{F}}_{(L,1)}|$ to increase relative to $[1+f_2]\to 1$, instead the opposite is true, as is seen in Fig.~\ref{fig:FL1_Ann}. 
Under the geometric series approximation $[1+f_2]\simeq1+O(e^{-x_2})$, $\hat{\mathcal{F}}_{(L,1)}$ becomes
\begin{align}
\hat{\mathcal{F}}_{(L,1)} &\simeq -\frac{2\alpha^2_s T^2}{9\sqrt{2}\pi^4}\int_0^\infty \frac{dx_b}{e^{x_b}+1} \Bigg[  \frac{x_b^2}{2x_a} \label{eq:FL1_x_b_integral}  \\
& + \frac{z^2}{x_a^2}\left\{x_b\arcsinh\left(\frac{x_a}{z}\right) - x_a\arcsinh\left(\frac{x_b}{z}\right)\right\}  \nonumber \\
& + \frac{x_b}{x_a}\left\{\sqrt{x_a^2+z^2} - \sqrt{x_b^2+z^2}\right\} \Bigg]+O\left(e^{-x_2}\right),\nonumber 
\end{align}
which, using Appendix~\ref{app:th_int}, gives
\begin{align}
\hat{\mathcal{F}}_{(L,1)} &\simeq -\frac{2\alpha^2_s T^2}{9\sqrt{2}\pi^4} \left[ \frac{3\zeta(3)}{4x_a} - \frac{A^+_1(z)}{x_a} -\frac{z^2B^+_0(z)}{x_a}\right. \label{eq:FL1_annih_cform} \\
& + \left. \frac{\pi^2}{12} \frac{\sqrt{x^2_a+z^2}}{x_a} + \frac{\pi^2z^2}{12x^2_a}\arcsinh\left(\frac{x_a}{z}\right) \right]+O\left(e^{-x_2}\right).\nonumber 
\end{align}
In the asymptotic limit $x_{a}\rightarrow \infty$, one gets
\begin{align}
\lim_{x_{a}\rightarrow \infty}\left[\hat{\mathcal{F}}_{(L,1)}\right]&=-\frac{\alpha^2_s}{54\sqrt{2}\pi^2}T^2.
\label{eq:FL1_asym_annihilation}
\end{align}
As was the case for $\langle \Delta E\rangle$, the longitudinal drag coefficient $\hat{\mathcal{F}}_{(L,1)}/(\alpha^2_s T^2)$, has a non-vanishing value in the infinite quark energy limit. The Peign\'e \& Peshier approach~\cite{Peigne:2007sd,Peigne:2008nd} gives
\begin{align}
\hat{\mathcal{F}}^{(PP)}_{(L,1)}\simeq-\frac{2\alpha_s^2T^2}{9\sqrt{2}\pi^4}\left[\frac{\pi^2}{12}-\frac{3\zeta(3)}{4x_a}\right]+O\left(\frac{z^2}{x_a}\right), 
\end{align}
while the leading logarithm yields
\begin{align}
\hat{\mathcal{F}}^{(LL)}_{(L,1)}\simeq-\frac{\alpha_s^2T^2}{9\sqrt{2}\pi^4}\ln^2(2)\left(1-\frac{z^2}{2x_a}\right),
\end{align}
where the plus-distribution regularization in Eq.~(\ref{eq:plus_dist}) was used to obtain the closed-form result above. The leading log, Peign\'e \& Peshier, as well as Eq.~(\ref{eq:FL1_annih_cform}) are all depicted in Fig.~\ref{fig:FL1_Ann}, along with the relative difference $\Delta \hat{\mathcal{F}}_{(L,1)}$ defined in Eq.~(\ref{eq:DeltaF}). 

\subsection{Longitudinal drag coefficient for Compton scattering}
\begin{figure}[H]
    \centering
    \includegraphics[width=1\linewidth]{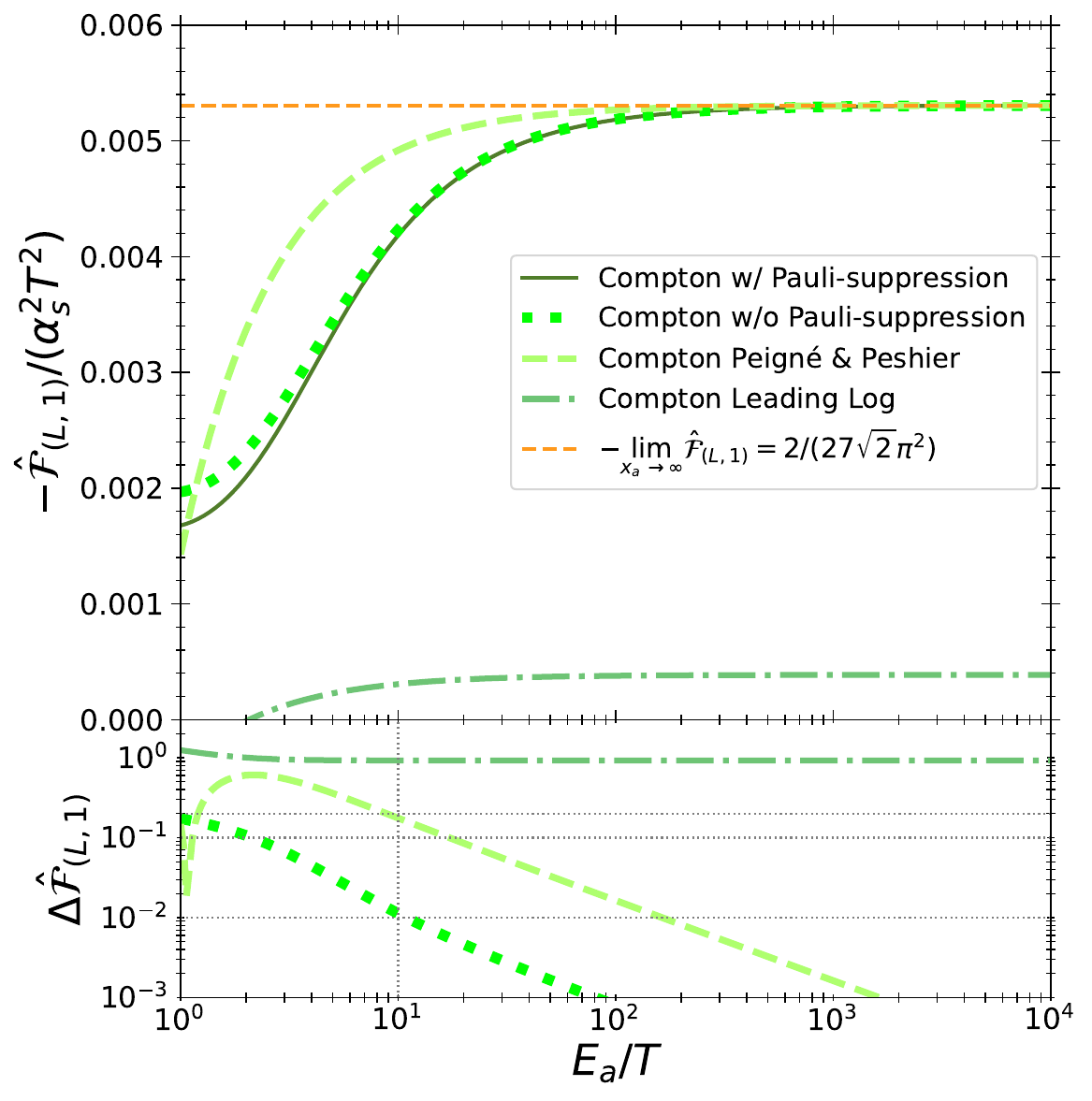}
    \caption{Top panel: $\hat{\mathcal{F}}_{(L,1)}$, as a function of $E_a/T$ for quark-gluon Compton scattering. Here, $m_\infty$ in Eq.~(\ref{eq:m_infty}) at $T=0.2$ GeV is used. Bottom panel shows the relative error obtained by Eq.~(\ref{eq:DeltaF}) for $(i)=(L,1)$.}
    \label{fig:FL1_compton}
\end{figure}

\begin{figure}[H]
    \centering
    \includegraphics[width=1\linewidth]{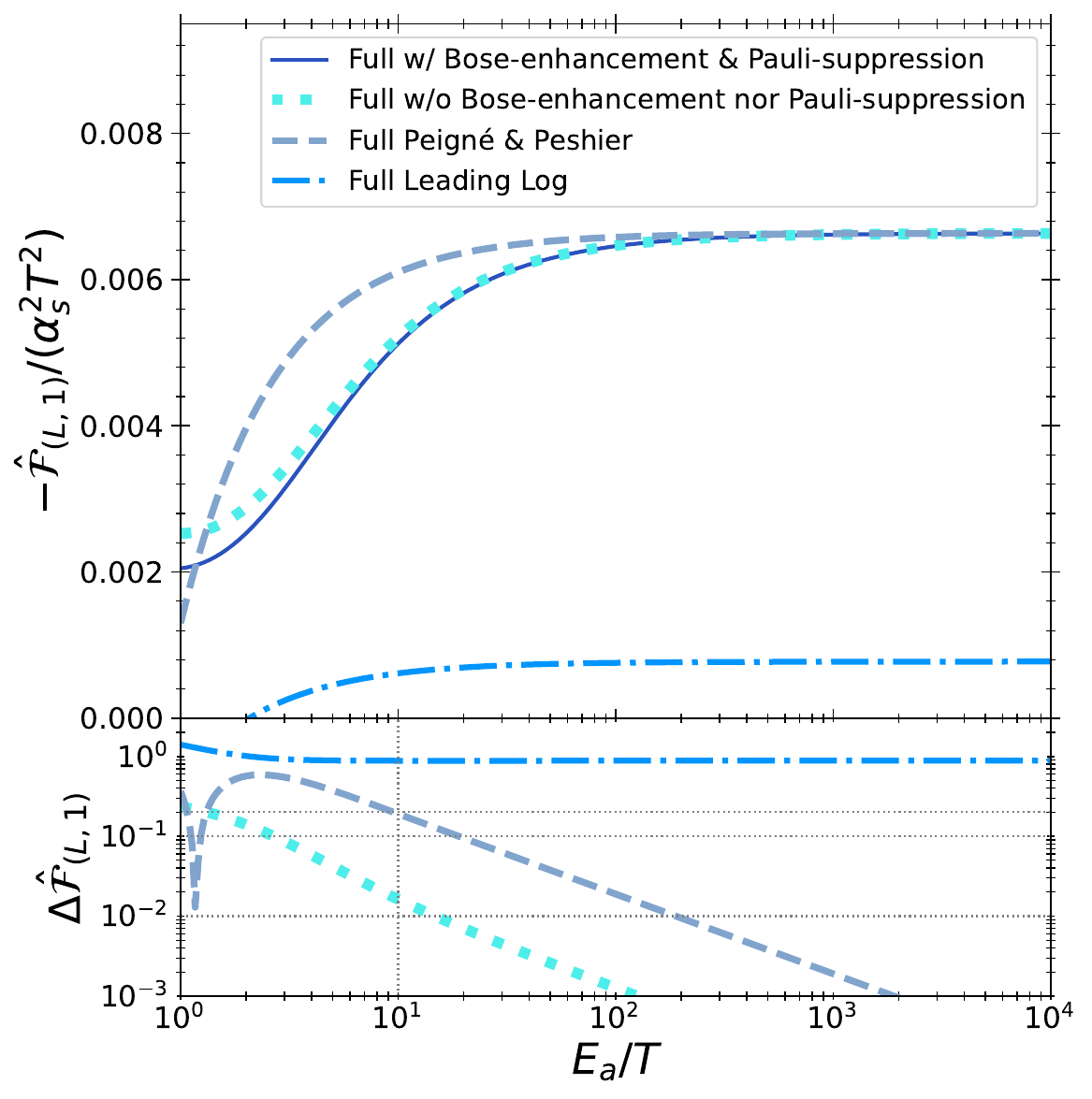}
    \caption{Top panel: $\hat{\mathcal{F}}_{(L,1)}$ from both annihilation and Compton scattering is shown against various approximations, where $m_\infty$ in Eq.~(\ref{eq:m_infty}) is evaluated at $T=0.2$ GeV. Bottom panel: the relative difference $\Delta \hat{\mathcal{F}}_{(L,1)}$ defined via Eq.~(\ref{eq:DeltaF}).}
    \label{fig:FL1_Full}
\end{figure}

\begin{figure}[H]
    \centering
\includegraphics[width=1.1\linewidth]{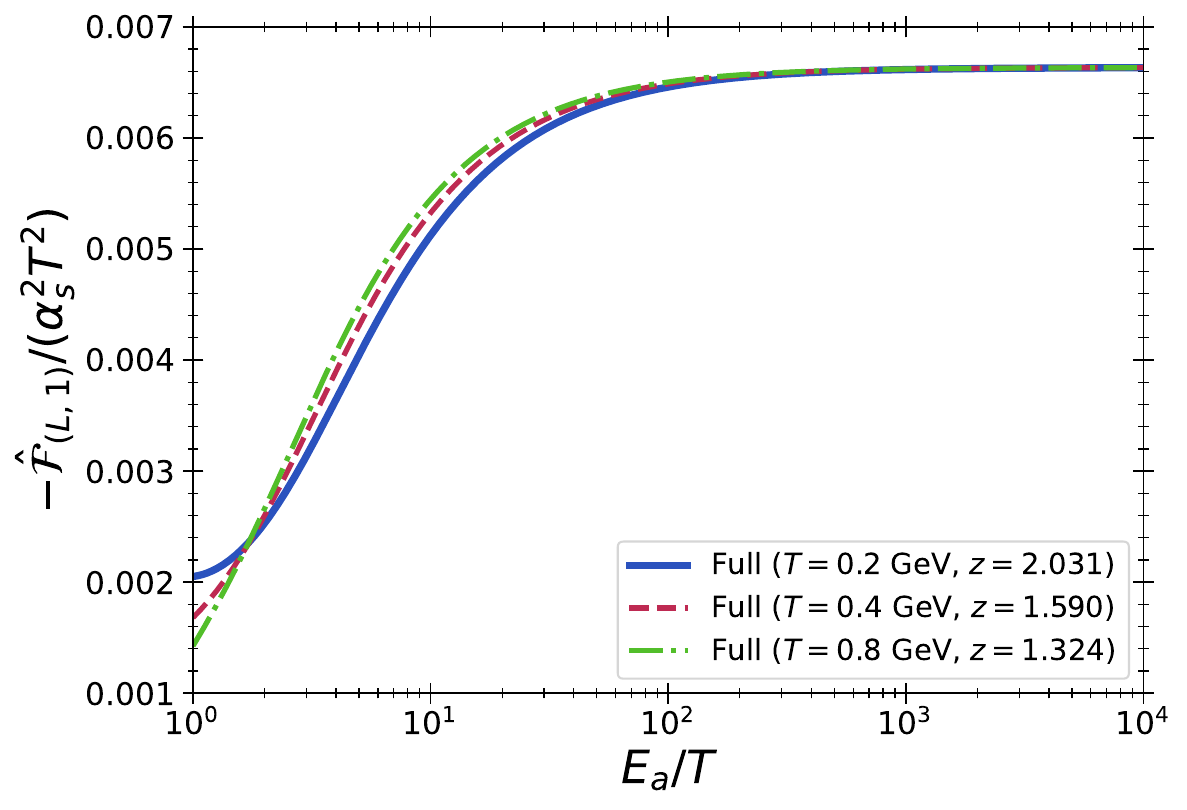}
    \caption{$\hat{\mathcal{F}}_{(L,1)}/(\alpha_s^2T)$ as a function of $x_a$ for $T=[0.2,0.4,0.8]$ GeV.}
    \label{fig:FL1_full_w_T}
\end{figure}

The squared matrix element for this process is given in Eq.~(\ref{eq:Compton_MSquare}), from which we construct its $ k^+$-weighted object 
\begin{align}
    k^{+} \overline{\lvert\mathcal{M}\rvert^2} &= -\frac{512\pi^2\alpha^2_s}{9\sqrt{2}E_a} \left[-4E_a\left(E_2-E_b\right)\frac{s}{t}+s\right],
    \label{eq:k+_weigh_compton}
\end{align}
which is needed to compute the drag coefficient. After integration over Mandelstam $t$ and $s$, $\hat{\mathcal{F}}_{(L,1)}$ is
\begin{align}
\hat{\mathcal{F}}_{(L,1)} &=-\frac{\alpha_s^2}{36\sqrt{2}\pi^5E_a^3}
\int_0^{\infty} \frac{dE_b}{e^{ \beta E_b}-1} \int_0^{E_a+E_b}  dE_2 \left[1-f_{2}\right] \nonumber \\
&\times \left\{-4E_a\left(E_2-E_b\right)\mathcal{G}\left[\frac{s}{t}\right]+\mathcal{G}\left[s\right]\right\},\label{eq:FL1}\\
\mathcal{G}[s]&=\frac{8\pi}{3} \lfloor E\rfloor \Big\{ \lfloor E\rfloor^2 + 3 \lfloor s\rfloor\Big\},\label{eq:G_s}
\end{align}
while $\mathcal{G}[s/t]$ is defined in Eq.~(\ref{eq:G_s_over_t}). The resulting transport coefficient is depicted in Fig.~\ref{fig:FL1_compton}, along with various approximations that are discussed below. Using the geometric series approximation $[1-f_2]\simeq 1+O\left(e^{-x_2}\right)$, one obtains
\begin{align}
\hat{\mathcal{F}}_{(L,1)}&\simeq-\frac{2\alpha^2_s T^2}{9\sqrt{2}\pi^4}\int_0^\infty\frac{dx_b}{e^{x_b}-1}\\
& \times \left[\left(\frac{1}{3}+\frac{x_b}{x_a}+\frac{4z^2}{3x_a^2}\right)\sqrt{x_a^2+z^2} \right. \nonumber\\
& -\left(\frac{x_b}{x_a}+\frac{x_b^2}{3x_a^2}+\frac{4z^2}{3x_a^2}\right)\sqrt{x_b^2+z^2} \nonumber\\ & +\frac{z^2(x_b-x_a)}{x_a^2}
\left\{\arcsinh\left(\frac{x_a}{z}\right)+\arcsinh\left(\frac{x_b}{z}\right)\right\} \nonumber \\ 
& + \frac{x_b^2}{x_a}+\sgn(x_a-x_b)\frac{4z^3}{3x_a^2} \nonumber\\
& \left.-\sgn(x_a-x_b)\right.
\frac{(x_a-x_b)^2+4z^2}{3x_a^2}\sqrt{\left(x_a-x_b\right)^2+z^2} \nonumber \\
& + \left. \frac{|x_a-x_b|z^2}{x_a^2}\arcsinh\left(\frac{x_a-x_b}{z}\right) \right]+O\left(e^{-x_2}\right).\nonumber
\end{align}
Finally, using Appendix  \ref{app:th_int} one gets
\begin{align}
    \hat{\mathcal{F}}_{(L,1)}  &\simeq -\frac{2\alpha^2_sT^2}{9\sqrt{2}\pi^4}\left[  \frac{\pi^2}{6x_a}\sqrt{x^2_a+z^2} - \frac{A_1^-(z)}{x_a}-\frac{A_2^-(z)}{3x_a^2}
     \right. \nonumber \\
    & -\left(\frac{1}{3} +\frac{4z^2}{3x^2_a} \right) \lim_{\epsilon\rightarrow 0^+}\left\{\Delta \alpha^{-}_{0}(\epsilon)+\sqrt{x^2_a +z^2} \ln(\epsilon)\right\} \nonumber \\
    & -\frac{4z^2}{3x_a^2}\lim_{\epsilon\rightarrow 0^+}\left\{A_0^-(\epsilon;z)+z\ln(\epsilon)\right\} \nonumber \\
    & +\frac{z^2B_1^-(z)}{x_a^2}-\frac{z^2B_0^-(z)}{x_a} + \frac{\pi^2 z^2}{6x^2_a}\arcsinh\left(\frac{x_a}{z}\right) \nonumber \\
    & + \frac{z^2}{x_a} \lim_{\epsilon\rightarrow 0^+}\left\{\Delta\beta^-_{0}(\epsilon) + \arcsinh\left(\frac{x_a}{z}\right)\ln(\epsilon)  \right\} \nonumber  \\
    & + \frac{2\zeta(3)}{x_a} + \frac{8z^3}{3x^2_a}\ln\left(1-e^{-x_a}\right) - \frac{z^2}{x^2_a}\Delta\beta^{-}_{1} - \frac{2}{3x_a}\Delta\alpha^{-}_{1} \nonumber \\
    & \left.  -\frac{1}{3x^2_a}\Delta\alpha^-_2 \right]+O(e^{-x_2}). 
    \label{eq:FL1_wo_pb}
\end{align}
In the asymptotic limit $x_{a}\rightarrow \infty$, the transport coefficient $\hat{\mathcal{F}}_{(L,1)}$, takes the following form:
\begin{align}
    \lim_{x_{a}\rightarrow \infty}\left[\hat{\mathcal{F}}_{(L,1)}\right] & =  -\frac{2\alpha^2_sT^2}{9\sqrt{2}\pi^4} \lim_{x_{a}\rightarrow \infty} \int^{x_{a}}_{0} dx_{b} \frac{2x_b}{e^{x_{b}}-1} \nonumber \\
    & =- \frac{2\alpha^2_s}{27\sqrt{2}\pi^2}T^2,
\end{align}
which is non-vanishing in the infinite quark energy limit. The Peign\'e \& Peshier approach~\cite{Peigne:2007sd,Peigne:2008nd} and the leading log yield
\begin{align}
\hat{\mathcal{F}}^{(PP)}_{(L,1)}&\simeq-\frac{2\alpha_s^2T^2}{9\sqrt{2}\pi^4}\left[\frac{\pi^2}{3}-\frac{2\zeta(3)}{x_a}\right]+O\left(\frac{z^2\ln(x_a)}{x_a}\right),\\
\hat{\mathcal{F}}^{(LL)}_{(L,1)}&\simeq-\frac{\alpha_s^2T^2}{9\sqrt{2}\pi^4}\ln^2(2)\left(1-\frac{z^2}{2x_a}\right)
\end{align}
Unlike the annihilation case, the inclusion of the Pauli-suppression in the Compton scattering dampens Eq.~(\ref{eq:FL1_wo_pb}), as shown in  Fig.~\ref{fig:FL1_compton}, which is the expected behaviour. 
Indeed, instead of magnifying the magnitude of both positive and negative regions, which drove the cancellations and a downward shift in the Bose-enhanced $q\bar q$ annihilation case, the $\left(1-f_2\right)$ factor suppresses the absolute weight of these contributions uniformly across phase space. Consequently, the cancellations in the $E_2<E_b$ domain are dampened rather than enhanced, yielding a final curve in Fig.~\ref{fig:FL1_compton} closer to the unsuppressed line. The precision of various approximations to reproduce Eq.~(\ref{eq:FL1}) is shown in the bottom panel of Fig.~\ref{fig:FL1_compton}. The sum of the contributions to $\hat{\mathcal{F}}_{(L,1)}$ from quark-antiquark annihilation and Compton scattering is shown in Fig.~\ref{fig:FL1_Full}, and is compared against various approximations. The pattern of which approximation performs better is the same as in Fig.~\ref{fig:FL1_Ann} and Fig.~\ref{fig:FL1_compton}. Lastly, as $z=m_\infty/T$ changes with temperature through Eq.~(\ref{eq:m_infty}), so does $\hat{\mathcal{F}}_{(L,1)}$ as is depicted in Fig.~\ref{fig:FL1_full_w_T} for three commonly reached temperatures in heavy-ion collisions. More details about the temperature dependence of jet-medium transport coefficients are presented in Section~\ref{sec:Fhat_w_T}. 
\begin{widetext}
\section{Transverse Momentum Broadening}
\label{sec:fhat_T2}
To calculate the coefficient $\hat{\mathcal{F}}_{(T,2)}$, it is convenient to first express the transverse momentum transfer $\pmb{k}_\perp^2$ in terms of the final phase-space integration variables: $s$, $t$, $E_b$, and $E_2$. As stated before, the $z$-axis is defined to be collinear with the initial quark's momentum, $\pmb p_a=p_a\hat z$; the square of the transverse momentum gained or lost by the quark-induced jet is
\begin{align}
    k_\perp^2=|\pmb{k}|^2\sin^2(\theta_{ak}),
    \label{eq:k_perp_sintheta}
\end{align}
where $\theta_{ak}$ is the angle between incoming jet quark momentum $\pmb{p}_a$ and exchanged (anti)quark three-momentum $\pmb{k}$. The angular dependence can be expressed in terms of the dot product $\pmb{p}_a\cdot \pmb{k}$ as follows
\begin{align}
    \cos(\theta_{ak})=\frac{\pmb{p}_a\cdot\pmb{k}}{\lvert\pmb{p}_a\rvert\lvert\pmb{k}\,\rvert}=\frac{\pmb{p}_a\cdot\pmb{p}_b - \pmb{p}_a\cdot\pmb{p}_2}{\lvert\pmb{p}_a\rvert\lvert\pmb{k}\,\rvert},
    \label{eq:k_perp_costheta}
\end{align}
where, $\pmb{k}=\pmb{p}_b - \pmb{p}_2$ has been used. The dot products in the numerator of Eq.~(\ref{eq:k_perp_costheta}) can be expressed in terms of Mandelstam variable $s$ and $t$ as follows
\begin{align}	
\pmb{p}_a\cdot\pmb{p}_2=\frac{2E_aE_2-s-t}{2},\quad \pmb{p}_a\cdot\pmb{p}_b=\frac{2E_aE_b-s}{2}.
\end{align}
To compute $\hat{\mathcal{F}}_{(T,2)}$, $\lvert\pmb{k}\rvert=\sqrt{\left(E_2-E_b\right)^2-t}$ is used to express momentum $k^2_\perp$ in terms of the parton energies and $t$, simplifying Eq.~(\ref{eq:k_perp_sintheta}) to
\begin{align}
	k_\perp^2&=-\frac{t^2+4E_aE_1t}{4E_a^2}.
    \label{eq:def_kperpSq}
\end{align}
%

\subsection{Quark-antiquark annihilation induced transverse momentum broadening}
Using Eq.~(\ref{eq:MSquare_annihilation}), the transverse momentum broadening coefficient is
\begin{align}
k^2_{\perp} \overline{\lvert\mathcal{M}\rvert^2}= \frac{1024\pi^2 \alpha^2_s}{36E^2_a} \Big[ - ut - 4E_a\left(E_a+E_b-E_2\right)u \Big] , 
\end{align}
where Eq.~(\ref{eq:def_kperpSq}) is used to simplify the expression. Substituting this result into the definition of the transverse momentum broadening coefficient $\hat{\mathcal{F}}_{(T,2)}$ in Eq.~(\ref{OHat}) and carrying out the Mandelstam integrals over $u$ and $t$, gives
\begin{align} \hat{\mathcal{F}}_{(T,2)}&=
\frac{\alpha^2_s}{72\pi^5E_a^4}\int_0^\infty\frac{dE_b}{e^{\beta E_b}+1}\int_0^{E_a+E_b}\! \! dE_2  \left[1+ f_2\right]\Big\{ -\mathcal{G} \left[ut\right] -4E_a\left(E_a+E_b-E_2\right)\mathcal{G} \left[u\right]\Big\}, 
\end{align}
where $\mathcal{G}\left[u\right]$ is given in Eq.~(\ref{eq:G_u}) while
\begin{align}
\mathcal{G} \left[ut\right] = \frac{32\pi}{15} \lfloor E\rfloor \left\{ 2\lfloor E\rfloor^4 +5\lfloor E\rfloor^2\left(\lfloor u\rfloor + \lfloor t\rfloor\right) + 10\lfloor u\rfloor \lfloor t\rfloor \right\},
\end{align}
with the various floor functions defined in Eq.~(\ref{eq:all_floors}). The two-dimensional integration over $E_b$ and $E_2$ is partitioned into six distinct kinematic regions. Within each region, the floor functions assume definite values, allowing the integrals to be evaluated explicitly. The corresponding phase-space boundaries and the expressions for the floor functions in each region are summarized in the Appendix~\ref{app:int_over_floor}. The transverse momentum broadening coefficient $\hat{\mathcal{F}}_{(T,2)}$ is given by 
\begin{align} 
&\hat{\mathcal{F}}_{(T,2)}= \frac{4\alpha^2_{s}T^3}{27\pi^4}\int_0^\infty\frac{dx_b}{e^{x_b}+1}\Bigg[x_b^2+\frac{x_b^3}{6x_a}-\frac{6x_b\polylog{2}(e^{-x_b})}{x_a}+\frac{6}{x_a}\Big\{\zeta(3)-\polylog{3}\left(e^{-x_b}\right)\Big\}+\frac{6x_b}{x^2_a}\Big\{\zeta(3)+3\polylog{3}\left(e^{-x_b}\right)\Big\}\nonumber\\
&+\frac{24}{x^2_a}\Big\{\polylog{4}\left(e^{-x_b}\right)-\zeta(4)\Big\}+\frac{6x_b}{x_a^3}\Big\{\polylog{4}\left(e^{-x_a-x_b}\right)-\polylog{4}\left(e^{-x_a}\right)\Big\}-\frac{6x_b}{x^3_a}\Big\{5\polylog{4}\left(e^{-x_b}\right)+3\zeta(4)\Big\}+\frac{48}{x^3_a}\Big\{\zeta(5)-\polylog{5}\left(e^{-x_b}\right)\Big\}\nonumber\\
&-\frac{24x_b}{x_a^4}\Big\{\polylog{5}\left(e^{-x_a}\right)+\polylog{5}\left(e^{-x_a-x_b}\right)\Big\}+\frac{24x_b}{x_a^4}\Big\{\polylog{5}(e^{-x_b})+\zeta(5)\Big\}+\frac{48}{x_a^4}\Big\{\polylog{6}\left(e^{-x_b}\right)-\zeta(6)\Big\} \nonumber \\
&+\frac{48}{x_a^4}\Big\{\polylog{6}\left(e^{-x_a}\right)-\polylog{6}\left(e^{-x_a-x_b}\right)\Big\}\Bigg],\label{eq:FT2_x_b_withBose_reordered}
\end{align}
where all the integrals admit a closed-form expression, given in Eq.~(\ref{eq:FT2_Ann_MultiZeta}). Simplifying Eq.~(\ref{eq:FT2_Ann_MultiZeta}) (c.f. Appendix~\ref{app:MultiLiZeta}), gives
\begin{align}
\hat{\mathcal{F}}_{(T,2)}&=\!\frac{4\alpha^2_{s}T^3}{27\pi^4} \Bigg[ \frac{3}{2}\zeta(3) + \frac{1}{x_a} \Bigg\{12\polylog{4}\left(\frac{1}{2}\right)-\frac{37\pi^4}{360}\Bigg\} +\frac{1}{2x_a} \Big\{21\zeta(3)\ln(2)- \pi^2\ln^2(2) + \ln^4(2)  \Big\} +\frac{1}{8x^2_a}\Big\{393\zeta(5) - 35\pi^2\zeta(3)\Big\} \nonumber \\
& +\frac{1}{x^3_a}\Bigg\{ 93\ln(2)\zeta(5)+\frac{9}{2}\zeta^2(3)-\frac{50567\pi^6}{393120} \Bigg\} +\frac{1}{13x^3_a}\Big\{3\ln^4(2)\pi^2-4\ln^2(2)\pi^4-3\ln^6(2)\Big\} \nonumber\\
& +\frac{1}{13x^3_a}\Bigg\{1728\polylog{6}\left(\frac{1}{2}\right)-243\polylog{6}\left(\frac{1}{4}\right)+16\polylog{6}\left(-\frac{1}{8}\right)\Bigg\} +\frac{1}{16x^4_a}\Big\{ 1899\zeta(7)-178\pi^2\zeta(5) \Big\}\nonumber\\
&-\frac{1}{2x_a^3}\Big\{\pi^2\polylog{4}\left(e^{-x_a}\right)+ 12\polylog{2,4}\left(-1,-e^{-x_a}\right)\Big\}-\frac{2}{x_a^4}\Big\{\pi^2\polylog{5}\left(e^{-x_a}\right)-12 \polylog{2,5}\left(-1,-e^{-x_a}\right)\Big\}\nonumber\\
& +\frac{48}{x_a^4}\Big\{\ln(2)\polylog{6}\left(e^{-x_a}\right)+  \polylog{1,6}\left(-1,-e^{-x_a}\right)\Big\} \Bigg],\label{eq:FT2_annihilation_full_MZV}
\end{align}
\end{widetext}
where $\zeta^2(3)\equiv\left[\zeta(3)\right]^2$ and $\polylog{n,m}(z_1,z_2)$ represents the multiple polylogarithm defined in Eq.~(\ref{eq:append_polylog}).

The transverse momentum broadening, without accounting for the Bose-enhancement factor for the outgoing thermal gluon, can be written in a closed-form expression as
\begin{align}
\hat{\mathcal{F}}_{(T,2)}&\simeq\frac{2\alpha^2_{s}T^3}{9}\left[\frac{\zeta(3)}{\pi^4}+\frac{7}{1080x_a}\right]+O\left(e^{-x_2}\right).\label{eq:FT2_annihilation_no_exp}
\end{align}
When comparing Eq.~(\ref{eq:FT2_annihilation_no_exp}) and Eq.~(\ref{eq:FT2_annihilation_full_MZV}), one realizes that two categories of terms are missing in Eq.~(\ref{eq:FT2_annihilation_no_exp}). While exponentially suppressed terms, e.g. $\polylog{n}\left(e^{-x_a}\right)$ and $\polylog{n,m}\left(-1,-e^{-x_a}\right)$, are expected to be absent from Eq.~(\ref{eq:FT2_annihilation_no_exp}), the lack of power-law corrections $x^{-2}_a$, $x^{-3}_a$ and $x^{-4}_a$ in Eq.~(\ref{eq:FT2_annihilation_full_MZV}) are not immediately anticipated. Note that while the $x^{-2}_a$, $x^{-3}_a$ and $x^{-4}_a$ terms may be small for the hardest partons within the core of the jet, as one goes towards softer, more peripheral jet partons, the $x^{-2}_a$, $x^{-3}_a$ and $x^{-4}_a$ corrections are not negligible. In fact, not including these higher-order corrections would bias Bayesian analysis of jet-medium interactions. As Eq.~(\ref{eq:FT2_annihilation_no_exp}) does not contain $x^{-2}_a$, $x^{-3}_a$ and $x^{-4}_a$ terms, we also inspect the approximation where Eq.~(\ref{eq:FT2_annihilation_full_MZV}) is devoid of $\polylog{n,m}\left(-1,-e^{-x_a}\right)$ terms.

\begin{figure}[H]
\vspace{0.2cm}
    \centering
    \includegraphics[width=1\linewidth]{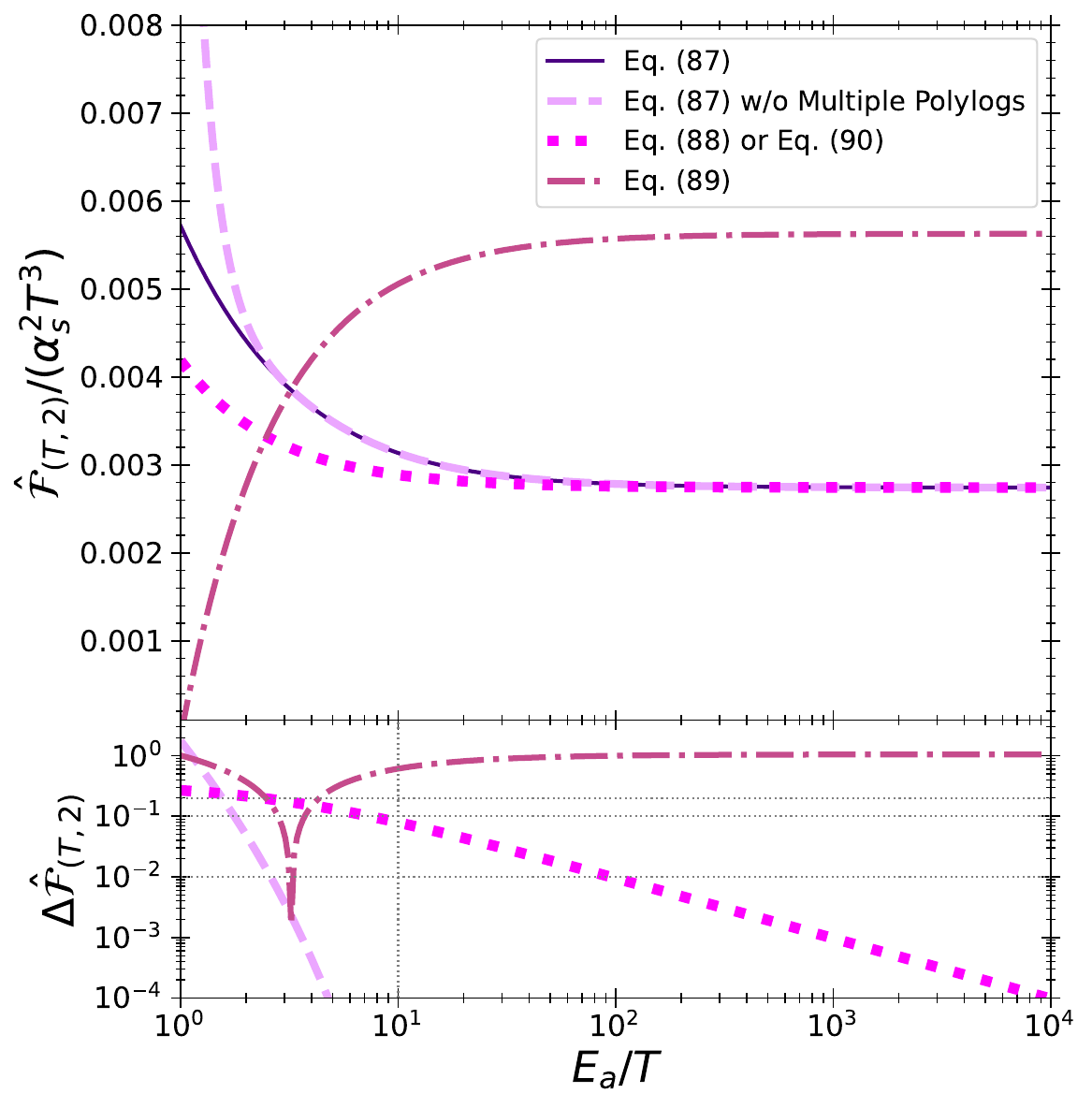}
    \caption{Top panel: Comparison of $\hat{\mathcal{F}}_{(T,2)}/(\alpha_s^2T^3)$ and its various approximations. Bottom panel: relative difference between approximations and the tree-level result defined via Eq.~(\ref{eq:DeltaF}).}
    \label{fig:FT2_Annihilation}
\end{figure}

The leading log approximation and Peign\'e \& Peshier approach~\cite{Peigne:2007sd,Peigne:2008nd}, $\hat{\mathcal{F}}_{(T,2)}$ simplifies to 
\begin{align}
\hat{\mathcal{F}}^{(LL)}_{(T,2)}&\simeq\frac{\alpha^2_s T^3}{18\pi^2}\left(1-\frac{z^2}{2x_a}\right),\label{eq:fsaLL_FT2_Annihilation} \\
\hat{\mathcal{F}}^{(PP)}_{(T,2)}&\simeq\frac{2\alpha_s^2T^3}{9\pi^4}\left[\zeta(3)+\frac{7\pi^4}{1080x_a}\right]+O\left(e^{-x_2}\right).\label{eq: FT2 Ann P&P}
\end{align}
The Peign\'e \& Peshier approximation gives the same result as Eq.~(\ref{eq:FT2_annihilation_no_exp}), as no cut-off enters into the $t$-channel of $\hat{\mathcal{F}}_{(T,2)}$. Figure~\ref{fig:FT2_Annihilation} compares various approaches and provides the relative error $\Delta \hat{\mathcal{F}}_i$ for each, as defined in Eq.~(\ref{eq:DeltaF}).
\begin{widetext}
\subsection{Transverse momentum broadening for Compton scattering}
In the Compton scattering process explored here, which proceeds through $t$-channel quark exchange as illustrated in Fig.~\ref{fig:fhat_diagrams}(c) and whose matrix element is given in Eq.~(\ref{eq:Compton_MSquare}), the transverse momentum broadening, i.e. the $\pmb{k}^2_{\perp}$ weighted matrix element squared, is
\begin{align}
    \pmb{k}_\perp^2  \overline{\lvert\mathcal{M}\rvert^2}=& \frac{256\pi^2}{ 9 E^{2}_{a} }\alpha^2_s \left[st+4E_a \left(E_a+E_b-E_2\right)s\right].
\end{align}
The transverse momentum broadening coefficient is 
\begin{eqnarray}
\!\!\! \hat{\mathcal{F}}_{(T,2)}&\!\!\!=\!\!\!&\frac{\alpha^2_s}{72\pi^5E_a^4} \int_0^\infty\frac{dE_b}{e^{\beta E_b}-1}\int_0^{E_a+E_b}dE_2 \left[1-f_{2}\right] \Big\{\mathcal{G}[st]+4E_a\left(E_a+E_b-E_2\right)\mathcal{G}[s]\Big\},
\end{eqnarray} 
where $\mathcal{G}[s]$ is defined in Eq.~(\ref{eq:G_s}), while $\mathcal{G}[st]$
\begin{align}
\mathcal{G}[st]=\frac{32\pi}{15}\left\lfloor E \right\rfloor\Big\{ 2\left\lfloor E \right\rfloor^4 + 5\left\lfloor E \right\rfloor^2 \left(\lfloor s \rfloor - \lfloor t \rfloor\right) - 10\left\lfloor s \right\rfloor \lfloor t \rfloor\Big\},\nonumber
\end{align}
The functions $\mathcal{G}[s]$ and $\mathcal{G}[st]$ depend on the floor functions, whose kinematically allowed integration range for $E_b$ and $E_2$ is divided into six different regions as illustrated in Appendix~\ref{app:int_over_floor}.  Within each region the integrals are evaluated explicitly, giving
\begin{align}
\hat{\mathcal{F}}_{(T,2)} &=\frac{4\alpha^2_{s}T^3}{27\pi^4}\int_0^\infty\frac{dx_b}{e^{x_b}-1}\Bigg[\frac{3x_b^2}{2} +\frac{x_b^3}{2x_a}-\frac{6x_b}{x_a}\polylog{2}(-e^{-x_b})-\frac{6}{x_a} \left\{ \polylog{3}\left(-e^{-x_b}\right) + \frac{3\zeta(3)}{4}\right\}\label{eq:FT2_Compton_withPauliFactor}\nonumber\\
&+\frac{3x_b}{x_a^2}\Big\{6\polylog{3}\left(-e^{-x_b}\right)+2\polylog{3}\left(-e^{-x_a}\right)-3\zeta(3)\Big\} +\frac{30}{x_a^2} \left\{ \polylog{4}\left(-e^{-x_b}\right) + \frac{7\pi^4}{720}\right\} -\frac{6}{x_a^2} \Big\{\polylog{4}\left(-e^{-x_a}\right)-\polylog{4}\left(-e^{-x_a-x_b}\right)\Big\} \nonumber \\
&+\frac{x_b}{x_a^3}\left\{\frac{49\pi^4}{120}-\frac{9\zeta(3)x_b}{2}-30\polylog{4}\left(-e^{-x_b}\right)\right\}+\frac{6x_b}{x_a^3}\Big\{x_b\polylog{3}\left(-e^{-x_a}\right)+\polylog{4}\left(-e^{-x_a-x_b}\right)\Big\}-\frac{6x_b}{x_a^3}\polylog{4}\left(-e^{-x_a}\right)\nonumber\\
&+\frac{12x_b^2}{x_a^4}\polylog{4}\left(-e^{-x_a}\right)+\frac{x_b}{x_a^4}\left\{\frac{7\pi^4x_b}{60}-45\zeta(5)+24\polylog{5}\left(-e^{-x_b}\right)\right\}-\frac{24x_b}{x_a^4}\Big\{2\polylog{5}\left(-e^{-x_a}\right)+\polylog{5}\left(-e^{-x_a-x_b}\right)\Big\} \nonumber \\ 
& -\frac{72}{x_a^3} \left\{ \polylog{5}\left(-e^{-x_b}\right)+\frac{15}{16}\zeta(5)\right\} + \frac{72}{x^4_a}\left\{ \polylog{6}\left(-e^{-x_b}\right) +\frac{31\pi^6}{30240}\right\}+ \frac{72}{x_a^4}\Big\{\polylog{6}\left(-e^{-x_a}\right)-\polylog{6}\left(-e^{-x_a-x_b}\right)\Big\}\Bigg].
\end{align}
Using the methodology from Appendix~\ref{app:MultiLiZeta}, the fully analytical result in Eq.~(\ref{eq:FT2_Compton_MultiZeta}) can be simplified into
\begin{align}
\hat{\mathcal{F}}_{(T,2)} &= \frac{4\alpha^2_{s}T^3}{27\pi^4} \Bigg[ 3\zeta(3)  +\frac{1}{x_a}\left\{ \frac{7\pi^4}{48}-12\polylog{4}\left(\frac{1}{2}\right)\right\}+\frac{1}{2x_a}\Big\{\ln^2(2)\pi^2 -\ln^4(2) -21\zeta(3)\ln(2)\Big\}+\frac{1}{8x^2_a}\Big\{393\zeta(5)-35\pi^2\zeta(3)\Big\}\nonumber \\
&+\frac{\pi^2}{x_a^2}\polylog{3}\left(-e^{-x_a}\right)+\frac{1}{x^3_a}\Bigg\{\frac{50567\pi^6}{393120}-93\ln(2)\zeta(5)-\frac{9\zeta^2(3)}{2} \Bigg\}+\frac{1}{13x^3_a}\Big\{4\ln^2(2)\pi^4+3\ln^6(2)-3\ln^4(2)\pi^2\Big\}\nonumber \\
&+\frac{1}{13x^3_a}\Bigg\{243\polylog{6}\left(\frac{1}{4}\right)-1728\polylog{6}\left(\frac{1}{2}\right)-16\polylog{6}\left(-\frac{1}{8}\right)\Bigg\}\nonumber+\frac{1}{16x^4_a}\Big\{ 1899\zeta(7)-178\pi^2\zeta(5) \Big\}-\frac{\pi^2}{x_a^3}\polylog{4}\left(-e^{-x_a}\right)\nonumber\\
&+\frac{24\zeta(3)}{x_a^4}\polylog{4}\left(-e^{-x_a}\right) -\frac{6}{x^2_a}\Big\{\polylog{5}\left(-e^{-x_a}\right) + \polylog{4,1}\left(-e^{-x_a},1 \right) \Big\}+\frac{6}{x_a^3}\Big\{2\zeta(3)\polylog{3}\left(-e^{-x_a}\right)+\polylog{2,4}\left(1,-e^{-x_a}\right)\Big\}\nonumber\\
&-\frac{8}{x_a^4}\Big\{\pi^2\polylog{5}\left(-e^{-x_a}\right)+3\polylog{2,5}\left(1,-e^{-x_a}\right)\Big\}+\frac{72}{x^4_a} \Big\{\polylog{7}(-e^{-x_a}) + \polylog{6,1}\left(-e^{-x_a},1\right) \Big\},
\label{eq:FT2_closed-form_compton_full_MZV}
\end{align}
with the multiple polylogarithm $\polylog{n,m}(z_1,z_2)$ defined in Eq.~(\ref{eq:append_polylog}). In the limit $[1-f_2]\simeq 1+O\left(e^{-x_{2}}\right)$, $\hat{\mathcal{F}}_{(T,2)}$ is 
\begin{eqnarray}
\hat{\mathcal{F}}_{(T,2)} &\simeq& \frac{4\alpha^2_{s}T^3}{9}\left[\frac{\zeta(3)}{\pi^4}+\frac{1}{90x_a}\right]+O\left(e^{-x_2}\right).\label{eq:FT2_Compton_no_Pb}
\end{eqnarray} 
The leading log (LL) approximation and Peign\'e \& Peshier approach~\cite{Peigne:2007sd,Peigne:2008nd} yield
\begin{align}
&\hat{\mathcal{F}}^{(LL)}_{(T,2)}\simeq\frac{\alpha^2_s T^3}{18\pi^2}\left(1-\frac{z^2}{2x_a}\right),\quad \\&\hat{\mathcal{F}}^{(PP)}_{(T,2)}\simeq\frac{4\alpha_s^2T^3}{9\pi^4}\left[\zeta(3)+\frac{\pi^4}{90x_a}\right]+O\left(e^{-x_2}\right), \label{eq:FT2_compton_LL_PP}
\end{align}
where the latter is the same as Eq.~(\ref{eq:FT2_Compton_no_Pb}).

\begin{figure}[H]
\centering
\begin{subfigure}[t]{0.48\textwidth}
  \centering
  \includegraphics[width=1\linewidth]{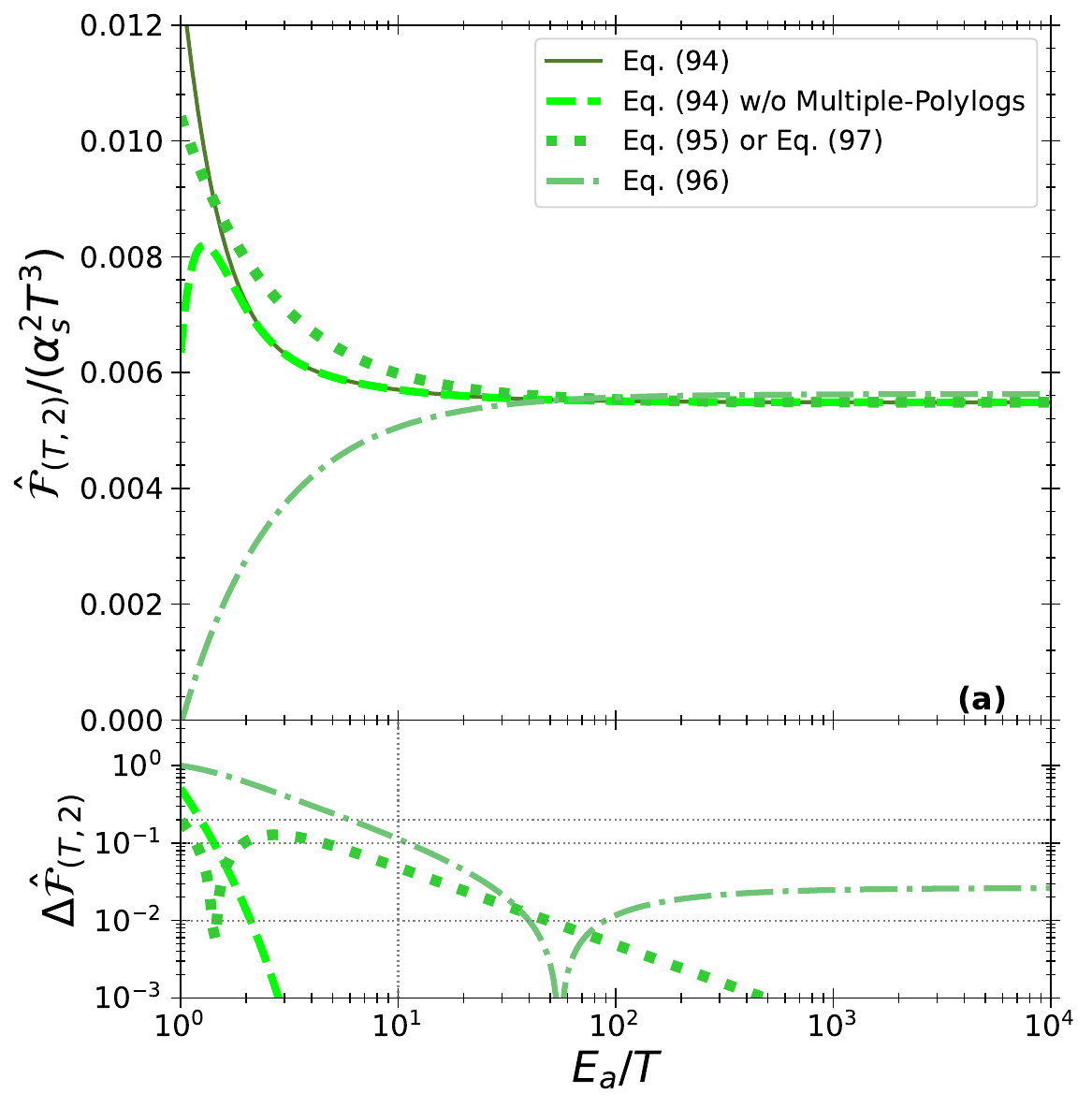}
  \caption{}
  \label{fig: FT2_Compto_1}
\end{subfigure}
~
\begin{subfigure}[t]{0.48\textwidth}
  \centering
  \includegraphics[width=1\linewidth]{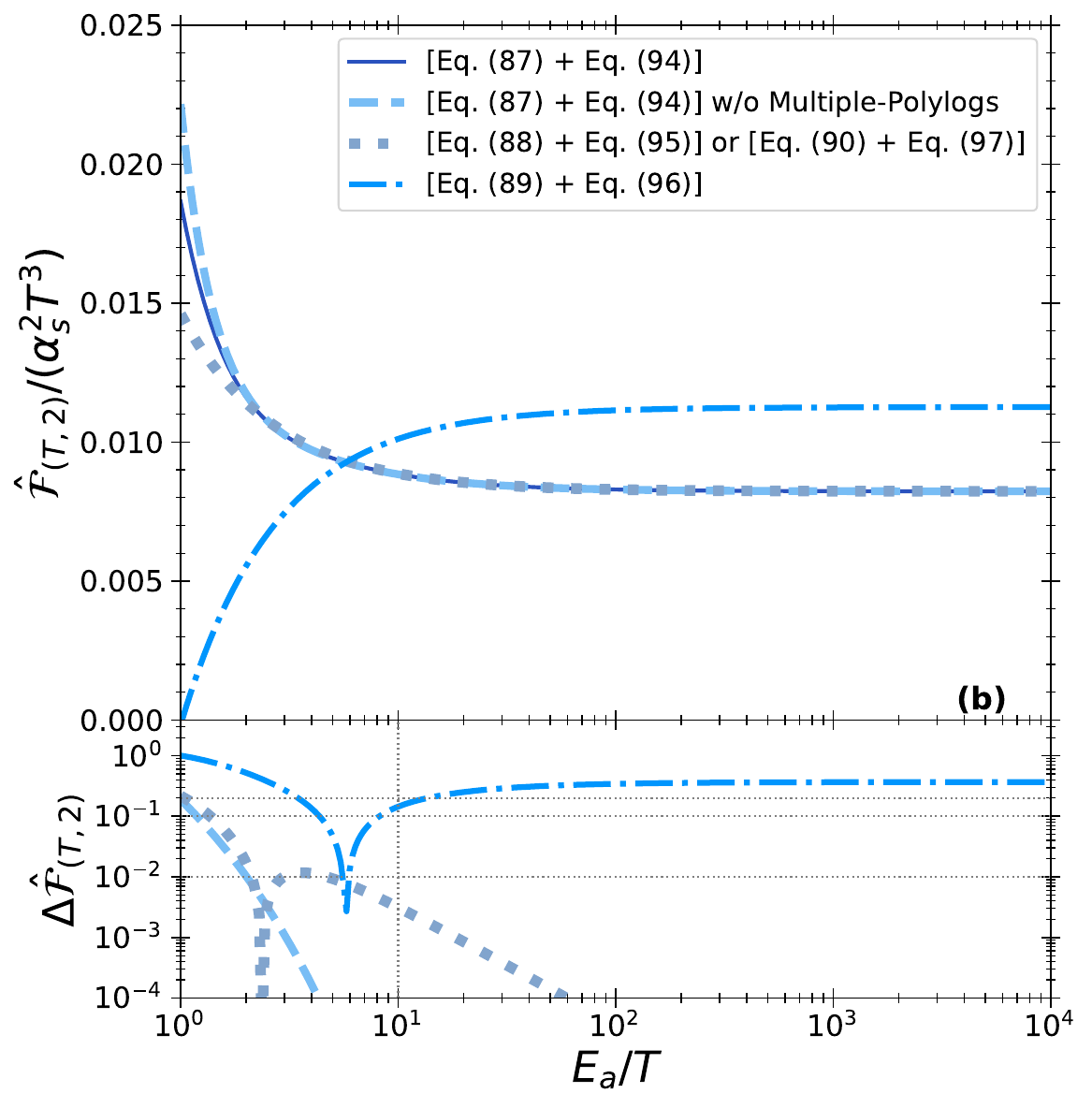}
  \caption{}
  \label{fig: FT2_Full_1}
\end{subfigure}
\caption{(a) Top panel: $\hat{\mathcal{F}}_{(T,2)}/(\alpha_s^2T^3)$ for quark-gluon Compton scattering from various approaches. Bottom panel: relative difference between various approaches, as defined by Eq.~(\ref{eq:DeltaF}). (b) Top panel: comparison of $\hat{\mathcal{F}}_{(T,2)}/(\alpha_s^2T^3)$ for Compton scattering and annihilation together, using different approaches. Bottom panel: same as (a).}
\label{fig:both}
\end{figure}
In Figs.~\ref{fig:FT2_Annihilation} and~\ref{fig: FT2_Compto_1}, the transverse momentum broadening coefficients $\hat{\mathcal{F}}_{(T,2)}$ are presented for both the annihilation and Compton channels, respectively. The Compton scattering channel contribution dominates over the annihilation channel by approximately a factor of two. For $x_a>10$, the inclusion of the Bose-enhancement or Pauli-blocking factors associated with the outgoing parton ``2" does not show significant effects on the final, combined result in Fig. \ref{fig: FT2_Full_1}, which is the sum of the annihilation and Compton scattering processes. While the Peign\'e \& Peshier approach provides an acceptable approximation to $\hat{\mathcal{F}}_{(T,2)}$, simply removing the $\polylog{n,m}(z_1,z_2)$  provides a remarkable approximation to $\hat{\mathcal{F}}_{(T,2)}$ that is also readily usable in jet-medium Monte Carlo simulations.
\section{Temperature dependence of transport coefficients}
\label{sec:Fhat_w_T}
This section presents the temperature dependence of jet-medium transport coefficients that are associated with the fermion-boson conversion processes for jet partons. In Fig.~\ref{fig:AllF-hat_Tdep_E100GeV}, we show the temperature dependence of transport coefficients $\hat{\mathcal{F}}_{0}/T$, $\hat{\mathcal{F}}_{(T,2)}/T^3$ and $\hat{\mathcal{F}}_{(L,1)}/T^2$ as a function of temperature for the temperature range $0.1$ GeV $<T<1$ GeV. The incoming jet quark energy is set to $E_a=100$ GeV, while the strong coupling constant $\alpha_s=0.3$ is fixed to allow for a direct comparison with JETSCAPE extraction of $\hat{q}$ \cite{JETSCAPE:2024cqe}.

The temperature dependence of transport coefficients is driven mainly by the dimensionless ratio $z=m_\infty/T$ via Eq.~(\ref{eq:m_infty}), while $x_a=E_a/T$ induces a modest temperature dependence. This temperature dependence is first shown in Figs.~\ref{fig:AllF-hat_Tdep_E100GeV} and \ref{fig:AllF-hat_Tdep_E10GeV} for a quark jet energy $E_a=100$~GeV and $E_a=10$~GeV, respectively. While the overall trends for $E_a=10$~GeV remain similar to the $E_a=100$~GeV case, the magnitude of $\hat{\mathcal{F}}_{(0)}/T^2$ increases by a factor of $\sim 10$ when the jet energy drops from 100~GeV to 10~GeV. $\hat{\mathcal{F}}_{(0)}/T$, displays the largest sensitivity to $T$, increasing logarithmically with temperature as depicted in Fig.~\ref{fig:f0_Ea_mult}. The transverse momentum broadening coefficient, $\hat{\mathcal{F}}_{(T,2)}/T^3$, exhibits the weakest temperature dependence of the three coefficients, as Eqs.~(\ref{eq:FT2_annihilation_full_MZV}) and (\ref{eq:FT2_closed-form_compton_full_MZV}) are independent of $z$. This results in the relatively smooth and flat curves seen in Figs.~\ref{fig:AllF-hat_Tdep_E100GeV} and \ref{fig:AllF-hat_Tdep_E10GeV}. Conversely, the drag coefficient $\hat{\mathcal{F}}_{(L,1)}/T^2$ demonstrates a stronger temperature dependence owing to its explicit $z$-dependence. These effects of $z$ become even more pronounced at lower jet energies, as shown in Fig.~\ref{fig: FL1_Ea_mult} for $E_a=10$~GeV. Figures~\ref{fig:F0_2Dplot} and \ref{fig:FL1_2Dplot} illustrate the non-trivial evolution of $\hat{\mathcal{F}}_{(0)}$ and $\hat{\mathcal{F}}_{(L,1)}$ as a function of $E_a$ and $T$.
\begin{figure}[H]
\centering
\begin{subfigure}[t]{0.48\textwidth}
\captionsetup[subfigure]{labelformat=empty}
  \centering
  \includegraphics[width=1\linewidth]{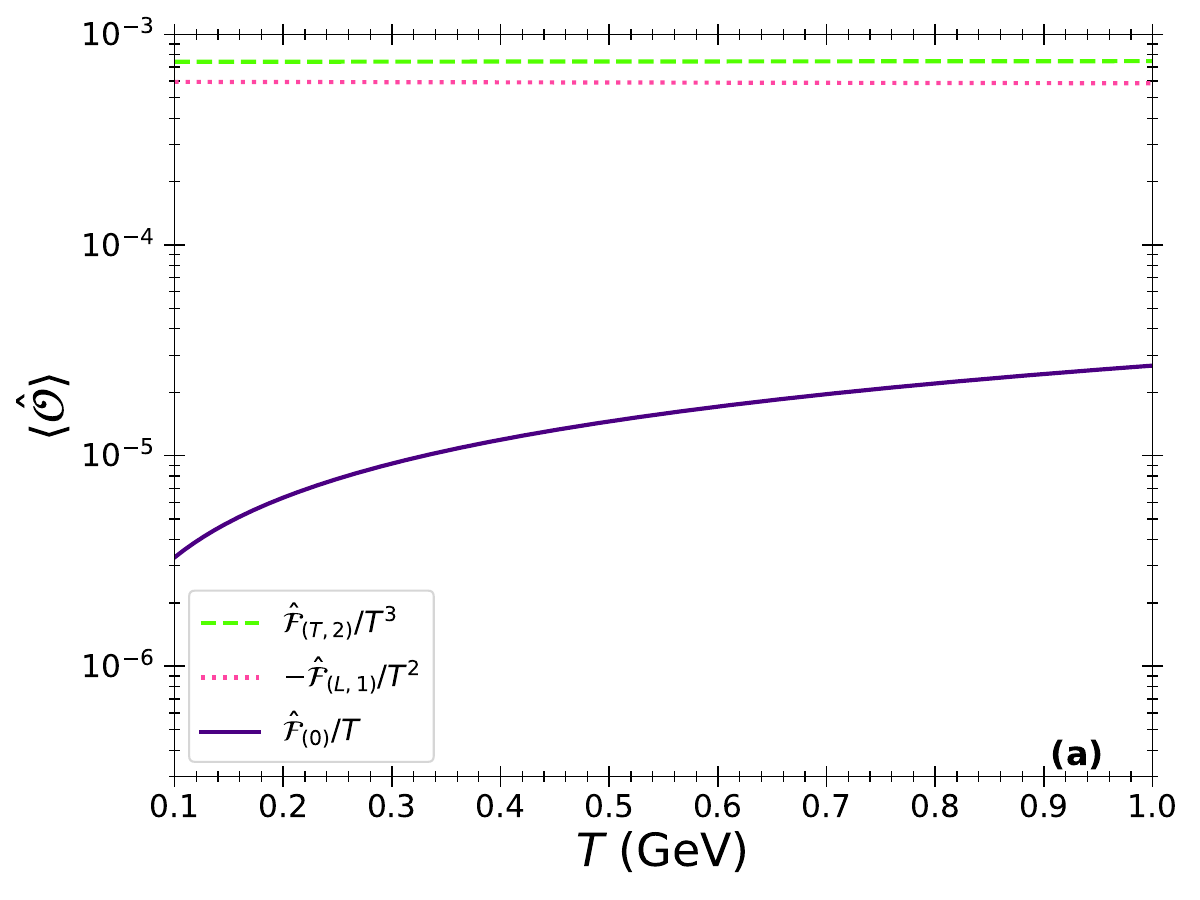}
  \caption{}
  \label{fig:AllF-hat_Tdep_E100GeV}
\end{subfigure}
~
\begin{subfigure}[t]{0.48\textwidth}
  \centering
  \includegraphics[width=1\linewidth]{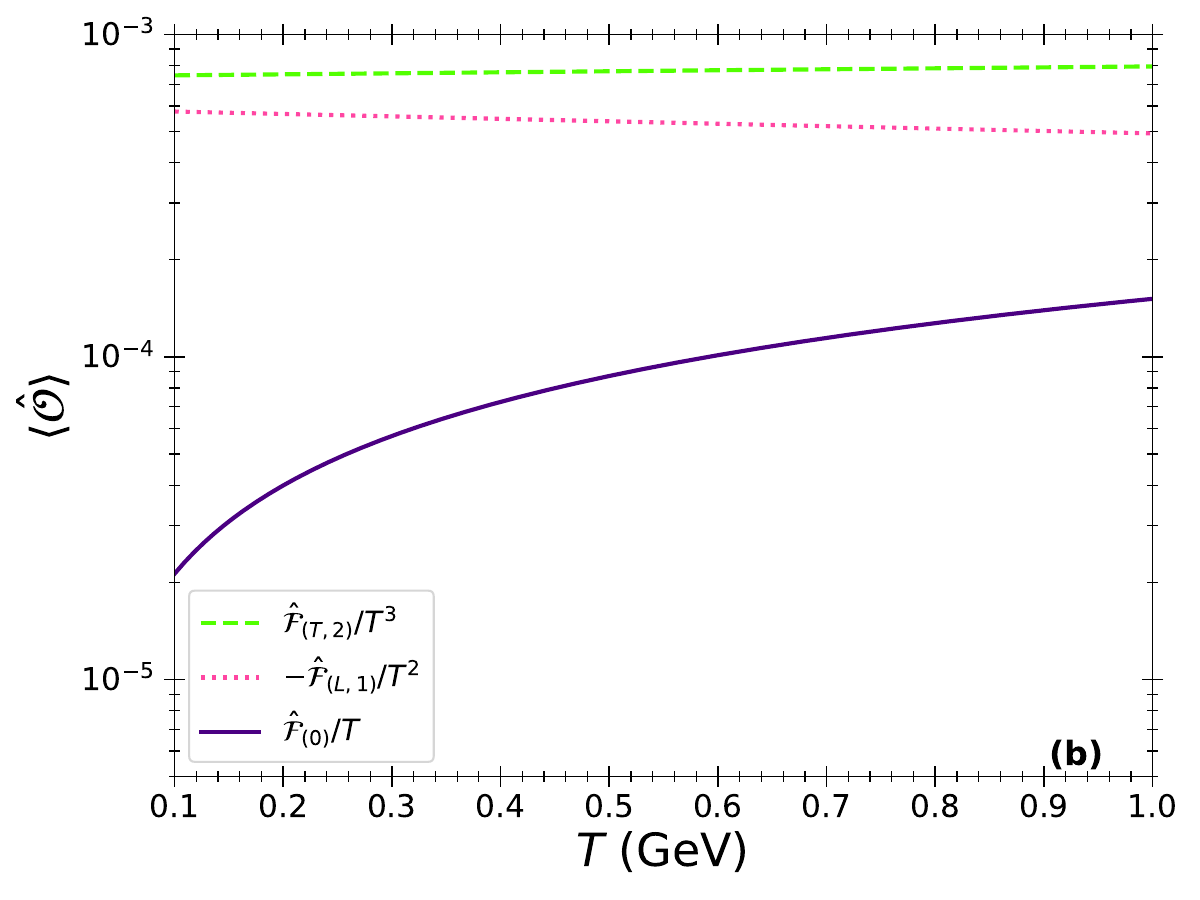}
  \caption{}
  \label{fig:AllF-hat_Tdep_E10GeV}
\end{subfigure}
~
\begin{subfigure}[t]{0.48\textwidth}
  \centering
  \includegraphics[width=1\linewidth]{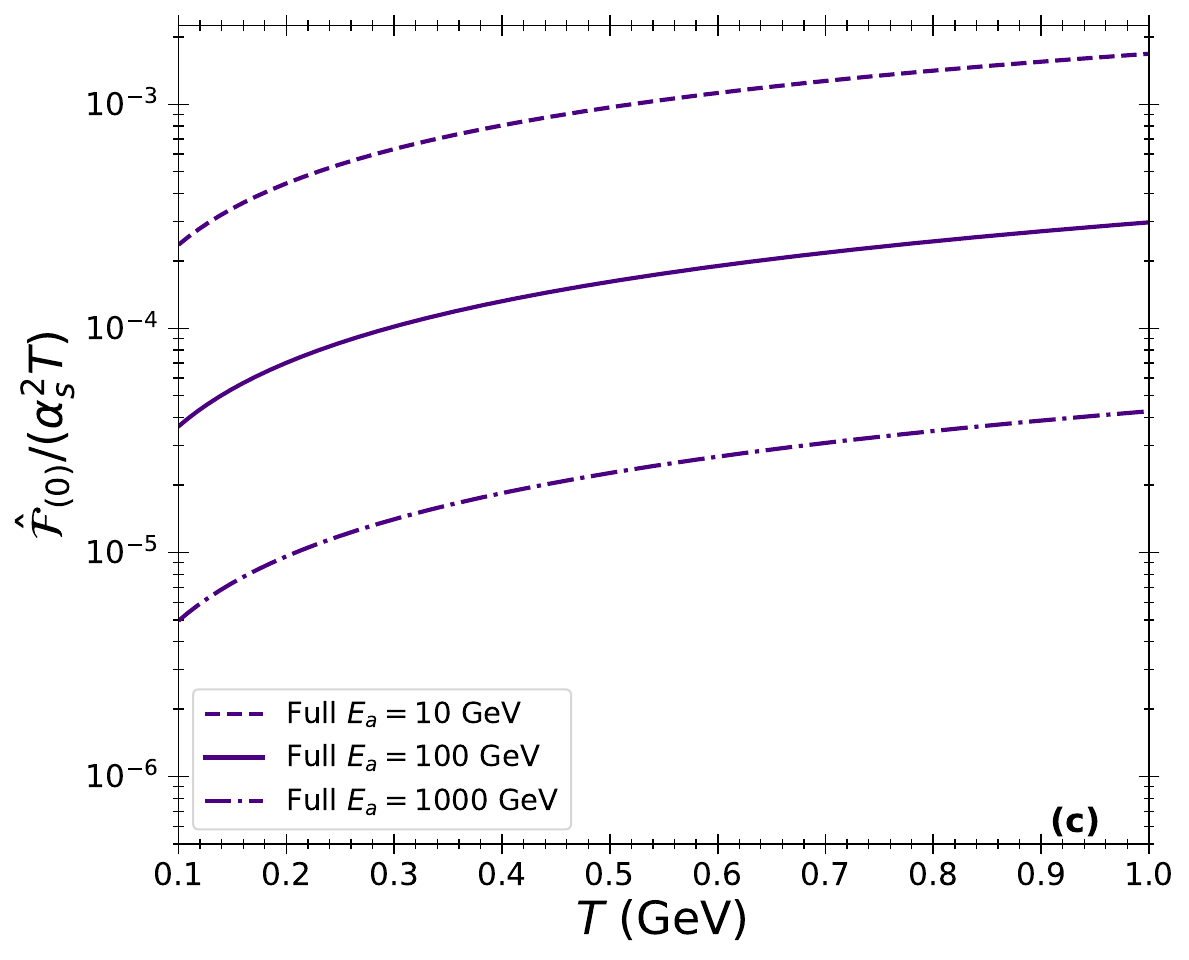}
  \caption{}
  \label{fig:f0_Ea_mult}
\end{subfigure}
~
\begin{subfigure}[t]{0.48\textwidth}
  \centering
  \includegraphics[width=1\linewidth]{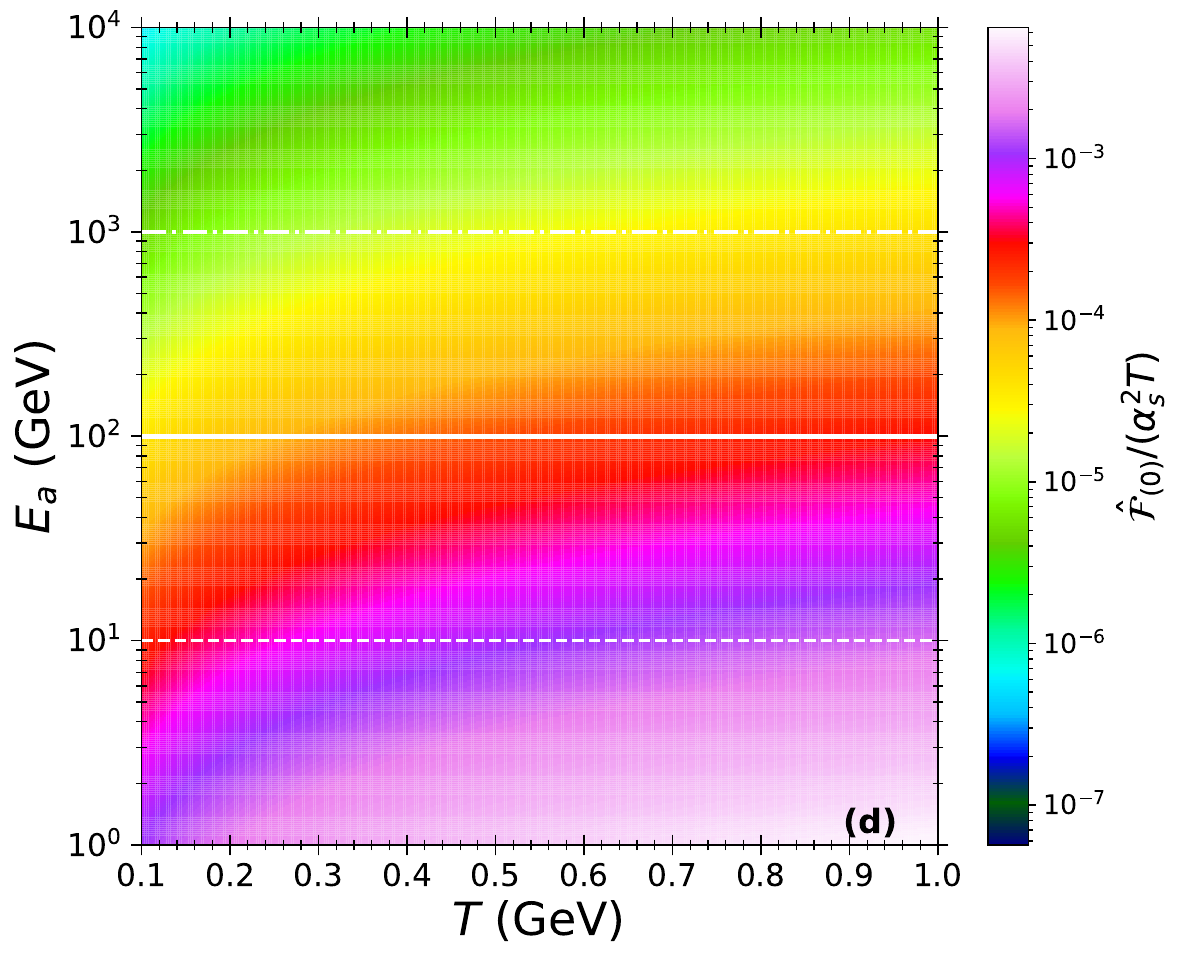}
  \caption{}
  \label{fig:F0_2Dplot}
\end{subfigure}
~
\begin{subfigure}[t]{0.48\textwidth}
  \centering
  \includegraphics[width=1\linewidth]{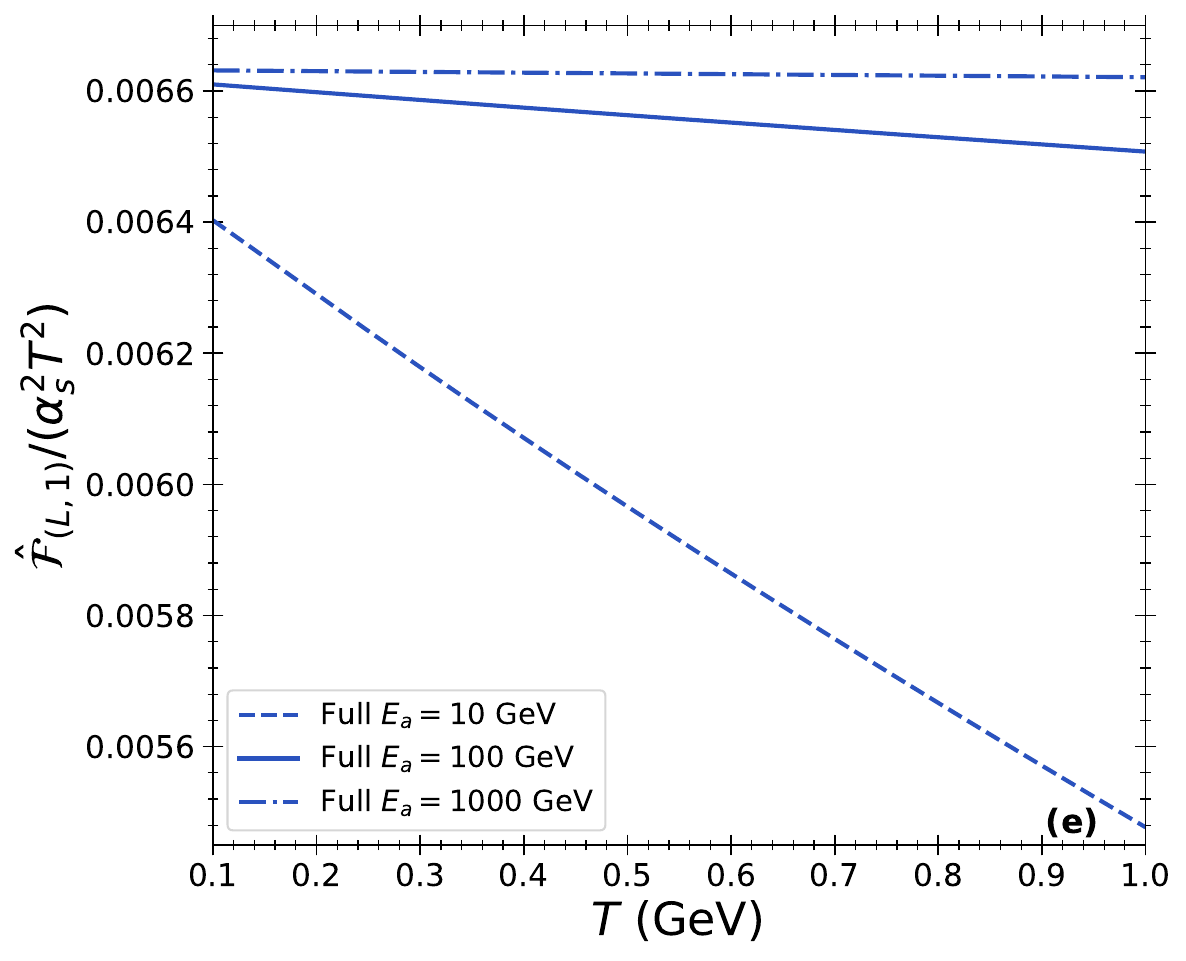}
  \caption{}
  \label{fig: FL1_Ea_mult}
\end{subfigure}
~
\begin{subfigure}[t]{0.48\textwidth}
  \centering
  \includegraphics[width=1\linewidth]{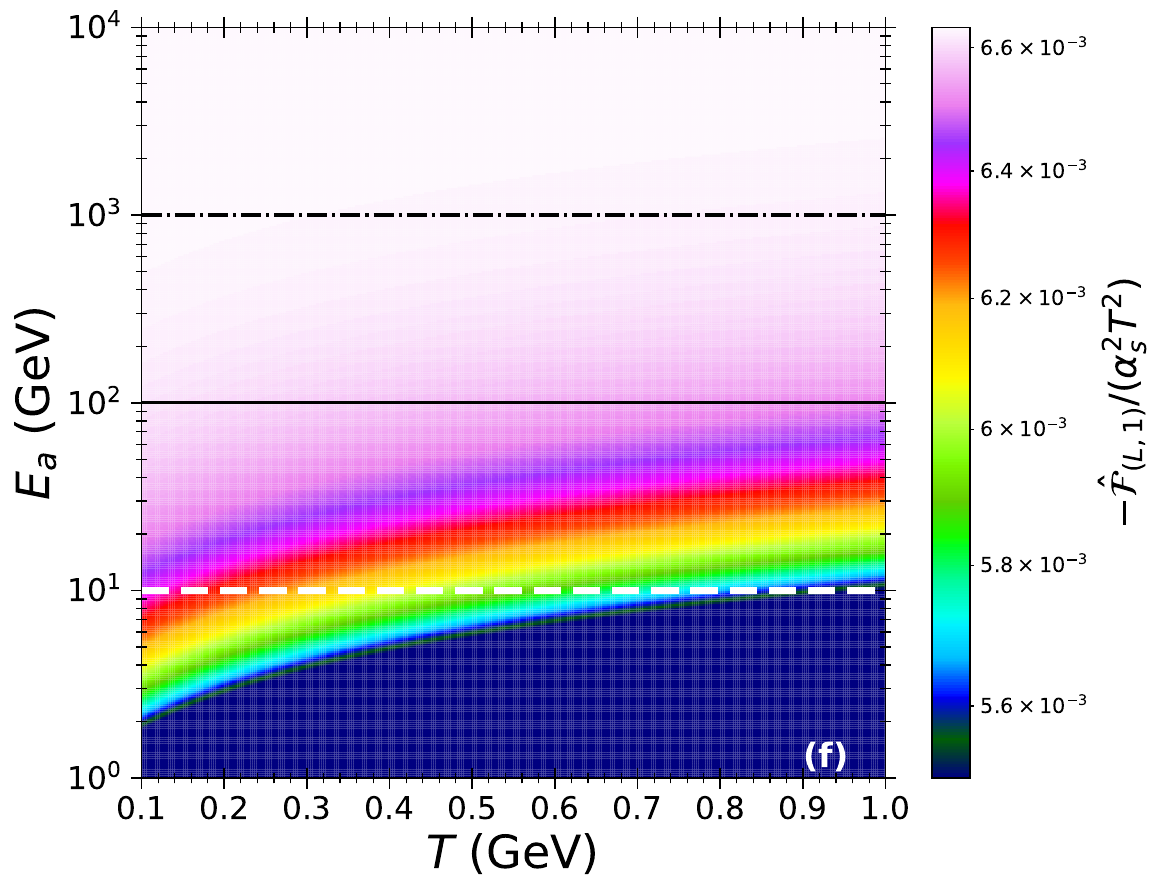}
  \caption{}
  \label{fig:FL1_2Dplot}
\end{subfigure}
\caption{Temperature dependence of jet-medium transport coefficients for quark-to-gluon conversion process at fixed coupling $\alpha_s=0.3$ with (a) $E_a=100$ GeV and (b) $E_a=10$ GeV, (c) $\hat{\mathcal{F}}_{(0)}/(\alpha_s^2T)$ and (e) $\hat{\mathcal{F}}_{(L,1)}/(\alpha_s^2T)$ as a function of $T$ for $E_a=10,100$, and $1000$ GeV. Temperature and energy dependence of (d) $\hat{\mathcal{F}}_{(0)}/(\alpha^2_s T)$ and (f) $\hat{\mathcal{F}}_{(L,1)}/(\alpha^2_s T)$.}
\label{fig:both}
\end{figure}

The calculation presented in Figs.~\ref{fig:AllF-hat_Tdep_E100GeV} and~\ref{fig:AllF-hat_Tdep_E10GeV} indicates that the transport coefficient $\hat{\mathcal{F}}_{(T,2)}/T^3$ is the dominant among the three transport coefficients explored which follow the hierarchy $\hat{\mathcal{F}}_{(T,2)}/T^3 \gg -\hat{\mathcal{F}}_{(L,1)}/T^2 \gg \hat{\mathcal{F}}_{0}/T$. The transverse momentum broadening coefficient $\hat{\mathcal{F}}_{(T,2)}/T^3\approx 7$--$8\times10^{-4}$ for the RHIC and LHC temperature range $T\in[0.2,0.4]$ GeV and quark jet energies $E_a\in[10,100]$ GeV. The longitudinal drag coefficient $\hat{\mathcal{F}}_{(L,1)}/T^2 \approx 6\times 10^{-4}$ and rate coefficient $\hat{\mathcal{F}}_{0}/T \in [3,60]\times10^{-6}$ for the temperature range $T\in[0.2,0.4]$ GeV and quark jet energies $E_a\in[10,100]$ GeV. Also, for a quark jet with 100 GeV energy, the most recent extraction based on the fit to leading hadrons and inclusive jets from the JETSCAPE collaboration constrains $\hat{q}/T^3 \in [4,12]$ (see Ref.~\cite{Ehlers:2024miy} for details).

\end{widetext}

\section{Summary and Outlook}
\label{sec:outlook}
In this paper, a first calculation of the jet-medium transport coefficients associated with the fermionic correlator is presented for a quark jet. A conversion process is considered at leading-order, where the incoming quark traversing a thermal medium exchanges a quark with the medium, resulting in the conversion of the quark into a gluon. The parton drag and diffusion coefficients associated with the conversion process are expressed in terms of two-point fermionic correlators. The transport coefficients are computed, and closed-form expressions are obtained within the tree-level $2\to2$ thermal perturbative QCD picture. A comparison of the full result with the leading-logarithmic approximation and the Peign\'e \& Peshier approach is provided to quantify the theoretical systematic uncertainties associated with the jet-medium transport coefficients. 

Within the higher-twist formalism, the quark jet energy loss at next-to-leading-order was derived in Ref.~\cite{Kumar:2025asj}. The calculation involved computing all possible medium-induced single-emission kernels, where interactions with the medium were modelled using the Glauber quark and gluon. The collinear twist expansion along longitudinal momentum and transverse momentum of kernels associated with the Glauber quark gives rise to a series of fermionic correlators such as $\hat{\mathcal{F}}_{0}$, $\hat{\mathcal{F}}_{(T,2)} $, and $\hat{\mathcal{F}}_{(L,1)}$. Such correlators represent transport coefficients for parton drag and diffusion, expressed as fermionic correlators in light-cone coordinates in Eqs.~(\ref{eq:F_hat_F0}-\ref{eq:F_hat_T2}).

The parton drag and diffusion coefficients associated with the conversion processes are formulated in terms of thermal $2\to2$ QCD scattering and expressed as moments of the tree-level, leading-order $2\to2$ scattering rate. Following the methodology presented in Appendix A of Ref.~\cite{Opitz:2026sgz}, as well as Chapter 14 of Ref.~\cite{Kapusta_Gale_2023}, the expectation value of an observable $\langle \hat{O}\rangle$ is initially formulated as a 9-dimensional integral and subsequently reduced to a 4-dimensional integral, involving two integrations over the energies of the thermal/recoil partons and two integrations over the Mandelstam variables $s$ and $t$. Closed-form expressions are derived for each transport coefficient, while incorporating Bose-enhancement or Pauli-suppression factors is done by numerical integration.

A comparison of the leading-logarithmic (LL) results for $\hat{\mathcal{F}}_0$ and $\hat{\mathcal{F}}_{(L,1)}$ with the corresponding closed-form expressions shows significant deviations over a broad range of kinematic conditions. This indicates that the LL approximation does not provide a quantitatively reliable description of the underlying scattering rates and transport coefficients in the regime considered here. Consequently, the use of LL expressions should be treated with caution, particularly when accurate predictions of parton energy loss transport coefficients are required. Moreover, a comparison with the Peign\'e \& Peshier (PP) approach is also provided, where the agreement with the full calculation is within 20-30$\%$ for $x_a=E_a/T > 10^{2}$. While this approach improves upon the leading log, we have found an even better closed-form approximation of leading-order tree-level scatterings and their jet-medium transport coefficients. In Monte Carlo simulations of parton propagation through a thermal QCD medium, LL approximations are commonly employed for computational efficiency and are often used to model the scattering rates and associated transport coefficients. Our results demonstrate that this approximation can introduce sizable systematic deviations from the full calculation. Incorporating the full closed-form expressions therefore provides a more accurate treatment of the conversion processes and can substantially improve the reliability of future Monte Carlo simulations. This would be highly important for phenomenological studies of jets, where transport coefficients enter as inputs controlling the evolution of the parton shower and medium-modified observables.

Our study indicates that the transport coefficient associated with transverse momentum broadening $\hat{\mathcal{F}}_{(T,2)}$ is the dominant among the three transport coefficients considered. In particular, the coefficients exhibit the hierarchy $\hat{\mathcal{F}}_{(T,2)}/T^3 \gg -\hat{\mathcal{F}}_{(L,1)}/T^2 \gg \hat{\mathcal{F}}_{0}/T$. For temperatures relevant to the RHIC and LHC environments, $T\in[0.2,0.4]$ GeV and quark jet energies $E_a\in[10,100]$ GeV, the transverse momentum broadening coefficient, $\hat{\mathcal{F}}_{(T,2)}/T^3$ is found to be $\approx 7-8\times 10^{-4}$. In the same temperature and energy range, the longitudinal drag coefficient $\hat{\mathcal{F}}_{(L,1)}/T^2 \approx 6\times 10^{-4}$, while the rate coefficient $\hat{\mathcal{F}}_{0}/T \in [3,60]\times 10^{-6} $. This hierarchy highlights that transverse momentum broadening provides the largest contribution to the medium modification to the parton shower in a thermal medium, while the conversion rate coefficient $\hat{\mathcal{F}}_{(0)}$ contributes the smallest among the three coefficients considered. An important and striking result of the study is the behaviour of the energy-loss coefficient, $\langle\Delta E\rangle$, in the high-energy limit. We show that $\langle\Delta E\rangle/(\alpha^2_s T^2)$ remains finite and non-vanishing as the quark energy approaches infinity. Thus, even in the asymptotic high-energy limit, the quark experiences a finite energy loss while propagating through the medium. This result highlights the non-trivial role of in-medium QCD interactions, indicating that the energy dependence of the energy loss coefficient does not simply vanish in the infinite-energy limit.

The medium-induced photon-emission kernel derived in Ref.~\cite{Kumar:2025egh} provides a novel framework for probing flavor hydrodynamization as the thermal medium evolves from the early-time Glasma dynamics towards a hydrodynamically expanding quark-gluon plasma. In particular, photon production induced by Glauber quark interactions with the medium can provide direct sensitivity to the evolution of quark occupation number during the pre-equilibrium stage. Implementing these kernels in a Monte Carlo framework requires a consistent treatment of fermion-to-boson conversion processes and the associated transport coefficients that serve as inputs to the full model calculation. The present study provides a determination of these transport coefficients, setting the stage for a future Monte-Carlo implementation. In a future study, the aim is to incorporate the Glauber-quark induced photon-emission scattering kernels, together with the fermion-to-boson conversion transport coefficients, inside a Monte Carlo event generator. Photon observables can then be used as sensitive probes of the pre-equilibrium dynamics, providing constraints on the evolution of the quark occupation number during the earliest stages of the heavy-ion collisions.

Furthermore, the present study provides quantitative estimates of the theoretical systematic uncertainties associated with the transport coefficients involving the exchange of a Glauber quark. These uncertainties should be incorporated into future Bayesian analyses, enabling a more robust extraction of the properties of the QGP using jet-medium interactions.
\section*{ACKNOWLEDGMENTS}
This work was supported by the Canada Research Chair under Grant Number CRC-2022-00146, the Natural Sciences and Engineering Research Council (NSERC) of Canada under Grant Number SAPIN-2023-00029, and the Canadian Foundation for Innovation John R. Evans Leaders Fund under Grant Number 44100. 
This work was also supported in part by the National Science Foundation (NSF) within the framework of the JETSCAPE collaboration, under Grant No. OAC-2004571 (CSSI:X-SCAPE).
\section*{Data Availability}
No data were created or analyzed in this study. 

\appendix
\begin{appendices}

\section{Thermodynamic integrals}
\label{app:th_int}
The most common integrals involving thermodynamics distributions that do not have a well-known closed form are defined in terms of the functions below
\begin{align}
\begin{split}
\alpha^\pm_n\left(c,\epsilon;x_a,z\right)&\equiv\int^c_\epsilon dx_b\frac{x^n_b\sqrt{\left(x_a-x_b\right)^2+z^2}}{e^{x_b}\pm1}\quad\forall n\in\mathbb{N}_0,\\
A^\pm_n\left(\epsilon;z\right)&=\alpha^\pm_n(\infty,\epsilon;0,z)\equiv\lim_{c\to\infty}[\alpha^\pm_n(c,\epsilon;0,z)],\\
A^\pm_n\left(z\right)&=\lim_{\epsilon\to0^+}\left[A^\pm_n(\epsilon;z)\right],\\
\beta^\pm_n(c,\epsilon;x_a,z)&\equiv\int^c_\epsilon dx_b\frac{x^n_b\arcsinh\left(\frac{x_a-x_b}{z}\right)}{e^{x_b}\pm1}\quad\forall n\in\mathbb{N}_0,\\
-B^\pm_n(\epsilon;z)&=\beta^\pm_n(\infty,\epsilon;0,z)\equiv\lim_{c\to\infty}[\beta^\pm_n(c,\epsilon;0,z)],\\
-B^\pm_n(z)&=\lim_{\epsilon\to0^+}[\beta^\pm_n(\infty,\epsilon;0,z)],\\
\Delta\alpha^\pm_n(\epsilon)&=2\alpha^\pm_n\left(x_a,\epsilon;x_a,z\right)-\alpha^\pm_n\left(\infty,\epsilon;x_a,z\right),\\
\Delta\alpha^\pm_n&=\lim_{\epsilon\to 0^+}\left[\Delta\alpha^\pm_n\left(\epsilon\right)\right],\\
\Delta\beta^\pm_n(\epsilon)&=2\beta^\pm_n\left(x_a,\epsilon;x_a,z\right)-\beta^\pm_n\left(\infty,\epsilon;x_a,z\right),\\
\Delta\beta^\pm_n&=\lim_{\epsilon\to 0^+}\left[\Delta\beta^\pm_n\left(\epsilon\right)\right].
\end{split}
\label{eq:alpha_beta_A_B_fct}
\end{align}
Throughout this paper, we use the above functions to express jet transport coefficients in a closed-form.

\section{Multiple polylogarithms and $\zeta$-functions}
\label{app:MultiLiZeta}
The multiple polylogarithm is defined as the nested series~\cite{Vollinga:2004sn}
\begin{align}
&\polylog{s_1,\dots,s_n}(z_1,\dots,z_n)\equiv\sum_{k_1>\dots>k_n>0}\frac{z_1^{k_1}\dots z_n^{k_n}}{k_1^{s_1}\dots k_n^{s_n}}\label{eq:append_polylog}\\
&=\sum_{k_1=1}^\infty\dots\sum_{k_n=1}^\infty\frac{z_1^{k_1+\dots+k_n}z_2^{k_2+\dots+k_n}\dots z_n^{k_n}}{\left(k_1+\dots+k_n\right)^{s_1}\left(k_2+\dots+k_n\right)^{s_2}\dots k_n^{s_n}},\nonumber
\end{align}
which is used to obtain a closed-form expression for integrals of the form
\begin{align}
    \int_0^\infty \frac{dx_b}{e^{x_b}-1}x_b^n\polylog{s}\left(e^{-x_a-x_b}\right)=n!\polylog{n+1,s}\left(1,e^{-x_a}\right),\nonumber\\
    \forall n\in\mathbb{N}
    \label{eq:MultiLiBose}
\end{align}
for bosons and
\begin{align}
    \int_0^\infty \frac{dx_b}{e^{x_b}+1}x_b^n\polylog{s}\left(e^{-x_a-x_b}\right)=-n!\polylog{n+1,s}\left(-1,-e^{-x_a}\right),\nonumber\\
    \forall n\in\mathbb{N}_0
    \label{eq:MultiLiFermi}
\end{align}
for fermions, where a series expansion of the polylogarithm and the distribution function is used as an intermediate step, before final resummation gives the multiple polylogarithms above. Further details regarding the intermediate steps that lead to the expressions in Eqs.~(\ref{eq:MultiLiBose}) and~(\ref{eq:MultiLiFermi}) have been presented in Ref.~\cite{Opitz:2026sgz}. While the fermionic result in Eq.~(\ref{eq:MultiLiFermi}) is valid for all $n\in\mathbb{N}_0$, for the bosonic case, the pole of the Bose-Einstein distribution function in Eq.~(\ref{eq:MultiLiBose}) poses an issue for $n=0$. In that case, divergences occur in pairs, which should be recombined as
\begin{align}
&\int_0^\infty dx_b\frac{\polylog{s}\left(e^{-x_a}\right)-\polylog{s}\left(e^{-x_a-x_b}\right)}{e^{x_b}-1}\nonumber\\
&=\polylog{s+1}\left(e^{-x_a}\right)+\polylog{s,1}\left(e^{-x_a},1\right).
\label{eq:MultiLiBosePole}
\end{align}
Note that Eqns.~(\ref{eq:MultiLiBose}-\ref{eq:MultiLiBosePole}) are also valid in the case $x_a=0$. 

Special cases of the multiple polylogarithm that are often encountered herein involve arguments $z_k=\pm1$. In such cases, the multiple polylogarithms reduce to multiple $\zeta$-functions, in the same spirit as the regular polylogarithm reduces to the Riemann $\zeta$-function via $\polylog{s}(1)=\zeta(s)$. The case $z_k=-1$ is similarly reducible using the Dirichlet $\eta$-function, where $\eta(s)$ is given by~\cite{Borwein:1995jr}
\begin{align}
    \eta(s)&\equiv-\polylog{s}(-1)=\sum_{k=1}^\infty\frac{(-1)^{k+1}}{k^s}\nonumber\\
    &=\begin{cases}
        \ln(2),\quad&s=1\\
        (1-2^{1-s})\zeta(s),\quad&s\neq1.
    \end{cases}
\end{align}
In the context of this paper, we only encounter double $\zeta$-functions, so there are four possible combinations of the two arguments, namely $z_{1,2}=\pm1$. Employing the common notation where a barred variable $\bar{s}$ implies a negative sign in the argument, leading to an alternating series representation, while non-barred ones use a positive argument (giving a non-alternating series), we have
\begin{align}
\begin{split}
    \zeta(s,t)&\equiv\polylog{s,t}(1,1)=\sum_{k_1=1}^\infty\sum_{k_2=1}^\infty\frac{1}{\left(k_1+k_2\right)^{s}k_2^{t}},\\
    \zeta(s,\bar{t})&\equiv\polylog{s,t}(1,-1)=\sum_{k_1=1}^\infty\sum_{k_2=1}^\infty\frac{(-1)^{k_2}}{\left(k_1+k_2\right)^{s}k_2^{t}},\\
    \zeta(\bar{s},t)&\equiv\polylog{s,t}(-1,1)=\sum_{k_1=1}^\infty\sum_{k_2=1}^\infty\frac{(-1)^{k_1+k_2}}{\left(k_1+k_2\right)^{s}k_2^{t}},\\
    \zeta(\bar{s},\bar{t})&\equiv\polylog{s,t}(-1,-1)=\sum_{k_1=1}^\infty\sum_{k_2=1}^\infty\frac{(-1)^{k_1}}{\left(k_1+k_2\right)^{s}k_2^{t}}.\label{eq:MultiZeta}
\end{split}
\end{align}
This can be generalized to arbitrarily many arguments using the general definition in Eq.~(\ref{eq:append_polylog}). The arguments of the multiple $\zeta$-functions herein are restricted to integer $s,t$ values, giving Multiple Zeta-Values (MZV). Various relations among MZV exist, the most important (and best-known) ones being Euler's reflection relations~\cite{Borwein:1995jr}
\begin{align}
\begin{split}
	\zeta(s,t)+\zeta(t,s)&=\zeta(s)\zeta(t)-\zeta(s+t)\\
	\zeta(\bar{s},\bar{t})+\zeta(\bar{t},\bar{s})&=\eta(s)\eta(t)-\zeta(s+t)\\
	\zeta(s,\bar{t})+\zeta(\bar{t},s)&=\eta(s+t)-\zeta(s)\eta(t)\label{eq:MZV_reflect}
\end{split}
\end{align}
which lets one transform any double zeta-value into its reflection, with arguments switched. Even more useful are reduction formulas in the form of linear integer relations, which reduce one double zeta-value to single polylogarithms $\polylog{n}(z)$ at specific points $z$ (including the simple Riemann-$\zeta$ or the Dirichlet-$\eta$). All of these relations therefore take the form
\begin{align}
    N\zeta(n,m)=\sum_kN_k\prod_j\polylog{l_{j,k}}(z_{j,k}),\label{eq:Zeta_Integer_Relation}
\end{align}
where the $N,N_k$ are integer coefficients, while $n+m=\sum_jl_{j,k}$. The integer relations in Eq.~(\ref{eq:Zeta_Integer_Relation}), as well as corresponding relations for the three alternating (i.e. barred) cases, namely $\zeta(\bar n, m)$, $\zeta(n, \bar m)$, and $\zeta(\bar n,\bar m)$, relate double zeta-values of so-called weight $n+m$ (the sum of its arguments) to a linear combination of products of single polylogarithms, where the total weight of each summand is equal to that of the double zeta-value. The simplest, and best known example, of such a reduction formula, discovered by Euler, is $\zeta(2,1)=\zeta(3)$, as a relation of weight 3.
\begin{widetext}
For specific cases, these integer relations have been proven, the simplest one being the cases $\zeta(s,s)$ and $\zeta(\bar{s},\bar{s})$, both of which follow from the reflection relations in Eq.~(\ref{eq:MZV_reflect}). In addition to that, for odd weights $n+m=2N+1$, one has~\cite{Borwein:1996yq}
\begin{align}
\begin{split}
	\polylog{n,m}(\sigma,\tau)&=\frac{(-1)^n}{2}\left[\binom{2N}{m-1}\polylog{2N+1}(\sigma)+\binom{2N}{b-1}\polylog{2N+1}(\tau)\right]-\frac{\polylog{2N+1}(\sigma\tau)}{2}+\frac{\left[1+(-1)^n\right]\left[1-\delta_1^\tau\delta_1^m\right]}{2}\polylog{n}(\sigma)\polylog{m}(\tau)\\
	&+(-1)^m\sum_{k=1}^{N-1}\left[\binom{2k}{n-1}\polylog{2k+1}(\sigma)+\binom{2k}{m-1}\polylog{2k+1}(\tau)\right]\polylog{2N-2k}(\sigma\tau)+\left[\delta_{-1}^\tau\delta_1^m-\delta_{-1}^\sigma\delta_1^n\right]\ln(2)\polylog{2N}(\sigma\tau)\label{eq:MZV_odd}
\end{split}
\end{align}
where $\sigma,\tau=\pm1$ includes all four cases, given in Eq.~(\ref{eq:MultiZeta}) and $\delta^i_j$ is the Kronecker-$\delta$. The four MZV relations obtained by substituting $\sigma,\tau=\pm1$ in Eq.~(\ref{eq:MZV_odd}) are inverted into Riemann-$\zeta$ functions by solving a system of linear equations of the kind shown in Eq.~(\ref{eq:MZV_reflect}). For double zeta-values of odd weights, the system of linear equations akin to Eq.~(\ref{eq:MZV_reflect}) can be inverted, while for even-weighted double zeta-values, one has more unknowns than equations, and thus no general way has been found to express these even-weighted double zeta-values in terms of single $\zeta$-functions. There still exist special cases of even-weighted double zeta-values for which an inversion formula is known; the simplest weight-4 case in Ref.~\cite{Broadhurst:1996az} is shown below
\begin{align}
    \zeta(\bar{3},\bar{1})=\frac{\pi^4}{180}+\frac{\pi^2\ln^2(2)}{12}-\frac{\ln^4(2)}{12}-2\polylog{4}\left(\frac{1}{2}\right).
\end{align}
The only double zeta-values still in need of simplification in our calculation are alternating double zeta-values of weight 6. To find their integer relation, we employ numerical methods, which were used historically to find many relations between MZVs of the same weight, some of which have been analytically proven thereafter. For the unproven relations, the underlying mathematical structure, as well as numerical evidence, strongly suggests them to be true (details are found in Ref.~\cite{Broadhurst:1996az}, for example).

The specific method used herein is the PSLQ algorithm~\cite{Bailey:1994em} (Partial Sums of squares using Lower triangular Orthogonal (Q) decomposition algorithm), which finds integer coefficients relating a set of irrational numbers, at a given working precision (i.e. 100 digits). Once the relationship is obtained at the 100-digit level, it was checked again at a higher precision (i.e. 1000 digits). The last step of increasing the precision and double-checking that a given result holds ensures that the relations obtained are not just forced at the initial (lower) precision. If a relationship holds solely at a lower precision, then going to a higher precision will not decrease the error accordingly. If, instead, the relation still holds at higher precision (up to round-off errors), then it is highly suggestive that the resulting relation is true, with a negligible probability of a mere coincidence.

We now provide the steps needed to obtain $\hat{\mathcal{F}}_{(T,2)}$ in Eqs.~(\ref{eq:FT2_annihilation_full_MZV}, \ref{eq:FT2_closed-form_compton_full_MZV}). In the annihilation case, one obtains (after integrating over $x_b$) the analytical result
\begin{align} 
&\hat{\mathcal{F}}_{(T,2)}= \frac{4\alpha^2_{s}T^3}{27\pi^4}\Bigg[ \frac{3}{2}\zeta(3)+\frac{1}{x_a}\left\{\frac{7\pi^4}{720}+6\ln(2)\zeta(3)+6\zeta(\bar{2},\bar{2})+6\zeta(\bar{1},\bar{3})\right\}+\frac{1}{x^2_a}\left\{\frac{\pi^2\zeta(3)}{2}-\frac{4\ln(2)\pi^4}{15}-18\zeta(\bar{2},\bar{3})-24\zeta(\bar{1},\bar{4})\right\}\nonumber\\
&+\frac{1}{x^3_a}\left\{48\ln(2)\zeta(5)-\frac{\pi^6}{60}+30\zeta(\bar{2},\bar{4})+48\zeta(\bar{1},\bar{5})\right\}+\frac{1}{x_a^4}\left\{2\pi^2\zeta(5)-24\zeta(\bar{2},\bar{5})-48\zeta(\bar{1},\bar{6})-\frac{16\ln(2)\pi^6}{315}\right\} \nonumber \\
&-\frac{1}{x_a^3}\left\{6\polylog{2,4}\left(-1,-e^{-x_a}\right)+\frac{\pi^2}{2}\polylog{4}\left(e^{-x_a}\right)\right\}+\frac{1}{x_a^4}\Big\{24\polylog{2,5}\left(-1,-e^{-x_a}\right)-2\pi^2\polylog{5}\left(e^{-x_a}\right)\Big\}\nonumber\\
&+\frac{48}{x_a^4}\Big\{\ln(2)\polylog{6}\left(e^{-x_a}\right)+\polylog{1,6}\left(-1,-e^{-x_a}\right)\Big\}\Bigg],\label{eq:FT2_Ann_MultiZeta}
\end{align}

while, in the case of Compton scattering, the result is
\begin{align}
\hat{\mathcal{F}}_{(T,2)} &=\frac{4\alpha^2_{s}T^3}{27\pi^4}\Bigg[3\zeta(3) +\frac{1}{x_a}\left\{\zeta(\bar{3},1)-\frac{\pi^4}{40}-6\zeta(2,\bar{2})\right\}+\frac{3}{x_a^2}\left\{6\zeta(2,\bar{3})+\frac{\pi^2}{3}\polylog{3}\left(-e^{-x_a}\right)-\frac{\pi^2}{2}\zeta(3)\right\}+\frac{30}{x_a^2}\left\{\frac{15\zeta(5)}{16}-\zeta\left(\bar{4},1\right) \right\}\nonumber\\
&-\frac{\pi^2}{x_a^3}\polylog{4}\left(-e^{-x_a}\right)+\frac{1}{x_a^3}\left\{72\zeta(\bar{5},1)-\frac{29\pi^6}{5040}-9\zeta^2(3)-30\zeta(2,\bar{4})\right\}+\frac{1}{x_a^3}\Big\{12\zeta(3)\polylog{3}\left(-e^{-x_a}\right)+6\polylog{2,4}\left(1,-e^{-x_a}\right)\Big\}\nonumber\\
& +\frac{1}{x_a^4}\left\{\frac{7\pi^4\zeta(3)}{30}-\frac{15\pi^2\zeta(5)}{2}+24\zeta(2,\bar{5})\right\}+\frac{24}{x^4_a}\left\{\frac{189\zeta(7)}{64}-3\zeta(\bar{6},1)+\zeta(3)\polylog{4}\left(-e^{-x_a}\right)\right\}\nonumber \\
&-\frac{8}{x_a^4}\Big\{\pi^2\polylog{5}\left(-e^{-x_a}\right)+3\polylog{2,5}\left(1,-e^{-x_a}\right)\Big\}-\frac{6}{x_a^2}\Big\{\polylog{5}\left(-e^{-x_a}\right)+\polylog{4,1}\left(-e^{-x_a},1\right)\Big\}\nonumber\\
&+\frac{72}{x_a^4} \Big\{ \polylog{7}\left(-e^{-x_a}\right) +  \polylog{6,1}\left(-e^{-x_a},1\right)\Big\}\Bigg].
\label{eq:FT2_Compton_MultiZeta}
\end{align}
To be able to simplify both expressions in Eqs.~(\ref{eq:FT2_Ann_MultiZeta}, \ref{eq:FT2_Compton_MultiZeta}), specific linear integer relations for alternating double zeta-values of weight 6 are needed. Employing the PSLQ algorithm, we are able to find the conjectured relations below
\begin{align}
    \zeta(\bar{5},\bar{1})&\overset{?}{=}\frac{19\pi^6}{42120}+\frac{3\zeta^2(3)}{8}+\frac{\pi^4\ln^2(2)}{156}-\frac{\pi^2\ln^4(2)}{208}+\frac{\ln^6(2)}{208}-\frac{36}{13}\polylog{6}\left(\frac{1}{2}\right)+\frac{81}{208}\polylog{6}\left(\frac{1}{4}\right)-\frac{1}{39}\polylog{6}\left(-\frac{1}{8}\right)\label{eq:zeta_5bar}\\
    \zeta(\bar{3},3)&\overset{?}{=}-\frac{4007\pi^6}{262080}+\frac{9\zeta^2(3)}{32}+\frac{93\ln(2)\zeta(5)}{8}-\frac{\pi^4\ln^2(2)}{26}+\frac{3\pi^2\ln^4(2)}{104}-\frac{3\ln^6(2)}{104}+\frac{216}{13}\polylog{6}\left(\frac{1}{2}\right)-\frac{243}{104}\polylog{6}\left(\frac{1}{4}\right)\nonumber\\
    &+\frac{2}{13}\polylog{6}\left(-\frac{1}{8}\right),\label{eq:zeta_3bar}
\end{align}
which hold up to 1000 decimal places, though these relations have not been rigorously proved. These two values complete the set of double zeta-values of weight 6 together with previously derived values, such as~\cite{Broadhurst:1996az}
\begin{align}
    \zeta(\bar{4},\bar{2})=\frac{97\pi^6}{90720}-\frac{3\zeta^2(3)}{4},
\end{align}
and other ones derived from known relations~\cite{Borwein:1995jr,Borwein:1996yq}. Together with the reflection relations from Eq.~(\ref{eq:MZV_reflect}), as well as relations from~\cite{Broadhurst:1996az}, all double zeta-values of weight 6 can be derived. With the results from Eqs.~(\ref{eq:zeta_5bar}, \ref{eq:zeta_3bar}), all double zeta-values from Eqs.~(\ref{eq:FT2_Ann_MultiZeta}, \ref{eq:FT2_Compton_MultiZeta}) now simplify to the expressions in Eqs.~(\ref{eq:FT2_annihilation_full_MZV}, \ref{eq:FT2_closed-form_compton_full_MZV}).
\end{widetext}
\section{Boundaries of integrals over Mandelstam variables}
\label{app:int_over_floor}
In $2\to2$ scattering, energy conservation gives
\begin{align}
	&E_a+E_b=E_1+E_2,\\
    &E_1=E_a+E_b-E_2\label{eq:E_1}
\end{align}
where $E_a$ and $E_b$ are the energies of the incoming particles, whereas the outgoing particle energies are $E_1$ and $E_2$. We define floor variables in terms of $E_a$, $E_b$, $E_1$ and $E_2$ as follows
\begin{align}
	\lfloor E\rfloor\equiv&\min(E_a,E_b,E_1,E_2),\nonumber\\
    \lfloor s\rfloor\equiv&\min(E_aE_b,E_1E_2),\nonumber\\
	\lfloor t\rfloor\equiv&\min(E_aE_1,E_bE_2),\nonumber\\
    \lfloor u\rfloor\equiv&\min(E_aE_2,E_bE_1),\label{eq:all_floors}
\end{align}
where $E_1$ is given by Eq.~(\ref{eq:E_1}). To perform the kinematic integrals for $E_2$ and $E_b$ yielding the rate and transport coefficients, the values of the floor variables are listed in Table~\ref{table:range_floor_var} for different ranges in $E_b$ and $E_2$.
\noindent
\begin{table}[H]
\begin{center}
\begin{tabular}{cc||ccccc}
\toprule

\makecell {Limits on $E_b$} & 
\makecell{Limits on $E_2$} & 
\makecell{\textbf{$\boldsymbol{\lfloor s\rfloor}$}} &
\makecell{\textbf{$\boldsymbol{\lfloor t\rfloor}$}} &
\makecell{\textbf{$\boldsymbol{\lfloor u\rfloor}$}} &
\makecell{\textbf{$\boldsymbol{\lfloor E\rfloor }$}} \\
\midrule

$\displaystyle \int_{0}^{E_a} dE_b$ & %
$\displaystyle \int_{0}^{E_b} dE_2$ &
\makecell{$E_1 E_2$} &
\makecell{$E_b E_2$} &
\makecell{$E_a E_2$} &
\makecell{$E_2$} \\
\midrule

$\displaystyle \int_{0}^{E_a} dE_b$ &
$\displaystyle \int_{E_b}^{E_a} dE_2$ &
\makecell{\textbf{$E_a E_b$}} &
\makecell{$E_b E_2$ } &
\makecell{$E_b E_1$} &
\makecell{\textbf{$E_b$}} \\
\midrule

$\displaystyle \int_{0}^{E_a} dE_b$ & 
$\displaystyle \int_{E_a}^{E_a+E_b} dE_2$ &
\makecell{$E_1 E_2$} &
\makecell{$E_a E_1$} &
\makecell{$E_b E_1$} &
\makecell{$E_1$} \\
\bottomrule

$\displaystyle \int_{E_a}^{\infty} dE_b$ & %
$\displaystyle \int_{0}^{E_a} dE_2$ &
\makecell{$E_1 E_2$} &
\makecell{$E_b E_2$} &
\makecell{$E_a E_2$} &
\makecell{$E_2$} \\
\midrule

$\displaystyle \int_{E_a}^{\infty} dE_b$ &
$\displaystyle \int_{E_a}^{E_b} dE_2$ &
\makecell{\textbf{$E_a E_b$}} &
\makecell{$E_a E_1$ } &
\makecell{$E_a E_2$} &
\makecell{\textbf{$E_a$}} \\
\midrule

$\displaystyle \int_{E_a}^{\infty} dE_b$ & 
$\displaystyle \int_{E_b}^{E_a+E_b} dE_2$ &
\makecell{$E_1 E_2$} &
\makecell{$E_a E_1$} &
\makecell{$E_b E_1$} &
\makecell{$E_1$} \\
\bottomrule
\end{tabular}
\end{center}
\caption{Value of floor variables based on the kinematic range of $E_b$ and $E_2$ in $a+b\rightarrow 1+2$ scattering. Once again, $E_1$ is defined in Eq.~(\ref{eq:E_1}).}
\label{table:range_floor_var}
\end{table}

\section{Compton scattering and annihilation rates for the quark jet}\label{app:compt_annih_rates}
Given that Ref.~\cite{Kumar:2025egh} discusses photon production, the photon production rate and associated transport coefficients for a quark jet traversing QGP are presented herein, for tree-level leading-order $2\to2$ scattering.

\begin{figure}[H]
\centering
\begin{subfigure}[t]{0.2\textwidth}
  \centering
  \includegraphics[width=1.25\linewidth]{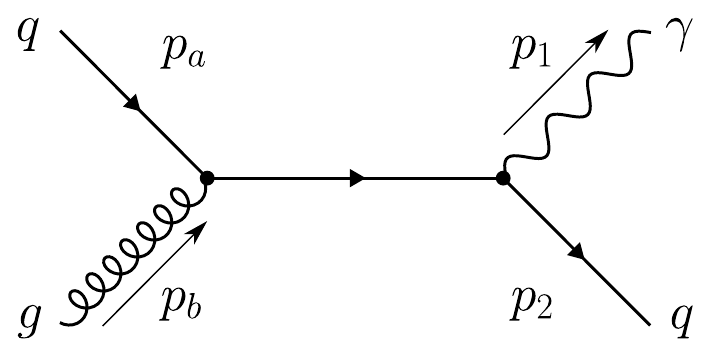}
  \caption{(a) $s$-channel}
  \label{fig:comp-s}
\end{subfigure}
~
\begin{subfigure}[t]{0.2\textwidth}
  \centering
  \includegraphics[width=0.7\linewidth]{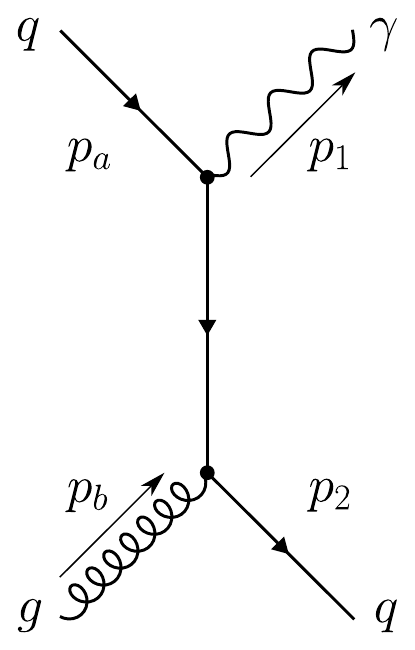}
  \caption{(b) $t$-channel}
  \label{fig:comp-t}
\end{subfigure}
\caption{Photon production via Compton scattering for a quark jet.}
\label{fig:compton diagram}
\end{figure}

\begin{figure}[H]
\centering
\begin{subfigure}[t]{0.2\textwidth}
  \centering
  \includegraphics[width=0.8\linewidth]{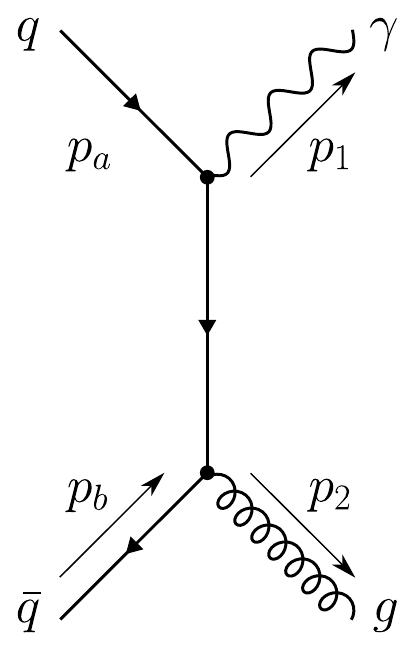}
  \caption{(a) $t$-channel}
  \label{fig:ann-t}
\end{subfigure}
~
\begin{subfigure}[t]{0.2\textwidth}
  \centering
  \includegraphics[width=0.8\linewidth]{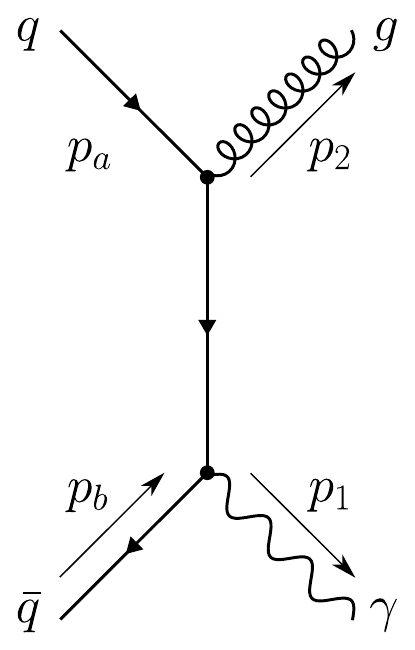}
  \caption{(b) $u$-channel}
  \label{fig:ann-u}
\end{subfigure}
\caption{Photon production via quark-antiquark annihilation.}
\label{fig:Annihilation diagram}
\end{figure}

\subsection{Compton scattering}
For $s$-channel Compton scattering, the photon emission rate is given by
\begin{align}
\frac{d^3R_s}{d^3p_a}&=\frac{N_q\alpha_s\alpha_{\mathrm{em}}T}{6\pi^4}\int_0^\infty\frac{dx_b}{e^{x_b}-1} \nonumber \\
& \times \left[\frac{x_b}{2x_a}-\frac{\pi^2x_b}{12x_a^2}+\frac{\polylog{2}(-e^{-x_a-x_b})}{x_a(x_a+x_b)} \right. \nonumber \\
& \left. \quad -\frac{\polylog{2}(-e^{-x_a})}{x_a^2}\right],
\label{eq:s_rate_compton_xb}
\end{align}
where $\alpha_{\rm em}$ is the fine structure constant with $N_q$ being the square of the fractional charge associated with the incoming jet quark. 
Equation~(\ref{eq:s_rate_compton_xb}) can be further simplified as
\begin{align}
\frac{d^3R_s}{d^3p_a}&=\frac{N_q\alpha_s\alpha_{\mathrm{em}}T}{6\pi^4} \left[\frac{\pi^2}{12}\left\{ \frac{1}{x_a}-\frac{\pi^2}{6x^2_a}\right\} \right. \\
& \left. + \int_0^\infty\frac{dx_b}{e^{x_b}-1}\left\{\frac{\polylog{2}\left(-e^{-x_a-x_b}\right)}{x_a\left(x_a+x_b\right)}-\frac{\polylog{2}\left(-e^{-x_a}\right)}{x_a^2}\right\}\right]. \nonumber 
\end{align}
Taking the first term of the geometric series $[1-f_2]\simeq 1+O\left(e^{-x_2}\right)$, the photon emission rate for $s$-channel is given by
\begin{align}
    \frac{d^3R_s}{d^3p_a}&\simeq\frac{N_q\alpha_s\alpha_{\mathrm{em}}T}{72\pi^2x_a}+O\left(e^{-x_2}\right).
\end{align}   
Similarly, the scattering rate for photon production via $t$-channel, using the first term of geometric series $[1-f_2]\simeq 1+O\left(e^{-x_2}\right)$, yields  
\begin{align}
\frac{d^3R_t}{d^3p_a}&\simeq\frac{N_q\alpha_s\alpha_{\mathrm{em}}T}{6\pi^4}\int_0^\infty\frac{dx_b}{e^{x_b}-1}\Bigg[\left(\frac{1}{2x_a}-\frac{x_b}{x_a^2}\right)\sqrt{x_a^2+z^2} \nonumber \\
&  +\left(\frac{x_b}{2x_a^2}-\frac{1}{x_a}\right)\sqrt{x_b^2+z^2} \nonumber \\ 
& +\left(\frac{x_b}{x_a}+\frac{z^2}{2x_a^2}\right) \left(\arcsinh\left(\frac{x_a}{z}\right)+\arcsinh\left(\frac{x_b}{z}\right)\right) \nonumber \\
& +\lvert x_a-x_b\rvert \left(\frac{z}{x_a^2}-\frac{1}{2x_a^2}\sqrt{\left(x_a-x_b\right)^2+z^2}\right) \nonumber \\
& -\sgn\left(x_a-x_b\right)\frac{z^2}{2x_a^2}\arcsinh\left(\frac{x_a-x_b}{z}\right) \Bigg]+O\left(e^{-x_2}\right).
\label{eq:t_rate_compton_photon_xb}
\end{align}
where $\sgn(x)\equiv2\Theta(x)-1$, while $\Theta(x)$ is the Heaviside function. The integrals in Eq.~(\ref{eq:t_rate_compton_photon_xb}) are identical to Eq.~(\ref{eq:F0_compton_xb}), owing to the fact that the photon emission diagram in Fig.~\ref{fig:compton diagram} is related to the gluon emission in Fig.~\ref{fig:fhat_diagrams}(c) via color and coupling factors being replaced as $N_q \alpha_{\rm em}\rightarrow 4\alpha_s /3$. Therefore, the closed-form expression is given by Eq.~(\ref{eq:F0_Compton_wo_Be_Pb}), provided that $(4\alpha_s/3)\rightarrow N_q \alpha_{\rm em}$.

Similarly, $\hat{\mathcal{F}}_{(T,2)}$ and $\hat{\mathcal{F}}_{(L,1)}$ for photon emission diagrams in Fig.~\ref{fig:compton diagram} are related to the gluon emission in Fig.~\ref{fig:fhat_diagrams}(c) via color and coupling factor being replaced as $N_q \alpha_{\rm em}\rightarrow 4\alpha_s /3$. Therefore, the closed-form expression is given by Eq.~(\ref{eq:FT2_closed-form_compton_full_MZV}) and Eq.~(\ref{eq:FL1_wo_pb}), provided that one replaces $4\alpha_s/3$ with $N_q \alpha_{\rm em}$.

\subsection{Quark-antiquark annihilation process}
The scattering rate for $t$-channel and $u$-channel photon production, without the Bose-enhancement factor, are given as 
\begin{align}
   \frac{d^3R_{t}}{d^3p_a} & =\frac{d^3R_{u}}{d^3p_a} \nonumber \\ &\simeq\frac{N_q\alpha_s\alpha_{\mathrm{em}}T}{6\pi^4}\int_0^\infty\frac{dx_b}{e^{x_b}+1} \nonumber \\
    & \times \left[ 
     \frac{z(x_a+x_b)}{x_a^2}-\frac{x_b\sqrt{x_a^2+z^2}}{x_a^2}-\frac{\sqrt{x_b^2+z^2}}{x_a} \right. \nonumber \\
    & + \left. \frac{x_b}{x_a}\left\{\arcsinh\left(\frac{x_a}{z}\right)+\arcsinh\left(\frac{x_b} {z}\right)\right\}\right]+O\left(e^{-x_2}\right).
\label{eq:rate_photon_annihilation_xb}
\end{align}
Using the auxiliary functions defined in Appendix \ref{app:th_int}, Eq.~(\ref{eq:rate_photon_annihilation_xb}) can be expressed in closed-form as
\begin{align}
  \frac{d^3R_t}{d^3p_a} & =\frac{d^3R_u}{d^3p_a} \nonumber \\ &\simeq\frac{N_q\alpha_s\alpha_{\mathrm{em}}T}{6\pi^4} \Bigg[ \frac{B^+_1(z)}{x_a} + \frac{\pi^2}{12x_a}\operatorname{arcsinh}\left(\frac{x_a}{z}\right)\nonumber\\
& - \frac{A^+_0(z)}{x_a} - \frac{\pi^2}{12x^2_a}\sqrt{x^2_a+z^2} \Bigg] +O\left(e^{-x_2}\right).
\label{eq:rate_photon_annihilation}
\end{align}
Note that $\hat{\mathcal{F}}_{(T,2)}$ and $\hat{\mathcal{F}}_{(L,1)}$  for photon emission diagrams in Fig.~\ref{fig:Annihilation diagram} are related to the gluon emission in Fig.~\ref{fig:fhat_diagrams}(b) via color and coupling factor being replaced as $N_q \alpha_{\rm em}\rightarrow 4\alpha_s /3$.

\end{appendices}


\bibliography{references}
\end{document}